\documentclass[fleqn,usenatbib]{mnras}

\usepackage{newtxtext,newtxmath}
\usepackage{caption}
\usepackage{subcaption}
\usepackage{multicol}
\usepackage{multirow}

\usepackage[T1]{fontenc}

\DeclareRobustCommand{\VAN}[3]{#2}
\let\VANthebibliography\thebibliography
\def\thebibliography{\DeclareRobustCommand{\VAN}[3]{##3}\VANthebibliography}

\usepackage{graphicx}	
\usepackage{amsmath}	

\usepackage{scalerel,tikz}
\usetikzlibrary{svg.path}
\definecolor{orcidlogocol}{HTML}{A6CE39}
\tikzset{orcidlogo/.pic={\fill[orcidlogocol] svg{M256,128c0,70.7-57.3,128-128,128C57.3,256,0,198.7,0,128C0,57.3,57.3,0,128,0C198.7,0,256,57.3,256,128z}; \fill[white] svg{M86.3,186.2H70.9V79.1h15.4v48.4V186.2z} svg{M108.9,79.1h41.6c39.6,0,57,28.3,57,53.6c0,27.5-21.5,53.6-56.8,53.6h-41.8V79.1z M124.3,172.4h24.5c34.9,0,42.9-26.5,42.9-39.7c0-21.5-13.7-39.7-43.7-39.7h-23.7V172.4z} svg{M88.7,56.8c0,5.5-4.5,10.1-10.1,10.1c-5.6,0-10.1-4.6-10.1-10.1c0-5.6,4.5-10.1,10.1-10.1C84.2,46.7,88.7,51.3,88.7,56.8z};}}
\newcommand\orcidicon[1]{\href{https://orcid.org/#1}{\mbox{\scalerel*{
\begin{tikzpicture}[yscale=-1,transform shape]\pic{orcidlogo};
\end{tikzpicture}}{|}}}}

\usepackage{newtxtext,newtxmath}
\DeclareSymbolFont{CMletters}{OML}{cmm}{m}{it}
\DeclareMathSymbol{\nu}{\mathord}{CMletters}{23}

\title[]{Lagrangian statistics of a shock-driven turbulent dynamo in decaying turbulence}

\author[J.K.J.~Hew \& C. Federrath]{
Justin Kin Jun Hew$^{\orcidicon{0000-0002-5238-6115}\,1,2}$\thanks{E-mail: u7322062@anu.edu.au},
Christoph Federrath$^{\orcidicon{0000-0002-0706-2306}\,1,3}$\thanks{E-mail: christoph.federrath@anu.edu.au},
\\
$^{1}$Research School of Astronomy and Astrophysics, Australian National University, Canberra, ACT~2611, Australia\\
$^{2}$Space Plasma Power and Propulsion Laboratory, Research School of Physics, Australian National University, Canberra, ACT~2601, Australia\\
$^{3}$Australian Research Council Centre of Excellence in All Sky Astrophysics (ASTRO3D), Canberra, ACT 2611, Australia
}

\date{Accepted XXX. Received YYY; in original form ZZZ}

\pubyear{2022}

\begin{document}
\label{firstpage}
\pagerange{\pageref{firstpage}--\pageref{lastpage}}
\maketitle

\begin{abstract}
Small-scale fluctuating magnetic fields of order $n$G are observed in supernova shocks and galaxy clusters, where its amplification is likely caused by the Biermann battery mechanism. However, these fields cannot be amplified further without the turbulent dynamo, which generates magnetic energy through the stretch-twist-fold (STF) mechanism. Thus, we present here novel three-dimensional magnetohydrodynamic (MHD) simulations of a laser-driven shock propagating into a stratified, multiphase medium, to investigate the post-shock turbulent magnetic field amplification via the turbulent dynamo. The configuration used here is currently being tested in the shock tunnel at the National Ignition Facility (NIF). In order to probe the statistical properties of the post-shock turbulent region, we use $384 \times 512 \times 384$ tracers to track its evolution through the Lagrangian framework, thus providing a high-fidelity analysis of the shocked medium. Our simulations indicate that the growth of the magnetic field, which accompanies the near-Saffman kinetic energy decay ($E_{\textrm{kin}} \propto t^{-1.15})$ without turbulence driving, exhibits slightly different characteristics as compared to periodic box simulations. Seemingly no distinct phases exist in its evolution, because the shock passage and time to observe the magnetic field amplification during the turbulence decay are very short ($\sim\!0.3$ of a turbulent turnover time). Yet, the growth rate is still consistent with those expected for compressive (curl-free) turbulence driving in subsonic, compressible turbulence. Phenomenological understanding of the dynamics of the magnetic and velocity fields are also elucidated via Lagrangian frequency spectra, which are consistent with the expected inertial range scalings in the Eulerian-Lagrangian bridge.
\end{abstract}




\section{Introduction}

Astrophysical gas flows in the interstellar medium (ISM) are often highly stratified and weakly magnetised \citep{zeldovich1983magnetic,tobias2002solar}, with fields of the order of $n$G to $10^2\mu$G, extending over large coherence length scales of the order of several kilo parsecs \citep{brandenburg1996magnetic,brandenburg2005astrophysical}. It is in these, often shock-dominated, 
compressible flows that the small-scale magnetohydrodynamic (MHD) turbulent dynamos can exist \citep{schober2012small,schleicher2013small,federrath2014turbulent,federrath2016magnetic,seta2022turbulent}, where small seed turbulent magnetic fields amplify into much larger ones in the presence of vorticity and turbulent fluctuations, which excites the field intermittently and sustains it by converting kinetic energy into magnetic energy \citep{batchelor1950spontaneous,mac2004control,federrath2011mach,brandenburg2018advances,AchikanathEtAl2021,seta2021saturation,KrielEtAl20221}. 

The primary effect of turbulence and anisotropy production is the amplification of the turbulent field through the transport terms in the MHD equations, which are governed by two dimensionless numbers called the magnetic Reynolds number $\text{Rm}_{\ell}$ and the hydrodynamic Reynolds number $\text{Re}_{\ell}$. These control the action of the magnetic field through the characteristic scales of turbulence, where $\ell$ is the characteristic length scale. This defines $\text{Re}_\ell = v\ell/\nu$, where $v$ is the turbulent velocity and $\nu$ is the kinematic viscosity. The turbulent magnetic resistivity $\eta$ defines the magnetic Reynolds number as $\text{Rm}_\ell = v\ell/\eta$ \citep{yokoi2013cross}. This further introduces the quantity called the magnetic Prandtl number, which is $\textrm{Pm}_\ell = \textrm{Rm}_\ell/\textrm{Re}_\ell$. Oftentimes in astrophysical flows, $\textrm{Re}_\ell$ and $\textrm{Rm}_\ell$ are very large and $\textrm{Pm}_\ell > 1$, leading to generation of large-scale vorticity, thus permitting the exponential amplification of a turbulent magnetic field, $B = B_0 \exp({\Gamma t})$, where $\Gamma$ is the growth rate, from below the viscous scale ($k_{\nu}$) to the resistive scale ($k_{\eta}$), such that $k _{\nu} < k < k_{\eta}$, but only up until the equipartition scale, $k \sim k_{\textrm{eq}}$, where the conversion between magnetic and kinetic energy slows down and the turbulent dynamo saturates \citep{schekochihin2002model}.

While substantial work has been done on the small-scale turbulent dynamo (SSD) through periodic box simulations, there are only a number of studies on this process in the context of post-shock turbulence.  The latter has been a subject of only a few numerical \citep{balsara2004amplification,vladimirov2006nonlinear,inoue2009turbulence,drury2012turbulent,2014downes,donnert2018magnetic,hu2022turbulent} and experimental studies \citep{sarma2002magnetic,meinecke2014turbulent,sano2021}. Some of these have been focussed on the amplification by shock compression and pre-shock pressure gradients only or on examining mixed pre- and post-shock turbulent media \citep{inoue2009turbulence,del2016turbulence,bohdan2021magnetic}, where the corrugated shock front interacts with density inhomogeneities \citep{giacalone2007magnetic,beresnyak2009turbulence}, inducing vorticity and turbulence transport enhancement. In most cases considered, two-dimensional (2D) numerical simulations were conducted with strong shock profiles emulating supernova blast and detonation waves, or heliospheric termination shocks, where magnetic flux lines are rapidly compressed and stretched, yielding orders of magnitudes of shock-induced amplification. For shock-driven turbulence, it has been suggested that the small-scale dynamo process likely contributed significantly to these amplifications \citep{mac2005distribution,federrath2014turbulent,federrath2016magnetic,mckee2020magnetic}. However, its impact is likely masked by the contribution from rapid shock compression \citep{balsara2004amplification,kim2006amplification}.

Moreover, we expect that two-dimensional numerical simulations conducted in prior works can significantly differ from their three-dimensional counterparts, since the development of three-dimensional coherent structures is not possible in the former, due to the topological constraints imposed in two-dimensional geometry. These have shown to play a crucial role in the turbulent dynamo process within post-shock turbulence \citep{inoue2013origin,2014downes,ji2016efficiency,hu2022turbulent} since purely 2D flows are unable to excite a dynamo according to \cite{zeldovich1957magnetic}'s anti-dynamo theorem.

Thus, motivated by the lack of studies in this particular area, we here propose to investigate the post-shock turbulent medium through the Lagrangian framework by studying the evolution of tracer trajectories in the moving volume behind a laser-driven shock front. This allows thorough analyses of the dynamical evolution of the turbulent dynamo in relation to its associated time scales, since the tracer trajectories follow the advected (co-moving) fluid parcels via \textit{streamlines}; thus providing a high-fidelity approach to studying the filamentary structures that compress or stretch the magnetic field lines in the flow, while avoiding amplifications caused directly by the shock front, or by stratified shear instabilities \citep{sano2012magnetic}. Such methods of injecting Lagrangian tracers have been applied by \citet{konstandin2012statistical} to establish the Lagrangian statistics of supersonic ISM turbulence with mixed solenoidal and compressive turbulence driving, and by \citet{homann2007lagrangian} and \citet{busse2010lagrangian} to the study of the Lagrangian structure functions and frequency spectra scalings in MHD turbulence. Lagrangian statistics for the Taylor-Green forced dynamo was also studied by \citet{homann2014structures}, where time evolution of the magnetic field was educed through the material frame with mass-averaged quantities, providing insight into the time scales of its evolution through a volume that is unaffected by advection due to the co-moving frame of reference . To our knowledge, there are no other studies applying the Lagrangian framework to quantify small-scale dynamo action, especially for shock-driven turbulence.

The rest of the paper is organised as follows. In Section \ref{sec:theory}, a theoretical background is given covering the details pertinent to our numerical experiment, including turbulent (small-scale) dynamos, Lagrangian statistics and decaying hydrodynamic and MHD turbulence. Then, in Section \ref{sec:sim} we describe our numerical model and setup. Finally, in Section \ref{sec:results} we provide the numerical results of our shock-driven dynamo simulations, and quantify the level of magnetic field amplification with quantitative comparisons to ISM dynamos. Section \ref{sec:conclusion} summarises the results and conclusions of the study.

\section{Theoretical Background} \label{sec:theory}

\subsection{The turbulent (small-scale) dynamo}

\subsubsection{Kinematic (exponential) growth phase}

In high $\textrm{Pm} = \nu/\eta$ plasmas ($\textrm{Pm} \gg 1$) such as in the ISM, there is little to no resistive decay ($\eta \sim 0$). The small-scale dynamo existing in the inner scales of hydrodynamic turbulence can grow exponentially from interactions with viscous eddies at the dissipation scale, $\ell_\nu \sim k_\nu^{-1}$ \citep{batchelor1950spontaneous,schekochihin2002model,kulsrud1992spectrum,XL2016}, such that when it reaches a stage where the magnetic excitation is so strong that at $k _{\nu} < k < k_{\eta}$ (kinematic regime), the magnetic energy spectrum in Fourier space, has a spatial distribution given by the resistive Green's function solution to the Kazantsev equation:
\begin{equation}\label{eqn:kazantsev}
M(k, t)=M_0 \exp \left(\frac{3}{4} \int \Gamma d t\right)k^{3/2}K_0\left(\frac{k}{k_\eta}\right),
\end{equation}
where $K_0$ is the Macdonald function, and the magnetic spectrum evolves as $M \sim k^{3/2}$ \citep{kazantsev1968enhancement,kulsrud1992spectrum,federrath2011mach}. Based on Kazantsev theory, one can also obtain a definition of the magnetic energy, via an integral over the magnetic energy spectrum,
\begin{equation}\label{eqn:magspec}
    E_{\textrm{mag}}=\frac{1}{2} v_{\mathrm{A}}^2=\frac{1}{2} \int_0^{k^{\prime}} M(k, t) dk,
\end{equation}
where $E_{\textrm{mag}}$ is the specific magnetic energy, and $v_{\mathrm{A}}$ is the Alfv\'en speed. Thus, the magnetic energy is dependent only on the viscous scale eddies, $k_{\nu} \sim \ell_{\nu} ^ {-1}$, and an amplitude term for the initial magnetic energy, $M_0 = \epsilon_0/k_{\nu}$, and $k^{\prime}$ is a reference scale,  where $k_{\nu} < k^{\prime} < k_\eta$.  Coupling this with the conducting limit of the MHD induction equation \citep{mckee2020magnetic,beattie2022growth}, we have
\begin{equation}\label{eqn:conducting}
    \frac{d{E}_{\textrm{mag}}}{dt}=2 \Gamma E_{\textrm{mag}}
\end{equation}
where the growth rate ($\Gamma$) is determined only by quantities at the dissipation scales,
\begin{equation}
    \Gamma(t)=\frac{\langle(\mathbf{B} \otimes \mathbf{B}
    ):(\nabla \otimes \mathbf{v})\rangle_{\nu}}{\left\langle B^2\right\rangle_{\nu}}.
\end{equation}
Thus, the magnetic energy $E_{\textrm{mag}}$ grows exponentially by $\exp({2\Gamma t})$ throughout the kinematic regime. Additionally, since the fundamental scales of this regime is governed by folds and random stretching at the diffusive scale, we have $\ell_\nu^2/ \nu \sim \ell_\eta^2/\eta$, then $k_\eta \sim k_\nu \textrm{Pm}^{1/2}$, as proposed by \cite{schekochihin2002model}, and confirmed recently by \cite{KrielEtAl20221} and \cite{2022axel}.

\subsubsection{Transition to saturation (non-linear stage)}

Now, we direct our attention towards the nonlinear stage of the dynamo, where the back-reaction by the Lorentz force is magnified enough that it is able to dampen the development of coherent structures; thus hindering the continual amplification of the field through the stretch-twist-fold-merge mechanism. Here we approach the peak scale of the magnetic spectrum, $k_{\textrm{peak}} = k^{\prime} \exp((3/5) \Gamma t)$. \cite{XL2016} argued that, by setting $E_{\textrm{mag}} \sim E_{\nu}$, where $E_{\nu}$ is the turbulent kinetic energy at the diffusive scale, we can account for the field growth near equipartition, since it is the eddies at the stretching scale, $\ell_{st} = k_{st}^{-1}$, where $k_{\textrm{inj}}<k_{st} \ll k_\nu$, that now dominate the interactions. Thus, we have
\begin{equation}
    E_{\textrm{mag}} = \frac{1}{2} (\nu \epsilon)^{1/2}
\end{equation}
where $\epsilon$ describes the kinetic energy dissipation rate at the inertial range, whose value is determined from the injection scales of Kolmogorov turbulence, $ \epsilon = k^{-1}_{\textrm{inj}} v^3_{\textrm{inj}}$, $k_{\textrm{inj}} = L_{\textrm{inj}}^{-1}$. It can be seen from Eqn.~\ref{eqn:magspec} that for eddies $k^{\prime} < k_{\textrm{peak}}  $, the dominant contribution of the magnetic energy always comes from the larger scales that seeded it, and no dependence is placed on the weaker fields whose contributions are negligible in the amplification process. Then, the magnetic energy amplifies until the peak of the power spectrum shifts to that of the viscous scale eddies, and one can eliminate the dependence on $k_{\nu}$, through the fact that there are only dependencies on the injection scales, $k_{\textrm{inj}}$ and $v_{\textrm{inj}}$ . By such dimensional arguments, one can then write
\begin{equation}\label{eqn:emag}
    E_{\textrm{mag}} = \frac{1}{2} (\nu \epsilon)^{1/2} \approx \frac{1}{2} v_{\textrm{inj}}^2,
\end{equation}
and finally, in the fully non-linear stage of the dynamo, we have minimal scale separation, such that $k_{\nu} \sim k_{\textrm{peak}}$. Thus, expanding the Macdonald function $K_0$ in  Eqn.~\ref{eqn:kazantsev}, for the low wavenumber limit, where $K_0\approx \ln(k_\eta) \sim \ln(k_\nu)$. One obtains a magnetic spectrum of the form \citep{XL2016,XL2017,XuLazarian2020}:
\begin{equation}\label{eqn:kazantsevlowk}
M(k, t)=M_0 \exp \left(\frac{3}{4} \int \Gamma d t\right)\left(\frac{k}{k_\nu}\right)^{3/2},
\end{equation}
Substituting this into Eqn.~\ref{eqn:magspec}, and taking the time derivative $d \ln(\dots) /dt$ we have:
\begin{equation}
\frac{d \ln(E_{\textrm{mag}})}{dt} \sim \frac{3}{4} \Gamma,
\end{equation}
and hence
\begin{equation}
    \frac{d E_{\textrm{mag}}}{dt} \sim \frac{3}{4} \Gamma E_{\textrm{mag}} \approx \frac{3}{8} \Gamma  v_{\textrm{inj}}^2,
\end{equation}
where $\Gamma \sim \alpha v_{\textrm{inj}}/L_{\textrm{inj}} $, and $\alpha$ is of order unity, which simplifies it to a linear differential equation of the form:
\begin{equation}\label{eqn:xu}
    \frac{d{E}_{\textrm{mag}}}{dt} = \beta \varepsilon.
\end{equation}
Using the earlier definition for the energy dissipation rate, \citet{XL2016} found directly that $ \beta = 3/38$ by accounting for the reconnection diffusion effect encountered in the nonlinear phase, where only a fraction of the total turbulent kinetic energy on the stretching to viscous scales contribute to the overall magnetic field amplification. The rest is dissipated via fast stochastic reconnection \citep{lazarian1999reconnection,eyink}, including natural mechanisms of viscous heating and turbulent diffusion \citep{kolmogorov1941local,kulsrud1992spectrum}. Similar scalings, with corresponding linear growth\footnote{Alternatively, consider simply that $v_{st}/\ell_{st} \sim \eta/\ell_{\eta}^2$, which gives $\ell_\eta \sim (\ell_{st} \eta/v_{st})^{1/2} \sim (\eta t)^{1/2}$. The selective decay mechanism suppresses high $k$-modes, which triggers a magnetic back-reaction when $B^2 \sim v_{st}^2$. Then, $dE_\mathrm{mag} /dt \sim v_{st} B^2/\ell_{st}  \sim v_{st}^3/\ell_{st} \sim \epsilon$, implying therefore that $E_{\textrm{mag}} \sim \epsilon t$ \citep{schekochihin2002model}. }, up until the suggested $k_\eta \sim k_{\textrm{inj}} \textrm{Pm}^{1/2} \textrm{Re}^{1/2} = k_{\textrm{inj}} \textrm{Rm}^{1/2}$ at saturation\footnote{According to this scenario, a quasi-static balance is achieved, where non-linear interactions arising from the injection (outer) scales of turbulence have dynamical time-scales comparable to folding at the resistive time-scales (i.e. $\tau_{\textrm{inj}} \sim \tau_{\eta}$). Hence, $\tau_{\textrm{inj}} \sim L_{\textrm{inj}}/v_{\textrm{inj}}\sim \tau_{\eta} \sim l_\eta^2 / \eta$, which yields the expected $k_{\eta} \sim \textrm{Pm}^{1/2}\textrm{Re}^{1/2} k_{\textrm{inj}}$ \citep{schekochihin2008mhd,2022galishnikova}. Note that this idealised relation does not consider the effect of tearing-mediated turbulence concentrated within anisotropic current sheets (e.g., \cite{2022galishnikova,beattie2022growth}).} have also been observed in numerous prior works \citep{kulsrud1992spectrum,schekochihin2002model,cho2009growth,beresnyak2009turbulence,beresnyak2012universal}. 

\citet{hu2022turbulent} applied this model to analyse the dynamo growth rate in shock-driven turbulence. Thus, we will also apply it for comparisons to our simulations. It should be noted upfront that the \citet{XL2016} model applies in the non-linear stage of the dynamo, i.e., when the Lorentz force has become strong, as discussed in this subsection. However, the simulations discussed below, have not reached this stage, as we will see, which makes a direct comparison to the \citet{XL2016} model difficult. Instead, our simulations here are in the exponential (often referred to as `kinematic' phase) growth stage of the dynamo.

\subsection{Lagrangian description of second-order statistics}
Similar to the Eulerian description of turbulence, one can describe two-point statistics such as the second-order structure function and the energy spectra through the Lagrangian framework. A unique advantage of this perspective is that it allows the treatment of point-like particle trajectories, which are co-moving in the direction of velocity streamlines, such that each particle has a time-dependent position, $\mathbf{X} =\mathbf{X}(\mathbf{X}_0,t_0) $, based on the Eulerian fixed-in-space velocity field $\mathbf{V}(\mathbf{X}(\mathbf{X}_0,t),t)$. Thus, trajectories in this frame of reference are not affected by advection, and therefore, each Lagrangian tracer particle represents a unique fluid/gas element that can be traced throughout the simulation. Through this, one can define the Lagrangian second-order structure function as
\begin{equation}
\mathcal{S}_2^{\Phi}(\Delta t)=\left\langle\left|\Phi_j(t+\Delta t)-\Phi_j(t)\right|^2\right\rangle
\end{equation}
where $\Phi$ is an arbitrary vector field and $j  = x,y $ are the longitudinal and transverse components of $\Phi$, over which we take its increments along each particle trajectory and average the values obtained as an ensemble of realisations. This quantity is spatially invariant in homogeneous turbulence and is also rotationally invariant in isotropic flow \citep{frisch1995turbulence}. In the inertial subrange $k_{\textrm{inj}}<k < k_{\nu}$, where $k_l$ is the injection scale or forcing scale, the energy spectrum follows an energy cascade. Thus, it can be shown that the Lagrangian K41 scaling for the second-order LSF with respect to Lagrangian frequency, $\omega \sim (\epsilon/ \nu)^{1/2}$, by the constant flux ansatz, has the form,
\begin{equation}  
                                \mathcal{S}_2(\Delta t)  \sim  (\Delta t)^p
\end{equation} 
up to small-scale intermittency corrections \citep{benzi1993extended,homann2007lagrangian,arneodo2008universal,benzi2010inertial,busse2010lagrangian,konstandin2012statistical,beresnyak2015parallel}. The velocity LSF follows a linear scaling ($p  = 1$) based on the Kolmogorov bridge relations as detailed below.

If we assume Kolmogorov (K41) \citep{kolmogorov1941local} or Goldreich-Sridhar (GS95) scaling \citep{sridhar1994toward,goldreich1995toward,goldreich1997magnetohydrodynamic}, which obtains both $E(k) \sim \epsilon^{2/3} k ^{-5/3}$ ($k = k_\perp$ for GS95) in Eulerian space, one can easily show that
\begin{equation}\label{eqn:kolm}
    E(\omega) \sim \epsilon \omega^{-2}
\end{equation}
is the expected scaling obtained for the kinetic energy spectrum \citep{inoue1951turbulent,corrsin1963estimates,tennekes1972first,tennekes1975eulerian,frisch1995turbulence}. We further note that in the three-dimensional incompressible MHD simulations of \citet{busse2010lagrangian}, excellent agreement was found for this scaling law given by Eqn.~\ref{eqn:kolm}, consistent with prior experimental \citep{mordant2004experimental} and numerical results \citep{yeung2006reynolds}. However, for two-dimensional simulations, it was found that $E(\omega) \sim \omega^{-3/2}$, in accordance with the Iroshnikov-Kraichnan (IK) phenomenology of turbulence, where $E(k) \sim k^{-3/2}$ \citep{iroshnikov1964turbulence,kraichnan1965inertial,kraichnan1977eulerian,gogoberidze2007nature} for the wavenumber spectra. Thus, on the basis that dynamical alignment at large scales \citep{mason2006} are dominated by Eulerian sweeping effects, \cite{busse2010lagrangian} suggested that the relevant timescale for the Lagrangian frequency spectrum should be the Eulerian correlation time. Therefore, following the Eulerian definition of a time spectra, with the ansatz of frequency-wavenumber self-similarity, i.e. $\omega E(\omega) \sim k E(k)$, an analogous IK scaling is found which is identical to the Eulerian time-frequency spectra \citep{tennekes1975eulerian,busse2010lagrangian}.

We note here also that for a Burgers' spectrum \citep{burgers1995mathematical}, $E(k) \sim k^{-2}$ occurring in shock-dominated, highly supersonic flows \citep{Federrath2013,FederrathEtAl2021}; since $v \sim \ell^{1/2}$, and assuming $t_{\textrm{ac}} \sim t_{\textrm{cas}}$, where $t_{\textrm{ac}}$ and $t_{\textrm{cas}}$ are autocorrelation and cascade timescales, respectively. We have $t_{\textrm{cas}} \sim \ell/v \sim \ell^{1/2}$, yielding $v_\textrm{cas}^2 \sim t_{\textrm{cas}}^2 \sim \omega^{-2}$. Thus,  the corresponding Lagrangian frequency spectrum should therefore scale as\footnote{This is a spectrum with no mathematically self-similar second-order structure function (SF2), since $\mathrm{SF}_2(\tilde{t}) = 2\left(v^2 - \int_{-\infty}^{\infty} E(\omega)\exp({i \omega \tilde{t}}) d \omega\right)= 2 \int_{-\infty}^{\infty} \left[1 - \exp({i\omega \tilde{t}})\right] E(\omega) d\omega $ using Wiener-Khinchin theorem, is conditionally convergent only when $E(\omega) \sim \omega^{-n}$  with $ n \in (1,3)$.}:
\begin{equation}
    E(\omega) \sim \omega^{-3}.
\end{equation}

\subsection{Decaying MHD turbulence}
In shock-driven turbulence without additional external turbulence driving, supersonic turbulence decays very rapidly on time scales of roughly one turnover time \citep{scalo1982dissipation,stone1998dissipation,mac1998kinetic,mac1999energy,federrath2012star}. Such time scales emphasise the importance of turbulence driving mechanisms \citep{mac2004control,schleicher2010small,federrath2016link,sur2019decaying}, which continuously supply kinetic energy into the system to allow for amplification of a small-scale seed magnetic field \citep{schober2012small,schleicher2013small,seta2020seed,seta2021saturation}. 

Numerical simulations with large-scale mean fields \citep{mac1998kinetic} and even seeded kinetic helicity ($H^k = \boldsymbol{v} \cdot (\nabla \times \boldsymbol{v})$) \citep{brandenburg2012kinetic,brandenburg2019dynamo} have shown that turbulent (or mean in the large-scale dynamo setting) magnetic fields can decay rapidly together with the kinetic energy, such that saturation or strong magnetic fields can never be achieved. The increased alignment of the velocity and magnetic fields associated with this process \citep{servidio2008depression}, suggests that even turbulence driven with a very strong shock, followed by a transient period of quiescence, will not be able to completely amplify small-scale magnetic fields. 

Here we also expect such phenomena to occur. Thus, the time-dependence of the energy flux will need to be quantified in this un-driven (decaying) turbulent configuration for accurate understanding of how magnetic fields can amplify in decaying ISM post-shock media. We note that in subsonic, incompressible turbulence, the energy flux follows a power-law decay, $E_{\textrm{kin}} \sim \langle v^2\rangle \propto t^{-n}$, where $n = 6/5$ if the Saffman integral is invariant \citep{saffman1967note}, and $n = 10/7$ if the Loitsyansky integral is conserved \citep{proudman1954decay} \citep[see also][]{davidson2000loitsyansky,krogstad2010grid,davidson2010decay}. In supersonic, isothermal turbulence, it has been found that $ 0.85 < n < 1.2$ \citep{mac1998kinetic,mac1999energy}, suggesting a decay much closer to that of the Saffman invariant.
\begin{figure}
\def\arraystretch{0.0}
\setlength{\tabcolsep}{0pt}
\begin{tabular}{cc}
\includegraphics[width=0.75\linewidth]{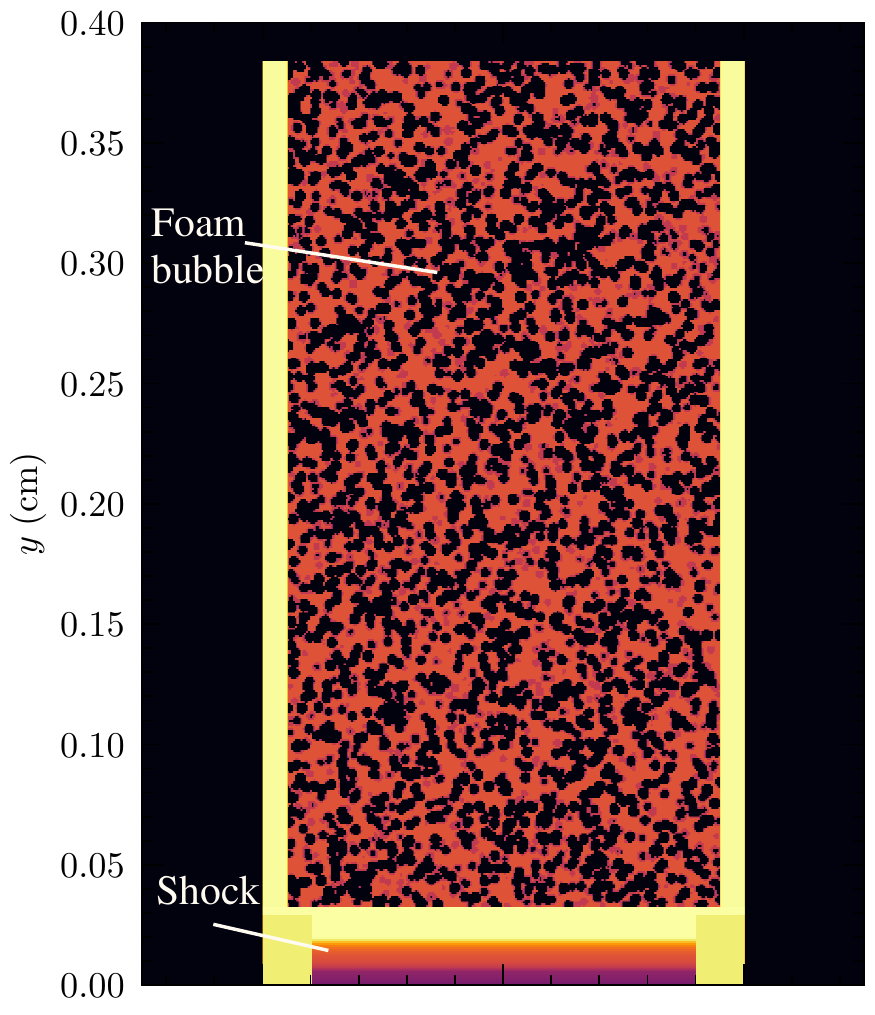} &
\hspace{-0.05cm}\multirow{2}{*}[7.43cm]{\includegraphics[width=0.2417\linewidth]{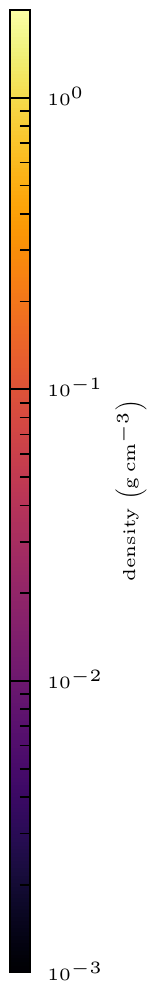}} \\
\hspace{-0.16cm}\includegraphics[width=0.762\linewidth]{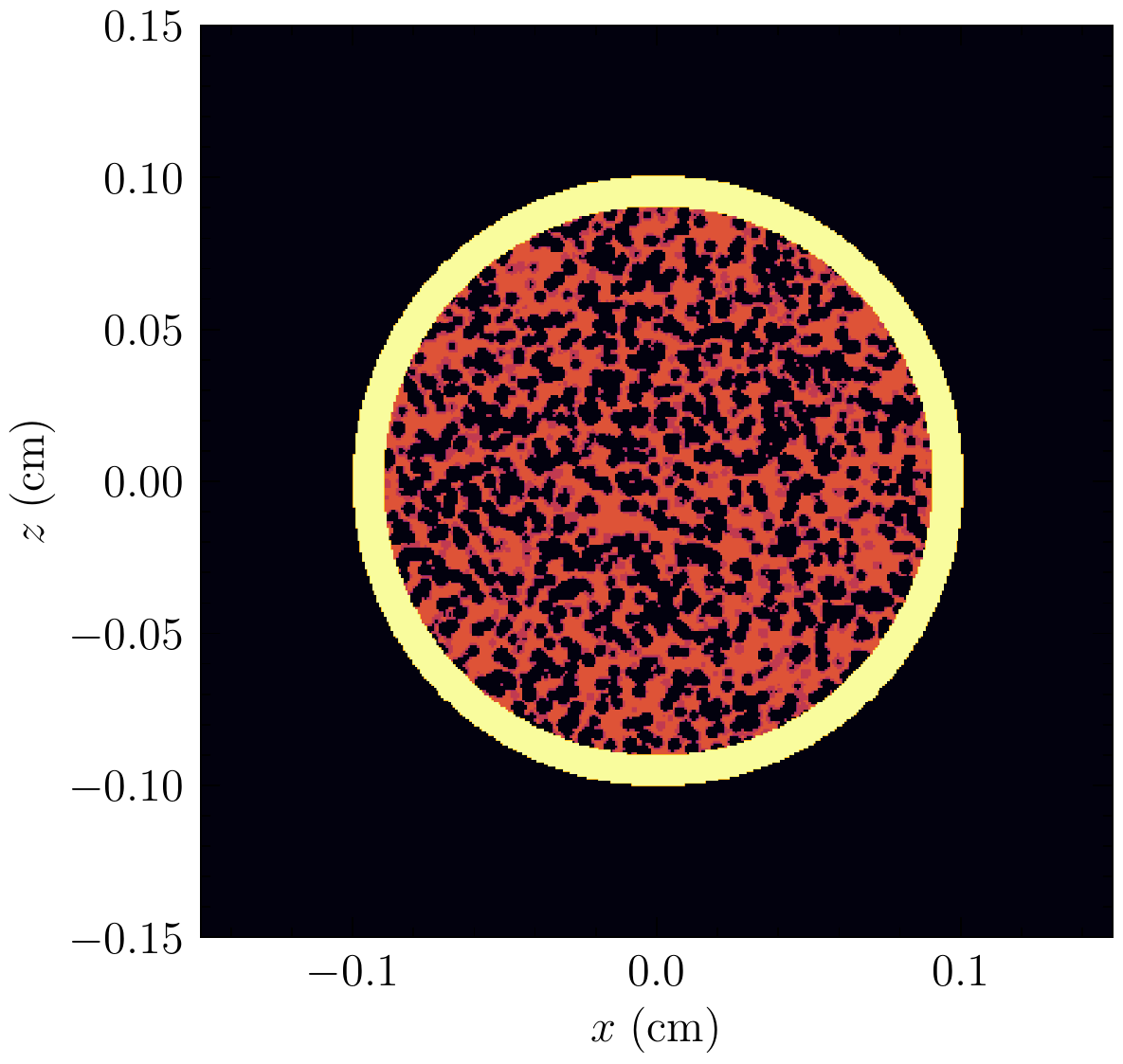} & 
\end{tabular}
\caption{Schematic of the geometrical configuration used in the present study. The physical system is identical to \citet{dhawalikar2022driving}, resembling the current experimental test setup at the NIF. The laser-driven shock hits the ablator at $y \approx 0.3$~mm, and propagates further through the cylindrical tube in the $y$-direction, subsequently interacting with the foam material, which are shown as black circles. The top panel shows a slice along the $z$-direction though the centre of the tube, while the bottom panel shows a slice along the $y$-direction, again at centre of the tube.}
\label{fig:geometry}
\end{figure}
Further numerical experiments \citep{biskamp1999decay,biskamp2000scaling,muller2000scaling,banerjee2004evolution,frick2010long,berera2014magnetic,brandenburg2015nonhelical,brandenburg2017classes,reppin2017nonhelical,sur2019decaying,bhat2021inverse} in three-dimensional non-helical\footnote{Non-helical in the sense of zero net helicity, but small-scale helical fluctuations are allowed under the assumption that they do not influence the large scale dynamics (see e.g., \cite{reppin2017nonhelical} for a quantitative discussion).} MHD turbulence also confirm scalings very close to the Saffman invariant, as well as the later known \citet{biskamp1999decay} scaling ($n = 1$) based on 2D anastrophy conservation\footnote{To clarify, in a recent work by \citet{hosking2021reconnection}, it was shown that the non-helical decay scaling should not be argued based on anastrophy conservation, but by the requirement of Hosking integral invariance. This still produces similar results, where with fast stochastic reconnection (i.e. $\epsilon_{\textrm{rec}} \sim \mathrm{const}$, for example in LV99 \citep{lazarian1999reconnection}),  $\langle B^2 \rangle \sim t^{-10/9} \sim \langle v^2 \rangle$, and with Sweet-Parker (SP) dominated reconnection (i.e. $\epsilon_{\textrm{rec}} \sim \tilde{S}_L^{-1/2})$, where $\tilde{S}_L$ is the Lundquist number, then  $\langle B^2 \rangle \sim t^{-20/17}$ and $\langle v^2 \rangle \sim t^{-19/17}$.}. 

Thus, here in our numerical experiment, we test these decay laws, and quantify the decay found in our simulations, suggesting how it may affect the dynamo growth rate over longer timescales.

\section{Numerical Simulations}\label{sec:sim}

\subsection{Governing Equations}
We use a modified version of the \texttt{FLASH} code \citep{FryxellEtAl2000}, with the HLL3R 3-wave approximate Riemann solver \citep{bouchut2010multiwave,waagan2011robust} to solve the fully three-dimensional, compressible MHD equations,
\begin{equation}
\frac{\partial \rho}{\partial t}+\nabla \cdot(\rho \boldsymbol{v})=0,
\end{equation}
\begin{equation}
\begin{aligned}
\rho\left(\frac{\partial}{\partial t}+\boldsymbol{v} \cdot \nabla\right) \boldsymbol{v} = \frac{1}{4 \pi}(\boldsymbol{B} \cdot \nabla) \boldsymbol{B} -\nabla\left(p_{\mathrm{th}} +\frac{B^2}{8 \pi}\right) \\
+\nabla \cdot(2 v \rho \mathcal{S})+\rho \boldsymbol{F}
\end{aligned}
\end{equation}
\begin{equation}
\frac{\partial \boldsymbol{B}}{\partial t}=\nabla \times(\boldsymbol{v} \times \boldsymbol{B})+\eta \nabla^2 \boldsymbol{B}
\end{equation}
\begin{equation}
\nabla \cdot \boldsymbol{B}=0
\end{equation}
where  $\rho$, $\mathbf{v}$, $p_{\textrm{tot}}=p_{\mathrm{th}}+(1 / 8 \pi)|\mathbf{B}|^2$, $\mathbf{B}$, and $e=\rho \epsilon_{\mathrm{int}}+(1 / 2) \rho|\mathbf{v}|^2+(1 / 8 \pi)|\mathbf{B}|^2$ denote the gas density, velocity, total pressure (sum of the thermal and magnetic), magnetic field, and energy density (sum of the internal, kinetic and magnetic), respectively.  $\mathcal{S}_{i j}=(1 / 2)\left(\partial_i v_j+\partial_j v_i\right)-(1 / 3) \delta_{i j} \nabla \cdot \mathbf{v}$ is the traceless rate of strain tensor, which is the symmetric part of the velocity gradient tensor that accounts for physical shear viscosity. Here  $\boldsymbol{F}$, the turbulence driving parameter is set to zero since we do not use any driven turbulence. The quantities $\nu$ and $\eta$ are the kinematic viscosity (dynamic viscosity divided by density), and the magnetic resistivity, respectively. Here we do not specify these dissipative terms, and instead use numerical viscosity and resistivity inherent in the Riemann flux functions as a subgrid-scale model for dissipation \citep{garnier1999use}. Thus, we perform implicit large-eddy simulations (ILES). We close the MHD equations with an equation of state (EOS) for an ideal monoatomic gas, i.e., $p_{\mathrm{th}} = \rho e_{\textrm{int}} (\gamma -1 ) $, where $\gamma = 5/3$ is the specific heat ratio. 

\subsection{Initial conditions and flow configuration}
Fig.~\ref{fig:geometry} displays the initial configuration used in the present study. The geometry is identical to that used in \cite{dhawalikar2022driving}, and corresponds also to the one currently being tested in the wind tunnel facility at the National Ignition Facility (NIF). The foam within the cylindrical domain is modelled as a CH-based polymer, and the foam voids with radius $r = 25$~mm are air bubbles contained within the foam, existing as the precursor small-scale density inhomogeneities to generate post-shock turbulence. Also, although the laser-driven blast wave propagating into the medium may inherently cause changes in the material chemistry, induce radiation via inverse Bremsstrahlung, as well as cooling effects, etc., we do not consider these properties, since the primary purpose of this setup is to study the turbulent dynamics of a post-shock medium generated by a shock running over a pre-structured medium. The thermodynamic properties are not a primary concern for this, as long as a reasonable turbulent density and velocity field results from the interaction, which is the case \citep{dhawalikar2022driving}.
Neglecting these effects will also allow us to make thorough comparisons of our numerical results to other studies of post-shock turbulence, as well as small-scale dynamo processes in the ISM. Thus, the simplified approach was taken for this purpose.
\begin{figure}
\def\arraystretch{0.0}
\setlength{\tabcolsep}{0pt}
\begin{tabular}{cc}
\includegraphics[width=0.75\linewidth]{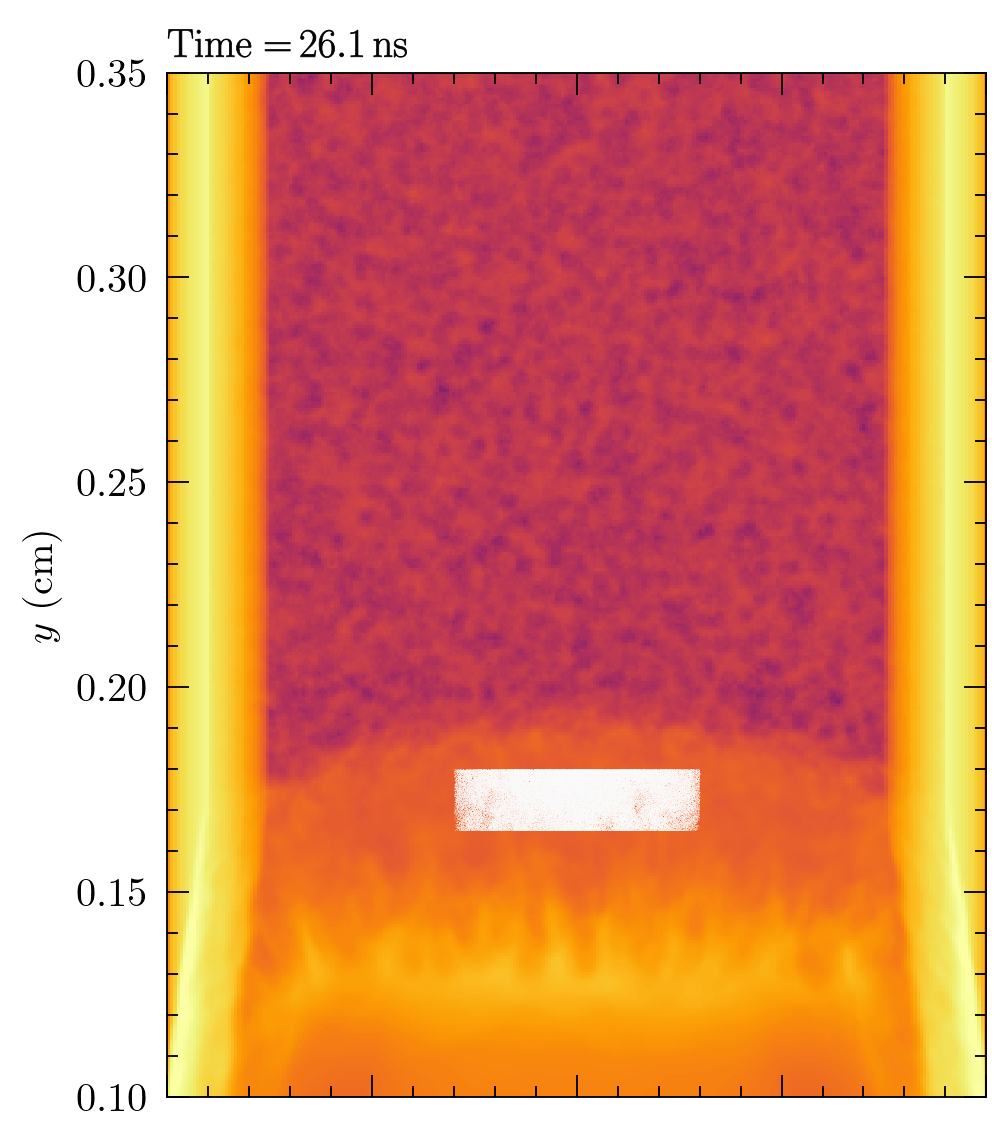} &
\hspace{-0.22cm}\multirow{2}{*}[7.06cm]{\includegraphics[width=0.2603\linewidth]{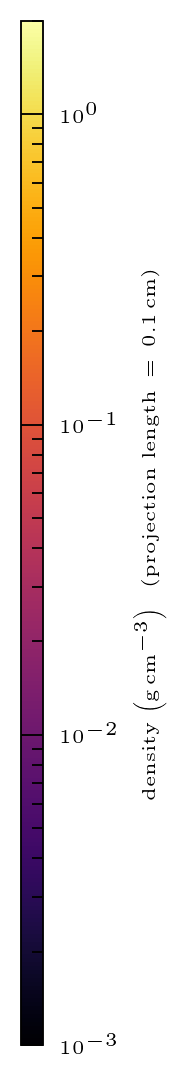}} \\
\hspace{-0.09cm}\includegraphics[width=0.8028\linewidth]{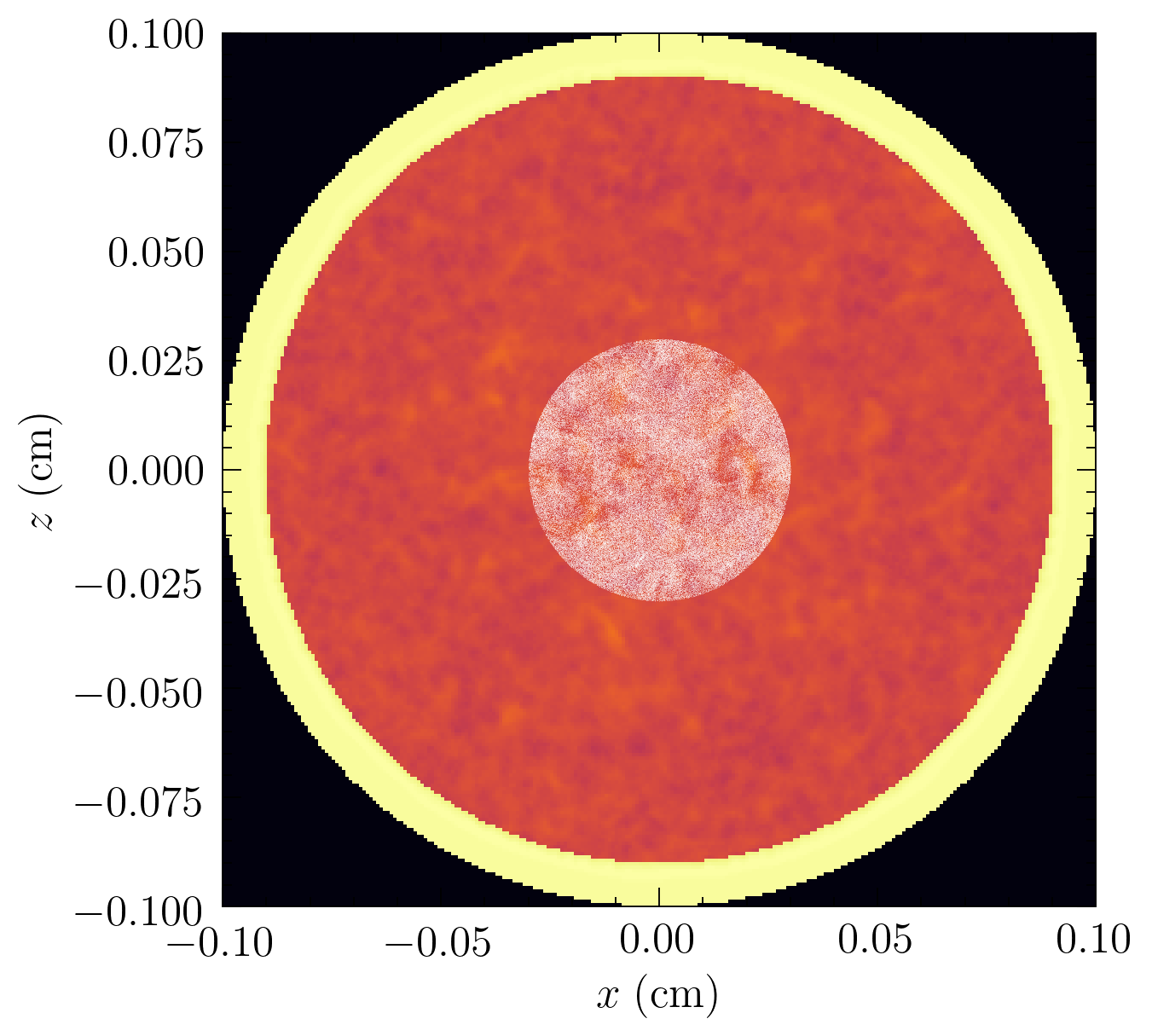} & 
\end{tabular}
\caption{Density distribution showing the initial Lagrangian volume chosen in the post-shock medium at $t = t_1 = 26.1$~ns, consisting of about $2 \times 10^5$ tracers. The volume chosen is a cylinder with radius $0.03$~cm, in accordance with the flow configuration itself. (a) $z$- projected density distribution, (b) $y$-projected density distribution, centred on the respective mid-plane of the shock tube. Tracer particles are shown as white points (note that each tracer technically corresponds to exactly the size of a grid cell, as we are using the cloud-in-cell particle-mesh interpolation scheme, i.e., while this graphical representation plots them as point particles, they actually occupy/trace the entire cylindrical volume in which they were initialised as a collective).}
\label{fig:lagvol}
\end{figure}
In order to study the growth of a turbulent magnetic field, we inject a very small-scale magnetic field of $B_{\textrm{turb}} = 5.5 \times 10^{-5}$~G, and also a mean guide field in the $y$-direction (streamwise) of that same value, corresponding to an initial plasma $\beta = 2 c_s^2/v_A^2 = 1 \times 10^{16}$. The turbulent field is initialised using Fourier modes, with an initial power law at large scales, $2 \leq kL/2 \pi \leq 20$ where $L$ is the 3D turbulent box size, and $k$ the wavenumber, containing a Kazantsev spectral scaling with a power-law  exponent of $3/2$ (see Sec.~\ref{sec:theory}). We also test a parabolic power with no mean field in the streamwise direction, with the magnetic field being injected at even larger scales, $1 \leq kL/2\pi \leq 3$, similar to that used in \cite{seta2020seed,seta2022turbulent}, and find negligible differences in the overall qualitative properties (i.e., the magnetic field amplification and other time-dependent properties remain the same). The turbulent initial magnetic fields were generated with the publicly available \texttt{TurbGen} code \citep{federrath2010comparing,FederrathEtAl2022ascl}. 
\begin{figure}
\def\arraystretch{0.0}
\setlength{\tabcolsep}{0pt}
\begin{tabular}{cc}
\includegraphics[width=0.75\linewidth]{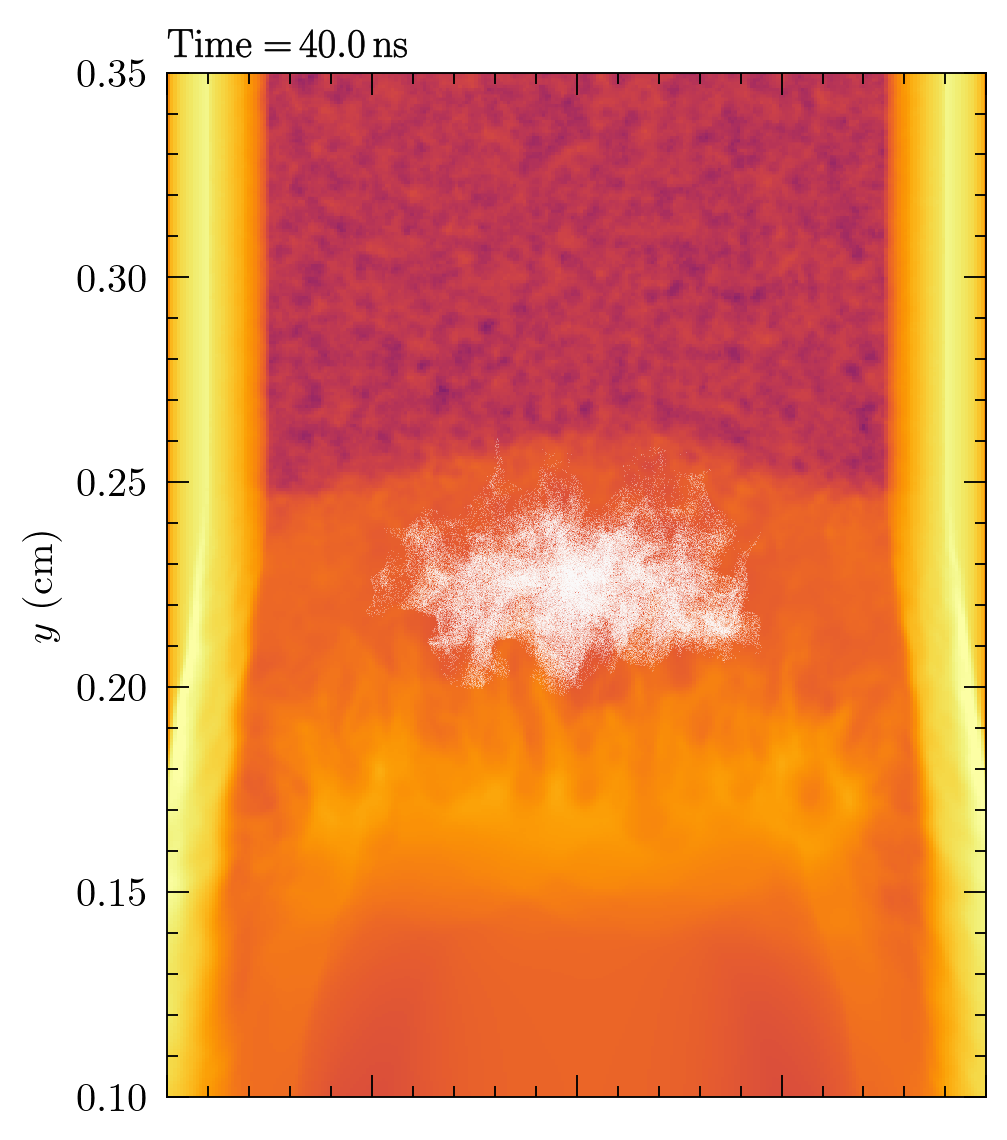} &
\hspace{-0.22cm}\multirow{2}{*}[7.06cm]{\includegraphics[width=0.2603\linewidth]{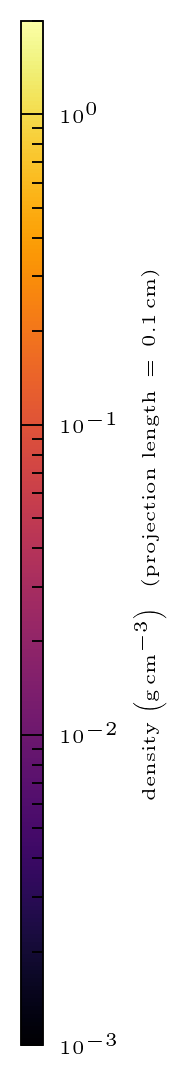}} \\
\hspace{-0.09cm}\includegraphics[width=0.8028\linewidth]{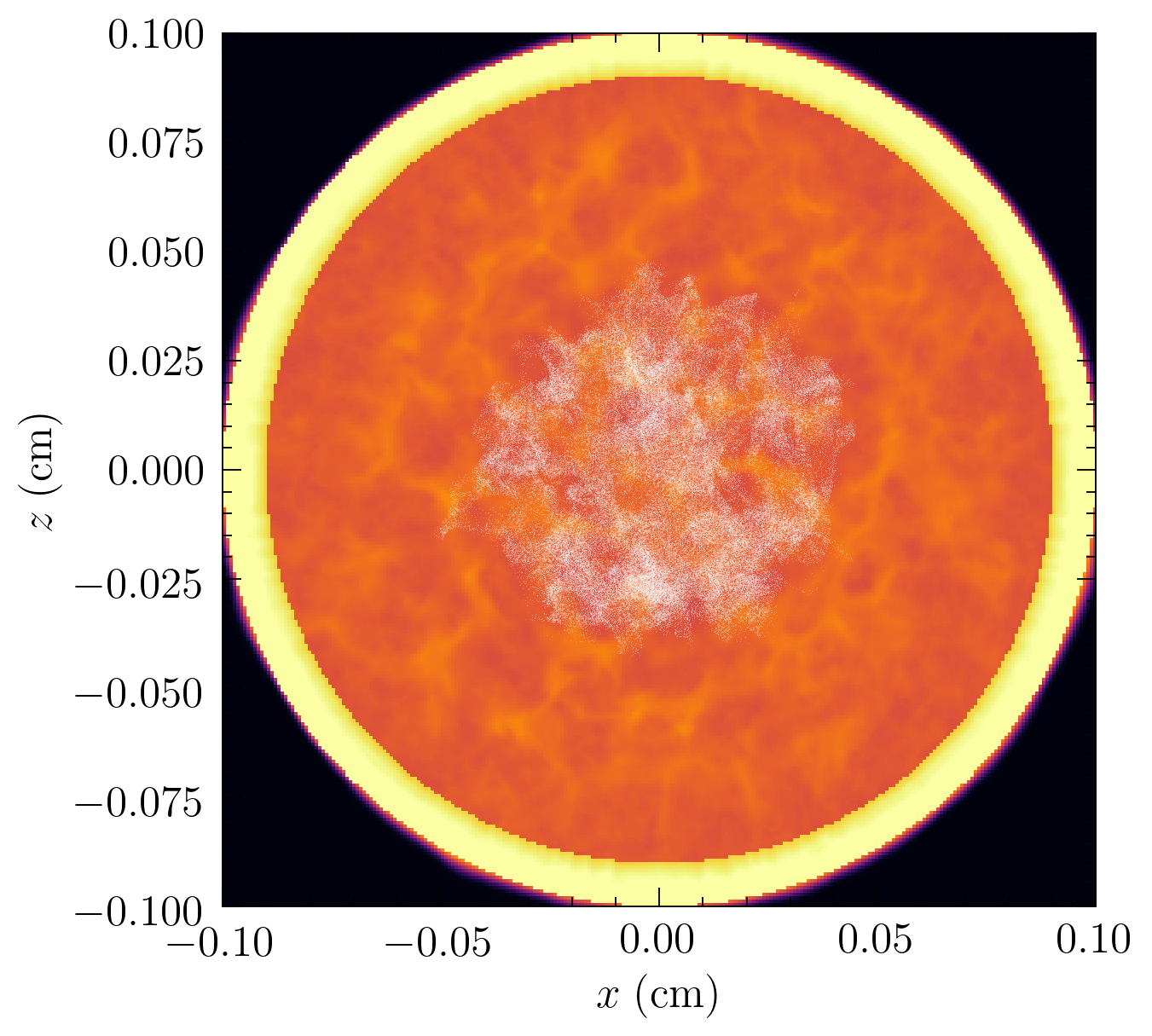} & 
\end{tabular}
\caption{Same as Fig.~\ref{fig:lagvol}, but at $t = t_2 = 40.0$~ns. The Lagrangian volume traced by the tracer particles has evolved into a complex structure. However, by the definition of the Lagrangian tracers, the collective of tracer particles still traces the same material as they were initialised in (cf., Fig.~\ref{fig:lagvol}), allowing us to study the magnetic field amplification and other turbulent properties, for exactly the same material at any given time.}
\label{fig:time40}
\end{figure}

\begin{figure}
\def\arraystretch{0.0}
\setlength{\tabcolsep}{0pt}
\begin{tabular}{cc}
\includegraphics[width=0.75\linewidth]{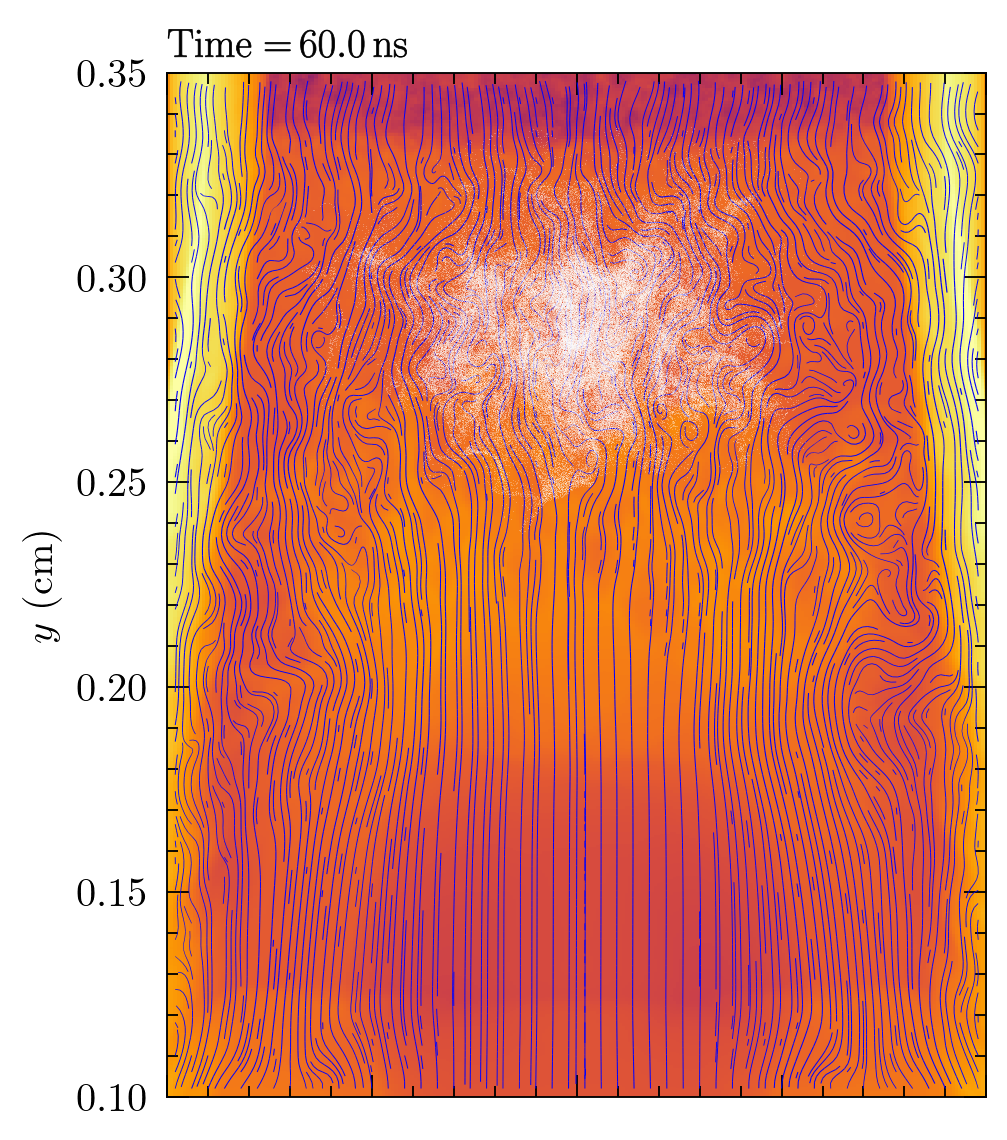} &
\hspace{-0.22cm}\multirow{2}{*}[7.06cm]{\includegraphics[width=0.2603\linewidth]{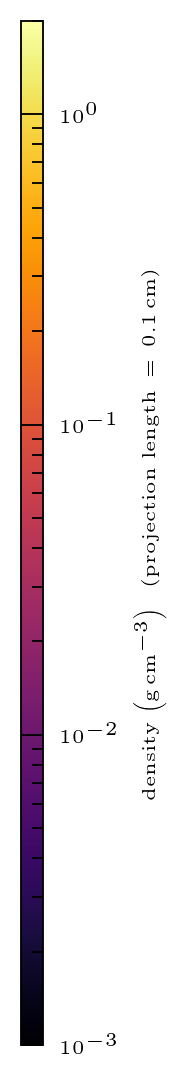}} \\
\hspace{-0.09cm}\includegraphics[width=0.8028\linewidth]{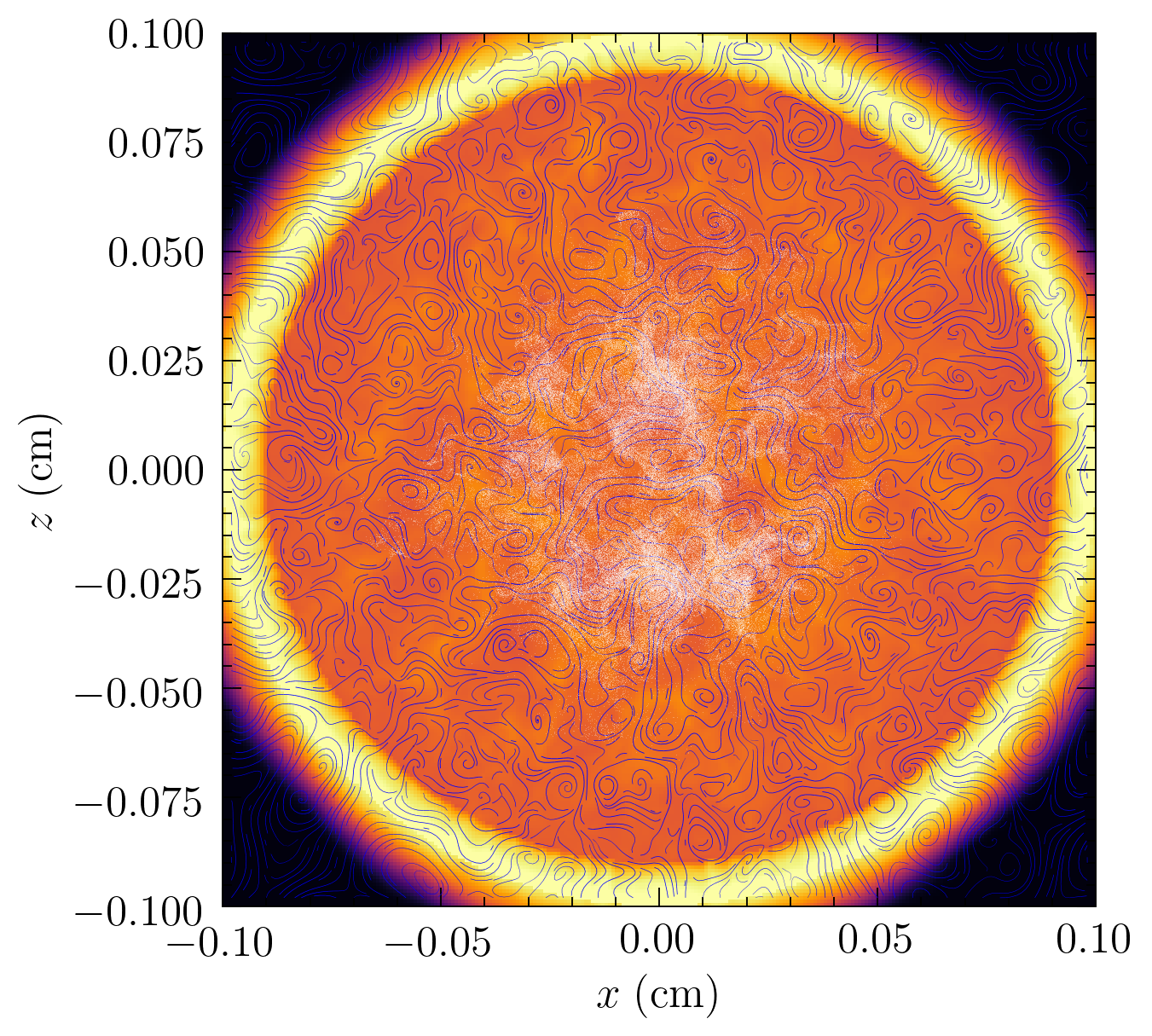} & 
\end{tabular}
\caption{Same as Fig.~\ref{fig:time40}, but at $t = t_3 = 60$~ns and with magnetic field lines (shown as blue streamlines) superimposed. The filamentary and tangled nature of the field is clearly visible. The collective of tracer particles is shown as white dots in these projections.}
\label{fig:t300}
\end{figure}

\subsection{Grid and Lagrangian statistics}

The simulation domain is a uniform grid with $384 \times 512 \times 384$ cells, with outflow boundary conditions \citep[as in][]{dhawalikar2022driving}.
For sampling the Lagrangian statistics, we initialise $384 \times 512 \times 384$ tracer particles (one in each grid-cell centre). This is comparable to the amount of tracers used in prior high-resolution periodic box simulations \citep{biferale2004multifractal,arneodo2008universal,benzi2010inertial,homann2007lagrangian,konstandin2012statistical}, thus allowing us to sample the time dynamics reliably.

In order to investigate the Lagrangian statistics specifically within the moving post-shock turbulent medium, we select a subset of tracer particles in the turbulent region behind the propagating shock front (i.e., where the shock has already passed), that is similar in size to the turbulence analysis region used in \cite{dhawalikar2022driving}. The cylindrical region chosen here (see Fig.~\ref{fig:lagvol}) is wide enough for such analyses, where we are able to sample over $2 \times 10^5$ tracers throughout the time evolution. This allows us to examine the growth rate of the magnetic field, while avoiding the domain boundaries, so as to avoid shock reflection (diffraction) effects or interactions with the ablator or pre-shock medium, which typically result in abrupt vorticity and magnetic field amplifications that are not associated with SSD action. We also ensure that the tracers do not sample the flow properties within the stratified shear instabilities, which only develop much further behind the shock front at the later stages of the time evolution.

\begin{table}
\begin{tabular}{|l|l|l|}
\hline
\textit{Post-shock parameters}    & Definition/Symbol                           & Mean                              \\ \hline
Mean Density             & $ \rho$                                     & $0.13$ g cm$^{-1}$                  \\ 
Turbulent Alfv\'en speed   & $v_{A} = |\mathbf{B}| / \sqrt{4 \pi \rho} $ & $1.76 \times 10 ^ {-9}$ cm s$^{-1}$ \\ 
Turbulent plasma beta    & $\beta = 2 c_s^2 /v_a^2 $                   & $ 2.29 \times 10^{15} $           \\ 
3D Turbulent Velocity    & $\sigma_{v,3\mathrm{D}}  = \sqrt{3} \sigma_v $      & $ 11.9 $ km $\textrm{s}^{-1}$     \\ 
Sound Speed              & $c_s = \sqrt{\gamma P/ \rho} $              & $ 20.0$ km $\textrm{s}^{-1}$      \\ 
Injection length scale & $ L_{\textrm{inj}}$                              & $0.14$ cm                          \\ 
Turbulent turnover time & $ L_{\textrm{inj}}/ \sigma_v$                              & $217$ ns                          \\ 
Alfven Mach number  &   $\mathcal{M}_A = \sigma_v/v_A$                                  & $3.89 \times 10^{14}$
\\
Mach number &    $\mathcal{M} = v/c_s$                                & 0.31
\end{tabular}
\caption{\label{tab:1} Calculated post-shock parameters in the post-shock turbulent medium. The large-scale turbulent turnover time, $T_{\textrm{ed}}$ is computed with the largest length scale that the Lagrangian volume occupies during the time evolution.}
\end{table}

\section{Results and discussion} \label{sec:results}
Table~\ref{tab:1} defines the computed mean values of the post-shock variables in the material volume traced throughout the time evolution. Crucially, the turbulent time (large-eddy turnover time) is calculated based on the largest length scale in the moving volume, which approximates the integral length scale in our simulations. This quantity is used throughout our time evolution analyses below. 
\subsection{Time evolution and probability distributions}

Fig.~\ref{fig:time40} displays the later stage of the time evolution of the density distribution with the Lagrangian tracers superimposed. It can be clearly seen that the tracers begin to disperse rapidly from its original position owing to the highly turbulent nature of the post-shock medium. As the shock front propagates further downstream, it clearly becomes corrugated in shape, similar to that observed in \citet{ji2016efficiency} and \citet{hu2022turbulent} due to interactions with the density inhomogeneities. Such changes in the global curvature of the shock further leads to enhanced vorticity production, particularly in the shock-parallel direction~\citep{kevlahan1997vorticity}. Furthermore, Fig.~\ref{fig:t300} clearly shows that the topology of the magnetic field lines are very tangled and filamentary in nature. This is an indicator of a turbulent dynamo mechanism \citep{federrath2016magnetic}.

\begin{figure}
	\includegraphics[width=\columnwidth]{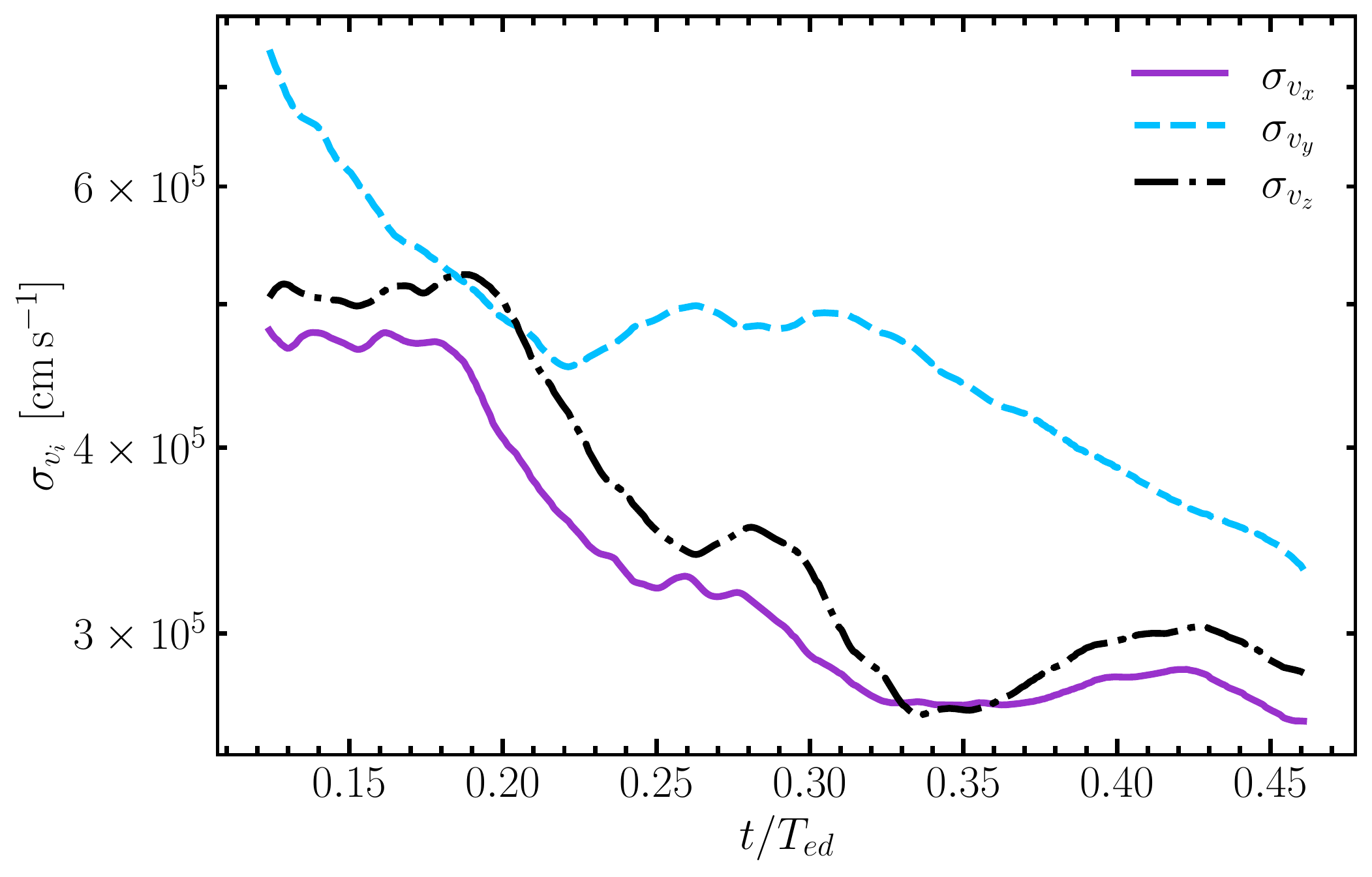}
    \caption{Time evolution of the Cartesian components of the turbulent velocity dispersion computed as an average across all tracers initially marked in Fig.~\ref{fig:lagvol}. The time is in units of the turbulent turnover time as defined in Tab.~\ref{tab:1}. We clearly see the decaying nature of the turbulence in the post-shock turbulent medium traced by the tracer particles.}
    \label{fig:sigmav}
\end{figure}

Fig.~\ref{fig:sigmav} shows the time evolution of the $x$, $y$ and $z$-components of the turbulent velocity dispersion (mass-weighted, as they were computed on the tracer particles) across all tracers in the moving post-shock volume. It can be seen that the initial velocity dispersion starts off at rather large values within the Lagrangian volume, of order $10^5\,\textrm{cm\,s}^{-1}$, with the streamwise component $\sigma_{v_y}$ always being slightly higher than the other two components, since it corresponds to the shock direction, where the shock profile was first injected. However, the values decay to almost half their value over less than half a turbulent turnover time. Such a behaviour cannot be purely explained by conversion of kinetic energy to magnetic energy, and is fundamentally indicative of decaying turbulence \citep{mac1998kinetic,mac1999energy}, where a fraction of the kinetic energy decays away as the corrugated shock front runs down the domain.

\begin{figure}
	\includegraphics[width=\columnwidth]{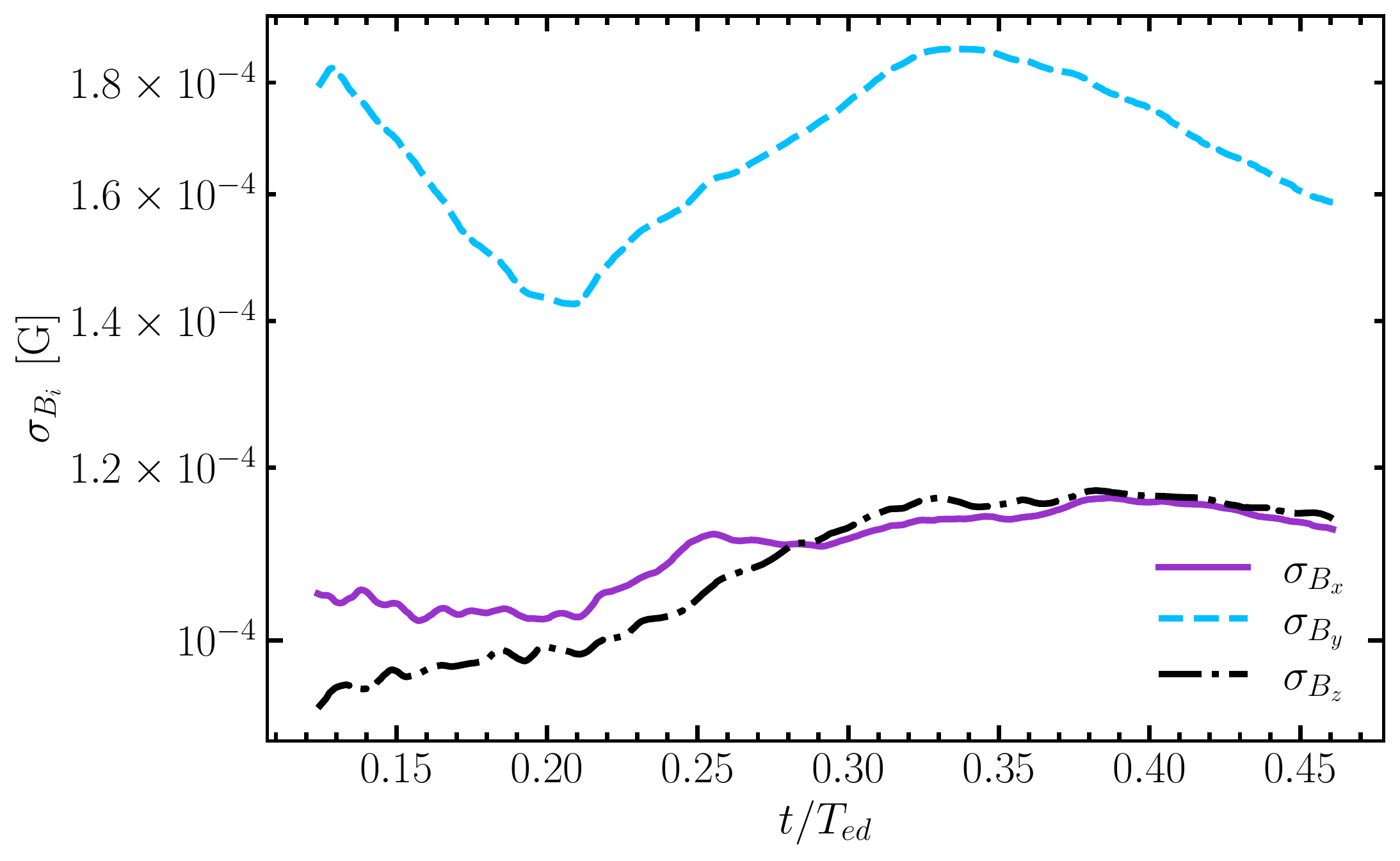}
    \caption{Same as Fig.~\ref{fig:sigmav}, but for the standard deviation of the magnetic field components.}
    \label{fig:sigmab}
\end{figure}

Fig.~\ref{fig:sigmab} shows the time evolution of the magnetic field, where we can notice substantial correlations with the corresponding velocity fields. The magnetic fields are gradually amplified over a short time scale, while the velocities decay. The streamwise field ($\sigma_{B_y}$) is always larger than the other components, likely owing to the additional amplification originating from shock compression. All values clearly indicate the anisotropic nature of the turbulent quantities, which crucially leads to the enhanced anisotropic nature of the vorticity. Our simulations further indicate that the magnetic field amplification by the turbulent dynamo effect does not even exceed an order of magnitude. This is similar to observations in prior numerical works \citep{giacalone2007magnetic,hu2022turbulent} with only slightly longer time evolution, where the seeded mean turbulent field amplifies by about a factor of 2 in half a turnover time. They however, primarily focussed on the maximum amplifications, we here consider the mass-averaged quantities through the Lagrangian framework, thereby removing compression effects from dynamo action. Moreover, the magnetic field amplification in our system is accompanied with a high degree of turbulent diffusion, so that no distinct phases or regimes can be observed in the averaged turbulent magnetic field evolution.

\begin{figure}
	\includegraphics[width=\columnwidth]{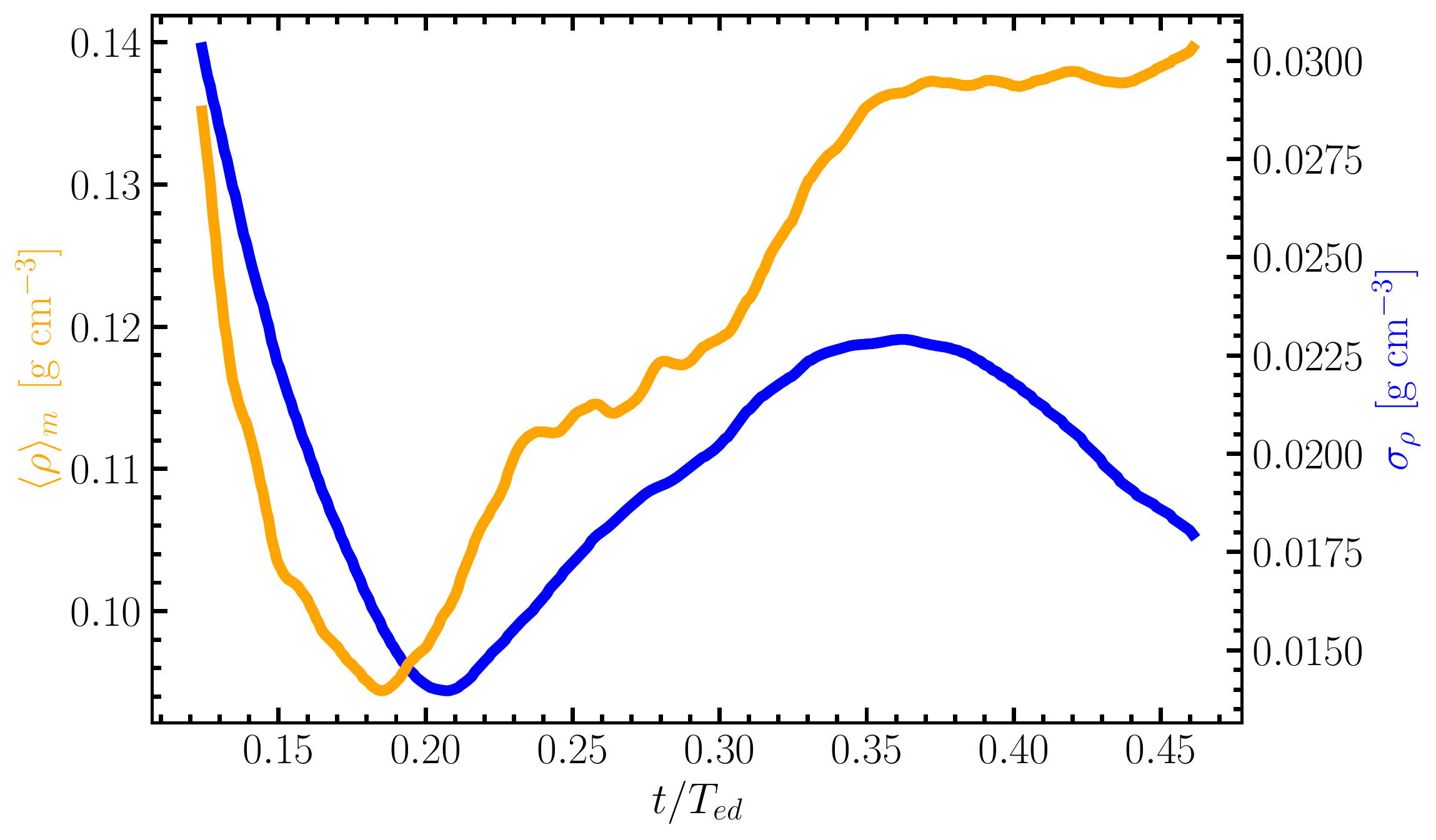}
    \caption{Time evolution of the mean density $\langle\rho\rangle_\textrm{m}$ and density dispersion ($\sigma_{\rho}$) of the tracers. This shows a relaxation stage until $0.2 \ t/T_{\textrm{ed}}$, followed by a compression phase between $0.20$ and $0.35\ t/T_{\textrm{ed}}$, which is the turbulent amplification phase we study in detail below, and finally another stage of dropping mean density, which is due to tracer particles beginning to break out at the top of the shock tube.  }
    \label{fig:rhoavg}
\end{figure}

In order to elucidate the effects of the shock compression and its influence on the magnetic field, we plot the mean density  and the density dispersion (Fig.~\ref{fig:rhoavg}). We note that at $t \approx 0.2 t/T_{ed}$, the density values begin to rise in both quantities, and display similar evolution with the magnetic field components (Fig.~\ref{fig:sigmab}). Such a result is typical of strongly compressive flows \citep{sur2010generation,federrath2011new}, where the magnetic field amplifies as $|\boldsymbol{B}| \sim \langle \rho \rangle^p $, where $p$ is some positive power and $\langle \rho \rangle$ is the mean density of the region of interest. Thus, in order to distinguish dynamo effect from shock compression-induced magnetic field amplification, the effect of the compression has to be corrected for, in order to isolate purely turbulent magnetic field amplification, i.e., dynamo action. A common strategy to account for the effect of compression is to divide the magnetic field by the density to some power \citep{sur2010generation,federrath2011new}. For instance, in a 3D medium in which the magnetic field is compressed in all three spatial directions, $B\sim \langle \rho \rangle^{2/3}$, because of mass and magnetic flux conservation during compression. 

\begin{figure}
	\includegraphics[width=\columnwidth]{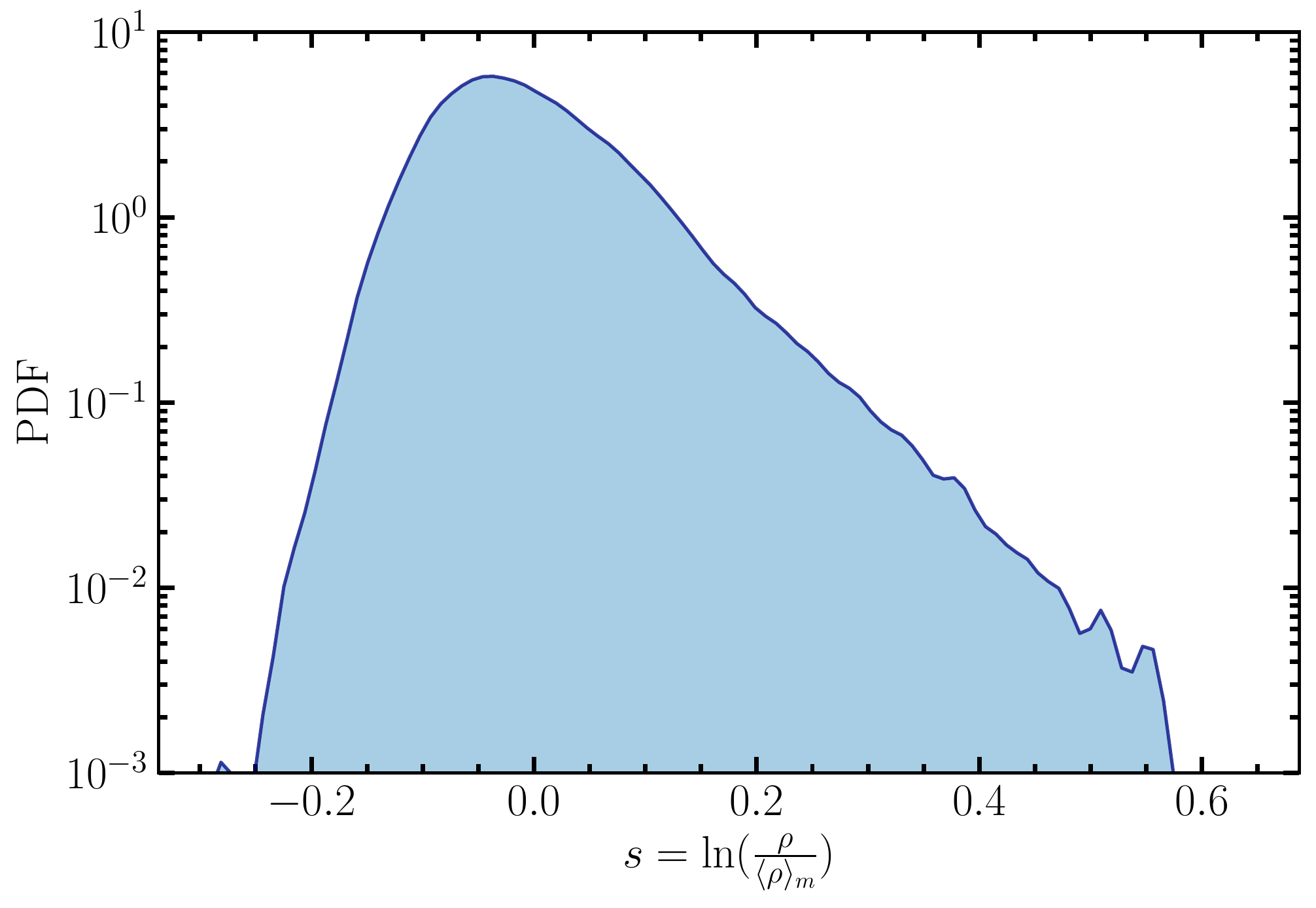}
    \caption{PDF of the logarithmic density contrast $s = \ln \rho/\langle \rho \rangle_m$, time-averaged across the tracer trajectories in the post-shock Lagrangian volume. The shape is similar to that analysed in \citet{dhawalikar2022driving}.}
    \label{fig:sm}
\end{figure}

\begin{figure}
	\includegraphics[width=\columnwidth]{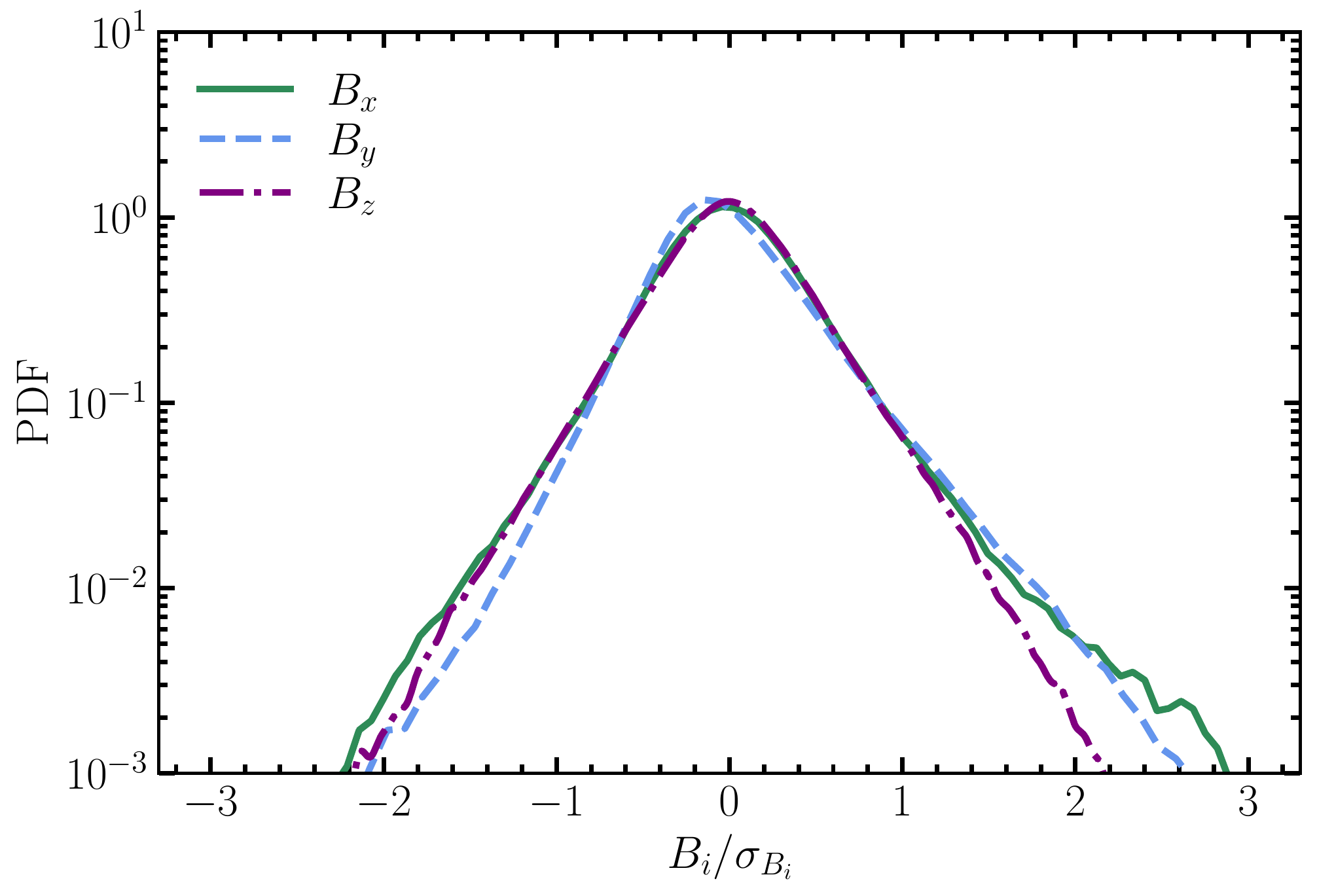}
    \caption{PDF of the turbulent magnetic field components $B_x$, $B_y$ and $B_z$ in the Lagrangian volume, based on the time-averaged trajectories of the tracer particles in the analysis box starting at $t \approx 0.13 T_{\textrm{ed}}$. The magnetic field components are non-Gaussian and display stretched tails due to the spatial intermittency naturally occurring in the kinematic stage of the dynamo as a result of Lagrangian chaos (random stretching of field lines) \citep{boldyrevske2001,schekokulsrud,seta2021saturation,seta2022turbulent}.}
    \label{fig:bpdf}
\end{figure}

\begin{figure}
	\includegraphics[width=\columnwidth]{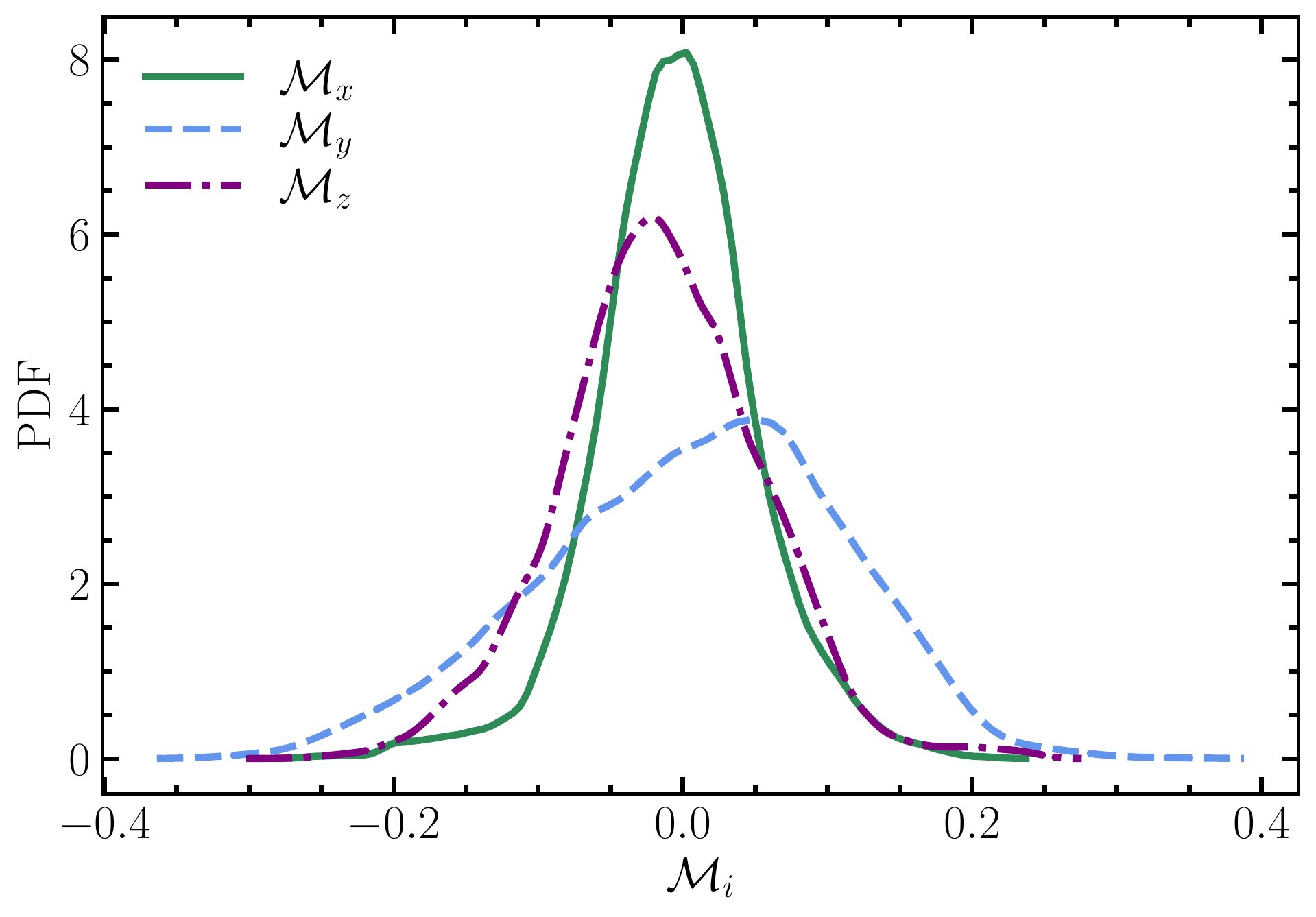}
    \caption{Same as Fig.~\ref{fig:bpdf}, but for the turbulent Mach number. $\mathcal{M}_y$ occupies a larger volume fraction compared to the other Mach number components, since it is in the shock direction. It therefore also displays somewhat more intermittent (non-Gaussian) features; similar to \citet{dhawalikar2022driving}.}
    \label{fig:mpdf}
\end{figure}

Further to this, we also find that the turbulent density dispersion (standard deviation of the density) amplifies by a factor of two, a value very similar to that observed in \cite{dhawalikar2022driving}, even with mass-averaged quantities. In order to quantify this, we show the probability distribution functions (PDFs) of the logarithmic density contrast $s = \ln(\rho/\langle \rho \rangle_m)$ time-averaged on the tracer particles within the Lagrangian volume in Fig.~\ref{fig:sm}, the magnetic field PDFs in Fig.~\ref{fig:bpdf}, and the Mach number PDFs in Fig.~\ref{fig:mpdf}. Here we notice that the density PDF displays salient characteristics similar to that found by \citet{dhawalikar2022driving}, with a log-normal for low to intermediate densities, and a power-law tail at high densities, despite the fact that we have utilised mass-averaged quantities, where it is known that substantial quantitative differences can exist \citep[see e.g.,][]{konstandin2012statistical}. The magnetic field PDFs in Fig.~\ref{fig:bpdf} show that the magnetic fields are spatially intermittent, with non-Gaussian stretched tails. This is consistent with the log-normality condition of the magnetic field PDF in the kinematic SSD  based on the white-in-time Fokker-Planck model \citep{boldyrevske2001,schekokulsrud,schekochihin2002c,schekochihin2004simulations} (i.e. the $B$ field components themselves will be non-Gaussian and spatially intermittent).  The $B_y$ component is slightly different than the rest, and occupies a slightly larger volume fraction. This is expected since the magnetic field in the shock direction is always larger than the other components, producing larger fluctuations compared to the $x$ and $z$ components. Nonetheless we note that the spatially intermittent character of the PDFs are indicators of the presence of the turbulent dynamo \citep{seta2021saturation,seta2022turbulent}, which has not yet reached saturation\footnote{At $v \sim B$ (saturated state), the log-normal magnetic field PDFs become increasingly Gaussian (non-intermittent), resembling then the quasi-normal velocity PDFs in a causal manner. This scenario is traced out nicely in \cite{,seta2021saturation,seta2022turbulent}.}. The Mach number PDFs (Fig.~\ref{fig:mpdf}) clearly illustrate a similar pattern as that observed for the magnetic ones, where the occupied volume in the shock direction is always larger due to the simple fact that it has larger variations near the shock front. They are, however, Gaussian, as expected for fully-developed turbulent flows \citep{Federrath2013,dhawalikar2022driving}. Overall, this highlights the role of the shock front in creating not only turbulent Mach number variations, but also turbulent magnetic field amplification as mentioned earlier, in the post-shock medium.


\begin{figure}
	\includegraphics[width=\columnwidth]{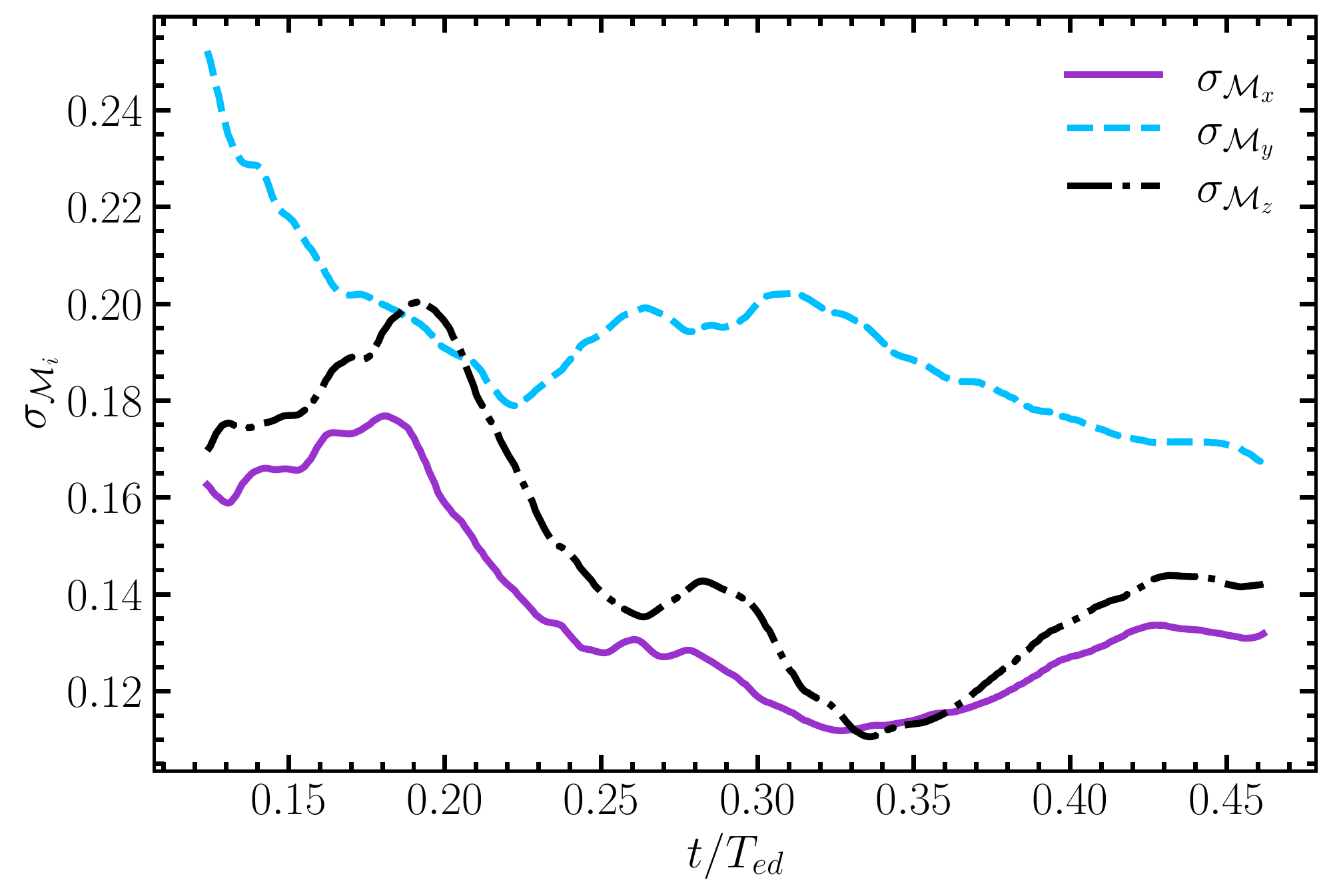}
    \caption{Time evolution of the Mach number components, $\mathcal{M}_x$, $\mathcal{M}_y$ ,$\mathcal{M}_z$, averaged across all the tracer trajectories.}
    \label{fig:machno}
\end{figure}
    %
         

\subsection{Vorticity evolution}
\begin{figure}
	\includegraphics[width=\columnwidth]{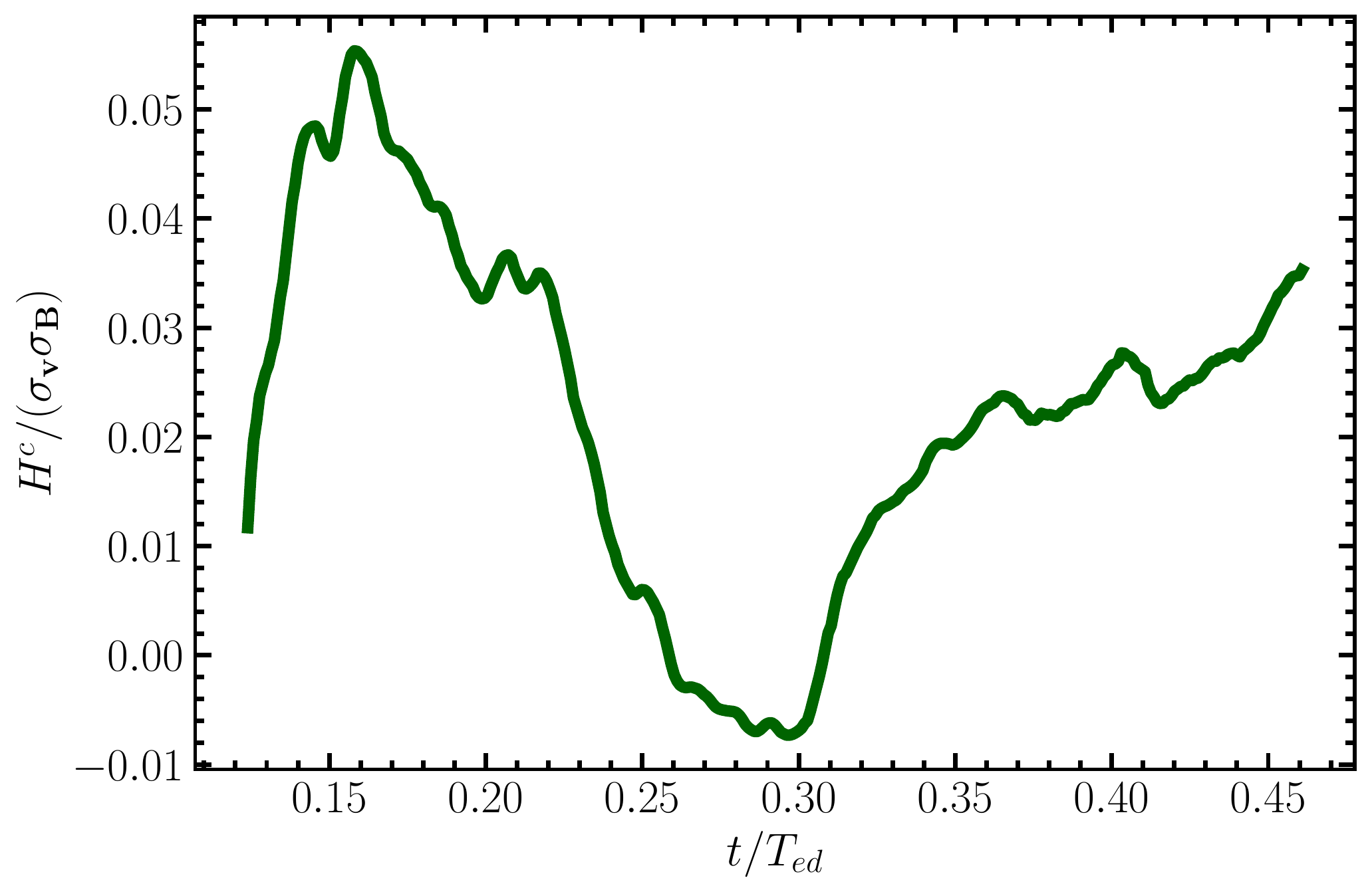}
    \caption{Time evolution of the normalised turbulent cross helicity across all tracer trajectores.}
    \label{fig:crosshel}
\end{figure}

The evolution of the small-scale dynamo is strongly influenced by the production of vorticity, which creates tangled field configurations and increase in the topological complexity of magnetic flux lines \citep{mee2006turbulence,federrath2011mach,seta2021saturation}. The quantities involved in this arise primarily from the non-linear term $\nabla \times ( \boldsymbol{v} \times \boldsymbol{B}$) in the MHD induction equation, which determines the electromotive force (e.m.f.) generation and consequently the magnetic field amplification. A quadratic invariant of the ideal MHD equations which quantifies the level of e.m.f.~production is the turbulent cross helicity ($H^c$), defined as
$H^c = \langle \mathbf{v} \cdot \mathbf{B} \rangle$, which defines the cross-correlation between the velocity and magnetic fields, and hence allows a quantitative measure of the degree of alignment between these two components \citep{yokoi1999magnetic,perez2009role,yokoi2013cross}. We show the normalised turbulent cross helicity $H^c/\sigma_{\mathbf{B}}\sigma{\mathbf{v}}$ in Fig.~\ref{fig:crosshel}. It can be seen that the 
turbulent cross helicity decreases in the initial time evolution up until $t = 0.3 t/T_{ed}$. This is associated with the gradual entanglement of the magnetic and velocity field lines, which explains the growth of the magnetic field during this period of the time evolution. Examining all component in Fig.~\ref{fig:sigmab}, we can observe an intimate connection between the cross helicity and the consequent decay of the magnetic fields at later time intervals. The associated increase of $H^c$ from $t \approx 0.3 t/T_{\textrm{ed}}$ leads to the increased alignment of $\mathbf{v}$ and $\mathbf{B}$, which inhibits the generation of the e.m.f. This explains the decay at late times in the magnetic fields.

\begin{figure}
	\includegraphics[width=\columnwidth]{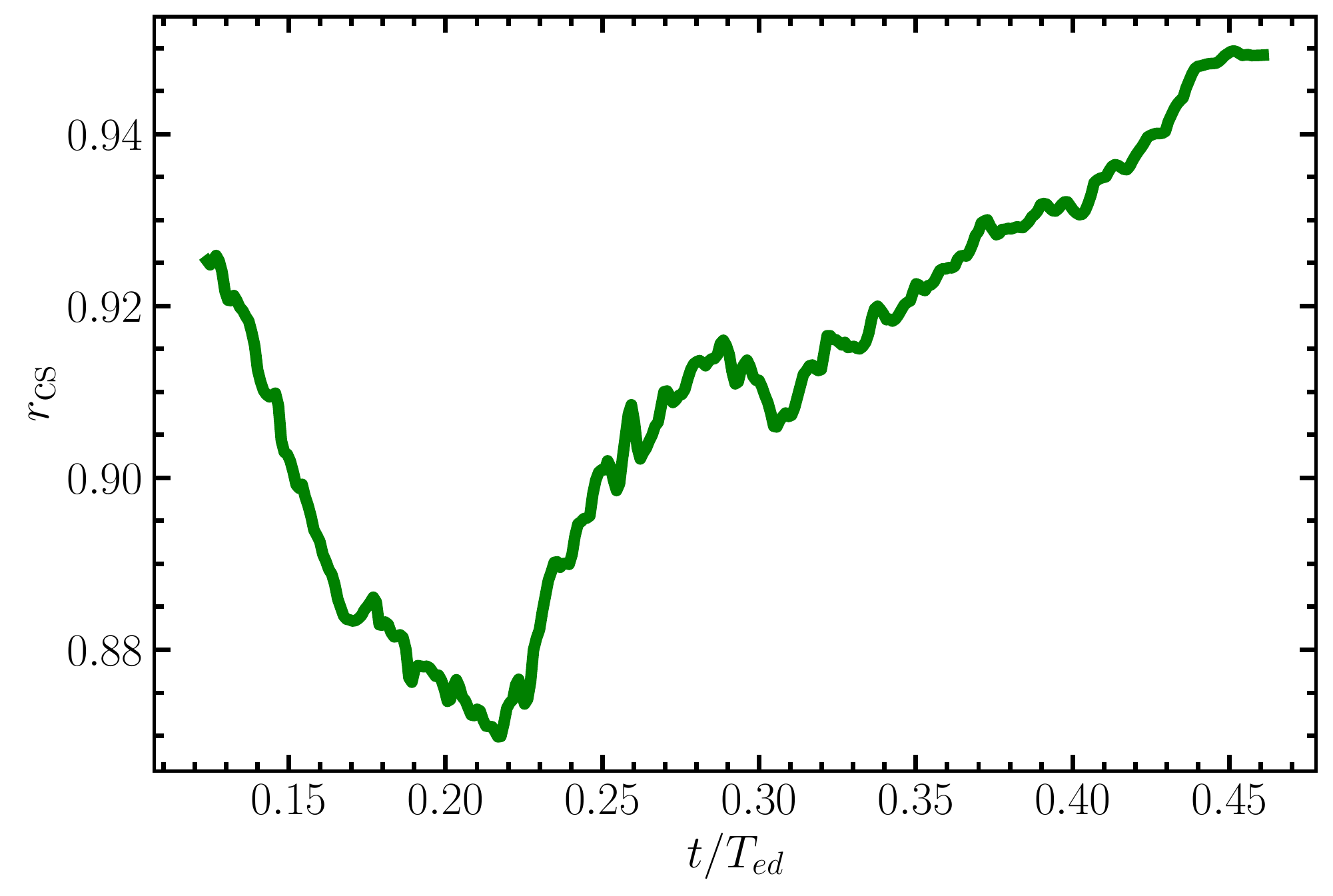}
    \caption{Time evolution of the small-scale solenoidal ratio as defined in Eqn.~\ref{eq:rcs}. This value is bounded in $[0,1]$ and therefore measures the relative strength of vorticity compared to the sum of vorticity and divergence (compression).}
    \label{fig:compratio}
\end{figure}

Further to this, in order to examine the contribution of small-scale solenoidal modes in the flow, we show the solenoidal ratio \citep{kida1990energy,kida1992energy,kritsuk2007statistics,federrath2010comparing,pan2016supernova}, defined as
\begin{equation}\label{eq:rcs}
    r_{\mathrm{cs}} \equiv \frac{\left\langle|\nabla \times \boldsymbol{v}|^2\right\rangle}{\left\langle|\nabla \cdot \boldsymbol{v}|^2\right\rangle+\left\langle|\nabla \times \boldsymbol{v}|^2\right\rangle},
\end{equation}
 which measures the contribution of the vorticity ($\boldsymbol{\omega} = \nabla \times \boldsymbol{v}$) relative to the full velocity field (sum of vorticity and divergence). This ratio is bounded in $[0,1]$, and thus provides a good indicator of the vorticity fraction in the local flow. Fig.~\ref{fig:compratio} displays this ratio, and shows that at the small scales for which this quantity is computed, the solenoidal modes ($\nabla \times \boldsymbol{v}$) are much larger than the contributions from compressive modes ($\nabla \cdot \boldsymbol{v}$). High values are expected in the case of post-shock turbulence \citep{kritsuk2007statistics,pan2016supernova}, since such drivers, while compressive in nature, still tend to induce high fractions of solenoidal modes in the flow \citep{federrath2010comparing,kritsuk2011comparing,federrath2013star}.
 
\begin{figure}
	\includegraphics[width=\columnwidth]{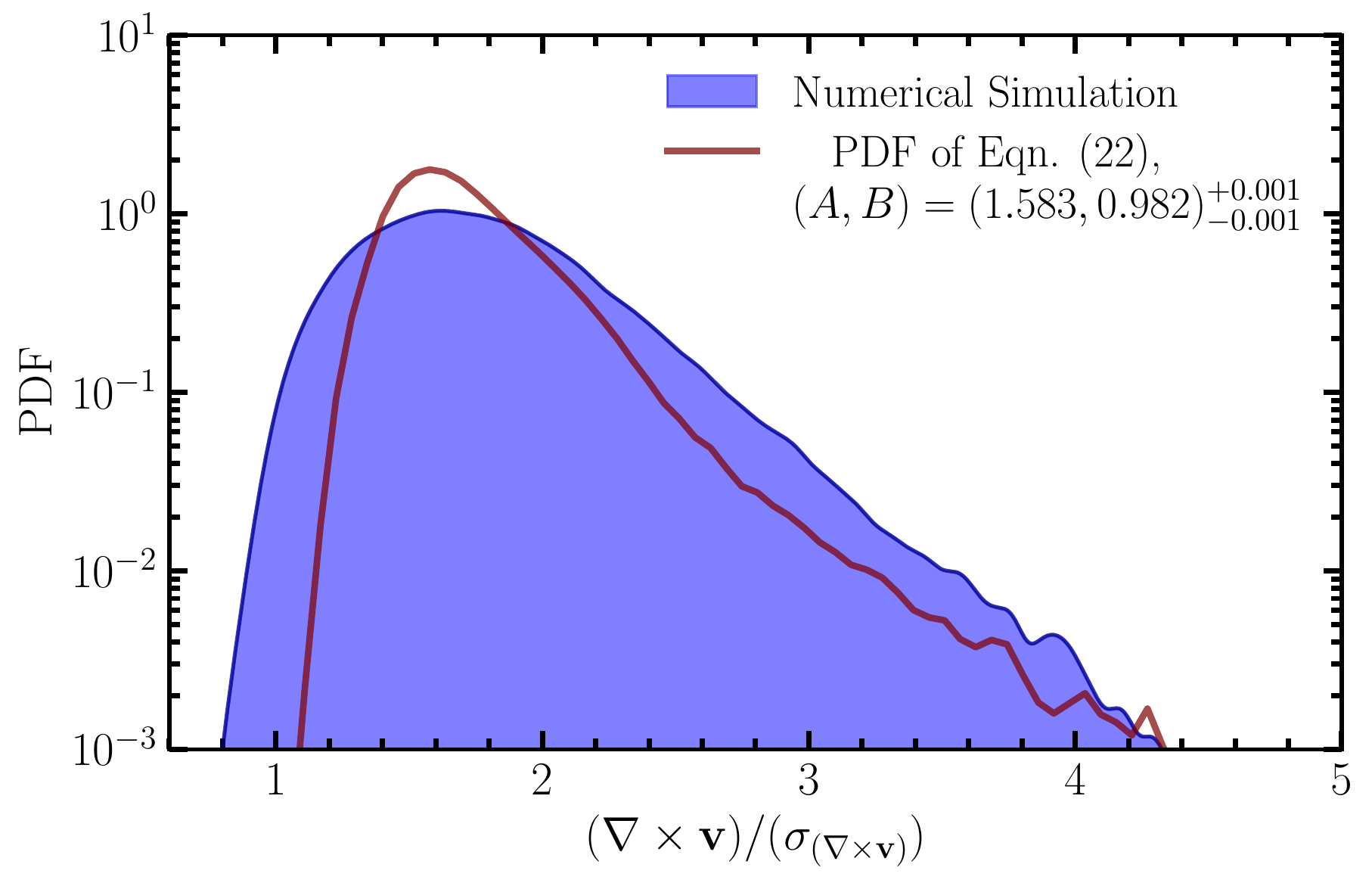}
    \caption{PDF of the vorticity, $\omega = \nabla \times \boldsymbol{v}$, normalised by its standard deviation. Similar to the log-normal density PDF (Fig.~\ref{fig:sm}), the vorticity PDF also shows a Gaussian plus power-law shape. Thus, we fit a semi-analytical model PDF that is directly related to the logarithmic density contrast ($s$), based on the vorticity generation behind a curved shock front (Eqn.~\ref{eq:pdfeqnvort}), assuming negligible baroclinicity, constant shock curvature and near self-similarity of the shock profile.  }
    \label{fig:vortpdf}
\end{figure}

 Fig.~\ref{fig:vortpdf}~displays the vorticity PDF, which shows a similar shape as the logarithmic density PDF (cf.~Fig.~\ref{fig:sm}), with a power-law tail at higher vorticity levels. We attribute this to the fact that not all regions in space have uniformly-distributed vorticity, and thus large-scale contributions only exist intermittently in space within the post-shock medium. Such structures may also explain the intermittency observed in the magnetic field PDFs (Fig.~\ref{fig:bpdf}), since intermittent magnetic field variations are strongly linked to vorticity production \citep{mee2006turbulence,federrath2011mach,seta2021saturation}. 

Furthermore, we show that the connection between the vorticity and logarithmic density contrast PDFs (Figs.~\ref{fig:vortpdf} and \ref{fig:sm}) lie in the fact that vorticity generation behind a three-dimensional curved shock front has an analytical relation that is related to the density perturbations \citep{kevlahan1997vorticity,kevlahan2009apj}:
\begin{equation}
\delta \omega=\frac{\mu^2}{1+\mu} \frac{\partial C_r}{\partial S}-\frac{\mu}{C_r}\left[\left(\frac{\mathrm{D} \boldsymbol{v}}{\mathrm{D} t}\right)_S+\frac{C_r^2}{1+\mu} \frac{1}{\rho} \frac{\partial \rho}{\partial S}\right]+\mu \boldsymbol{\omega} 
\end{equation}
where $C_r$ is the velocity in the shock-normal frame, $\mu$ is the normalised density jump across the shock, $\partial /\partial S$ is the tangential component of the directional derivative and $S$ denotes the shock tangential surface. For the sake of simplicity, we assume that the flow ahead of the shock is initially uniform, which reduces it to a well-known result
\citep{hayes1957vorticity,kanwal1959curved}, given by
\begin{equation}
\delta \omega \boldsymbol{b}=-\frac{\mu^2}{1+\mu} \boldsymbol{n} \times\left(\boldsymbol{v}_\mathrm{shock} \cdot \boldsymbol{K}+\frac{\partial C_r}{\partial S}\right)_S
\end{equation}
where $\boldsymbol{b}$ and $\boldsymbol{K}$ denote the shock-tangential direction and shock curvature, respectively and $\boldsymbol{v}_\mathrm{shock}$ is the shock velocity. Since $\mu \sim \exp({s}) - 1$, we have:
\begin{equation}\label{eq:pdfeqnvort}
    \delta \omega \sim \frac{\mu^2}{1+ \mu} \simeq \frac{A \left[ \exp(s) -1 \right]^2}{1 + B \left[\exp({s}) -1\right]}
\end{equation}
if we assume a mostly pseudo-stationary (pseudo-steady) shock (i.e.,~$v_\mathrm{shock} , ~\partial C_r/\partial S \simeq~\mathrm{const}$) as well as constant shock curvature ($ \boldsymbol{\lvert K \rvert} \simeq \mathrm{const}$), which leaves behind the free parameters $A$ and $B$. Taking the PDF of Eqn.~\ref{eq:pdfeqnvort} in the moving post-shock frame, we find reasonably close agreement between the model and the vorticity PDF (Fig.~\ref{fig:vortpdf}), bearing in mind the aforementioned assumptions. This therefore shows the strong connection between the logarithmic density contrast $s$ and the vorticity generation behind a shock. While the model PDF we derive here also neglects vorticity contribution from the baroclinic term, which generates vorticity through the misalignment between pressure and density gradients ($\nabla p_{\textrm{th}} \times \nabla \rho)$, the fact that it still suffices to predict the overall shape of the long-tailed intermittent distribution suggests that baroclinicity may not play a crucial role in highly subsonic, post-shock turbulence, as has already been reported previously \citep{mee2006turbulence,federrath2011mach,livescu2016,federrath2016magnetic,tian2019,AchikanathEtAl2021}; while such effects, are usually magnified in pre-shock, supersonic turbulence \citep[e.g., cosmic-ray pressure gradients; see][]{beresnyak2009turbulence,drury2012turbulent,2014downes}. Moreover, the close agreement between the PDFs elucidate that shock curvature effects play a pre-dominant role in vorticity generation within post-shock turbulence, and also further solidifies that we have successfully isolated the turbulence generation behind a shock front by employing the Lagrangian frame of reference.

\subsection{Dynamo amplification}
With the analyses above, we have established that dynamo action is present in the post-shock turbulent medium in our simulations. Here we educe the magnitude of its amplification, and compare it to values obtained for dynamos in the literature \citep{federrath2011mach,XL2016}. Firstly, we conduct two additional simulations with the exact same parameters, but only vary the seed for the foam void distribution, and subsequently take the average of the values from all three of them. The different seeds were also found to not influence the overall dynamics of the system, which gives confidence to the numerical results. Averaging over these additional seeds is merely to improve the statistical significance of our results and to allow for a more accurate determination of the growth rate of the dynamo in the post-shock medium.

\begin{figure}
	\includegraphics[width=\columnwidth]{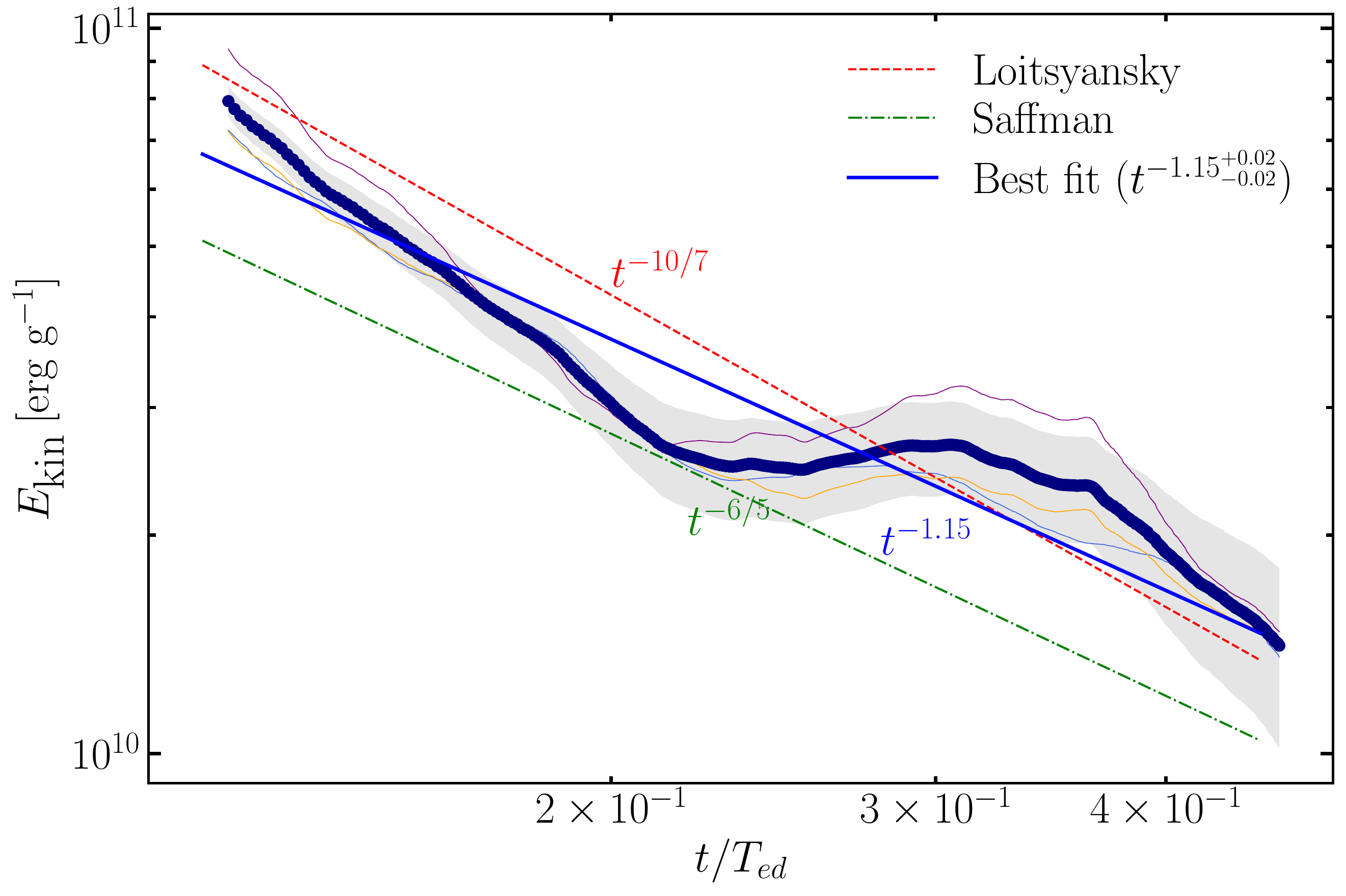}
    \caption{Time evolution of the kinetic energy of the simulation data (thick black line), with best-fit line and scaling parameters obtained as $t^{-1.15}$ shown as the blue solid line. The scalings obtained for the Loitsyansky and Saffman invariants are shown for comparison, as the red dotted and green dash-dotted lines, respectively. Thin lines show individual simulations with three different random seeds for the foam, which are used to obtain the averaged line (thick black line) with the 1-sigma band shown as the shaded grey region.}
    \label{fig:ekin}
\end{figure}

We further examine the level of turbulent diffusion by plotting $E_{\textrm{kin}}$, as shown in Fig.~\ref{fig:ekin}. It can be clearly seen that in less than half a turnover time, the kinetic energy drops by an about a factor of 6, as reflected also in the turbulent velocity components. We fit the scaling of $E_{\textrm{kin}}$ in our simulations, averaged across the three different seeds, and find that $E_{\textrm{kin}} \sim t^{-1.15 \pm 0.02}$. This value of the power-law exponent of the decay is very close to the Saffman integral invariant, which goes as $t^{-6/5}$. Interestingly, this value is also very similar to that observed by \citet{mac1998kinetic} for their subsonic case, which had a scaling of $t^{-1.1}$. This is consistent with scaling expected in kinetically dominated turbulence. As mentioned earlier, many numerical experiments, \citep{biskamp1999decay,biskamp2000scaling,christensson2001inverse,banerjee2004evolution,frick2010long,berera2014magnetic,brandenburg2015nonhelical,brandenburg2017classes,reppin2017nonhelical,sur2019decaying,bhat2021inverse} have also observed scalings between the range of the Saffman integral and that of \citet{biskamp1999decay}, where the exact decay law should depend on whether $v \sim B$, $v \ll B$ or $v \gg B$. Thus, we find that the system undergoes significant turbulence decay, and the dynamo effect will most likely no longer be sustained after a long time evolution, at least not at the same intensity as compared to early times when the turbulence is still strong. This is consistent with previous works. It also shows that in such a decaying system, the dynamo growth rate is time dependent, at least when quantified over a significant amount of time, due to the time-dependence of the large-scale turbulent turnover time. Such an observation, has also been made for helical large-scale $\alpha^2$-dynamos \citep{brandenburg2019dynamo}.

Now, in order to fully capture the dynamo-induced magnetic field amplification, we note that the shock-normal streamwise field always has higher amplifications than the rest. This is attributed to the compression at the shock front, and primarily a result of the large-scale systematic stretching of field along the shock propagation direction. Thus, we neglect this contribution, because we want isolate the truly turbulent amplification process, and therefore only calculate the density-normalised magnetic energy for components parallel to the shock front ($B_x$ and $B_z$).

\begin{figure}
	\includegraphics[width=\columnwidth]{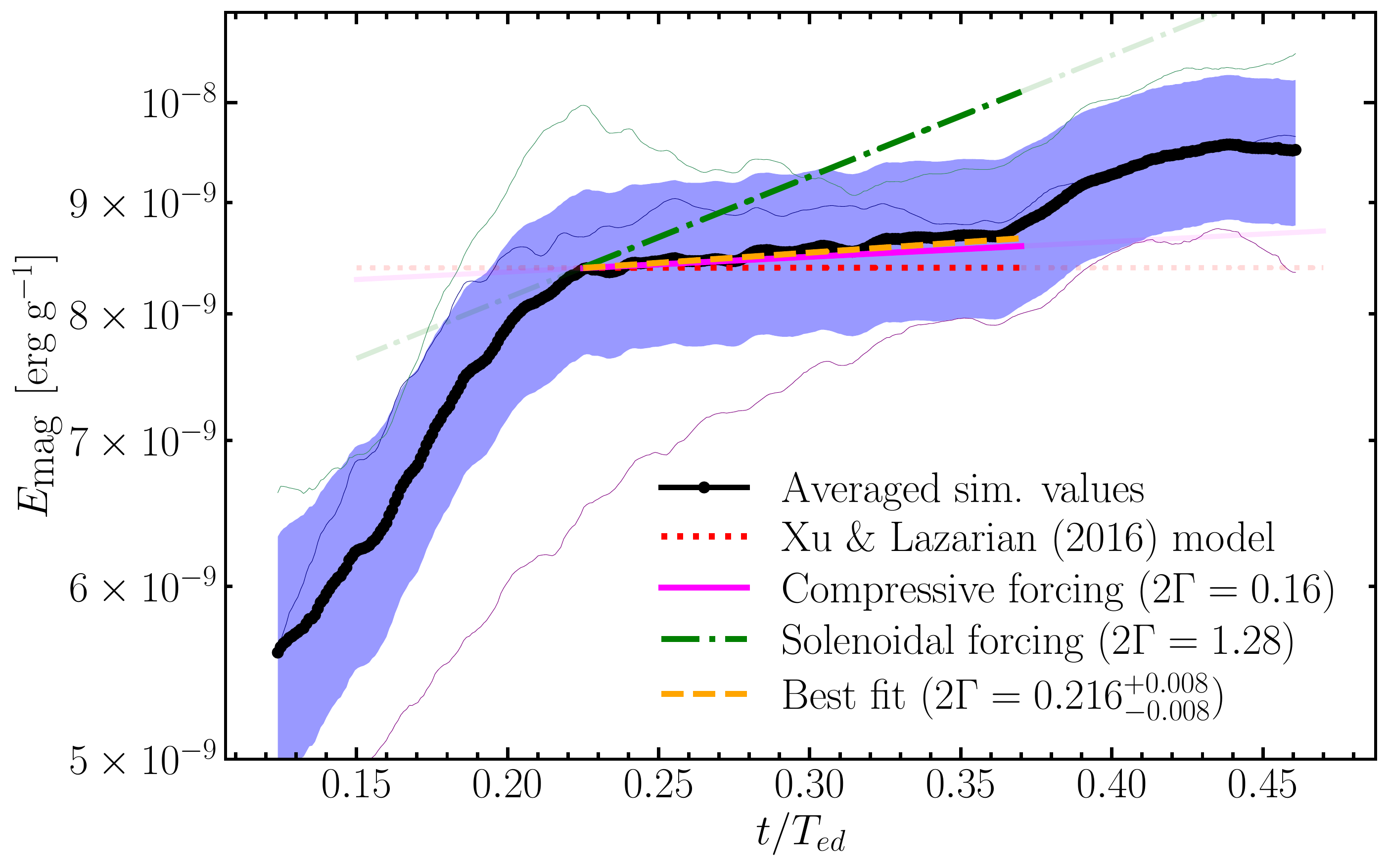}
    \caption{Time evolution of the specific magnetic energy ($E_{\textrm{mag}} = 1/2 V_A^2 $), averaged across the three different seeds for the foam void distribution. We compare the growth rates in the region where an exponential growth is observed, with rates expected for compressive and solenoidal turbulence driving mechanisms \citep{federrath2011mach}, as well as the analytical model of \citet{XL2016}.}
    \label{fig:emagcompre}
\end{figure}

Fig.~\ref{fig:emagcompre} shows the magnetic energy as a function of time. As mentioned before, there are seemingly no distinct phases or stages for the evolution of the magnetic energy, because the time to observe dynamo amplification during the onset of decaying turbulence originating from turbulent (numerical) diffusion is very short, only $\sim$ 0.3 of a turbulent turnover time. We find that in the intermediate range of time scales at $t\approx 0.195 - 0.380 t /T_{\textrm{ed}}$, the growth is very close to exponential. We attribute the initial growth of the field to a numerical transient, where the field experiences a sudden growth at early stages of its evolution due to the prior strong shock compression. The later stages are also neglected in consideration that many of the tracer trajectories have exited the medium with the propagating shock, and thus may not be able to capture the full temporal dynamics of the magnetic energy. 

Thus, we fit the growth rate in this time window, where $2 \Gamma  = 0.216 \pm 0.008$ is the best fit obtained. The time-averaged Mach number is $\mathcal{M} = 0.31$ (Fig.~\ref{fig:machno}). Based on measurements of the growth rate in driven turbulence box simulations by \citet{federrath2011mach} and \citet{AchikanathEtAl2021}, purely solenoidal driving would yield a growth rate near unity, while purely compressive driving would yield $2 \Gamma_{\textrm{comp}} = 0.16$, close to what we find for the present shock-induced simulations.

For purposes of further comparisons with dynamos where clear, distinct phases can be observed (kinematic, nonlinear, saturated), we also show the prediction of the \citet{XL2016} non-linear phase model (Eq.~\ref{eqn:xu}),\begin{equation}
    E_{\textrm{mag}} = E_{\textrm{initial}} + \frac{3}{38} \epsilon (t-t_{\textrm{initial}}),
\end{equation}
where $E_{\textrm{initial}}$ and $t_{\textrm{initial}}$ correspond to the initial magnetic energy and time where the dynamo process begins. Here, we find that the model is able to predict the growth of the magnetic field we observed in the averaged data from all three of our numerical simulations with reasonable accuracy, although we must emphasise that it applies only in a non-linear phase, with the assumption of Kazantsev-Kraichnan phenomenology for solenoidally forced (not decaying) turbulence. Thus, in the presence of compressive driving, we do not expect that the non-linear growth phase to be well-captured by the analytical model.  

To further educe the overall growth rate, we use the semi-empirical estimate provided by \cite{kulsrud2005plasma} (see also~\cite{fraschetti2013}~and Appendix~\ref{appen:b} in this work, where we provide a derivation), which assumes homogeneity and isotropy of the velocity two-point correlator to obtain a relation between the growth rate, $\Gamma$ (in units of $T_{\textrm{ed}}^{-1}$) and the vorticity induced downstream of a shock, $\lvert \boldsymbol{\omega} \rvert$ as:
\begin{equation}\label{eqn:21}
\Gamma \approx \frac{\pi}{3} \lvert \boldsymbol{\omega} \rvert T_{\textrm{ed}} 
\end{equation}
\begin{figure}
	\includegraphics[width=\columnwidth]{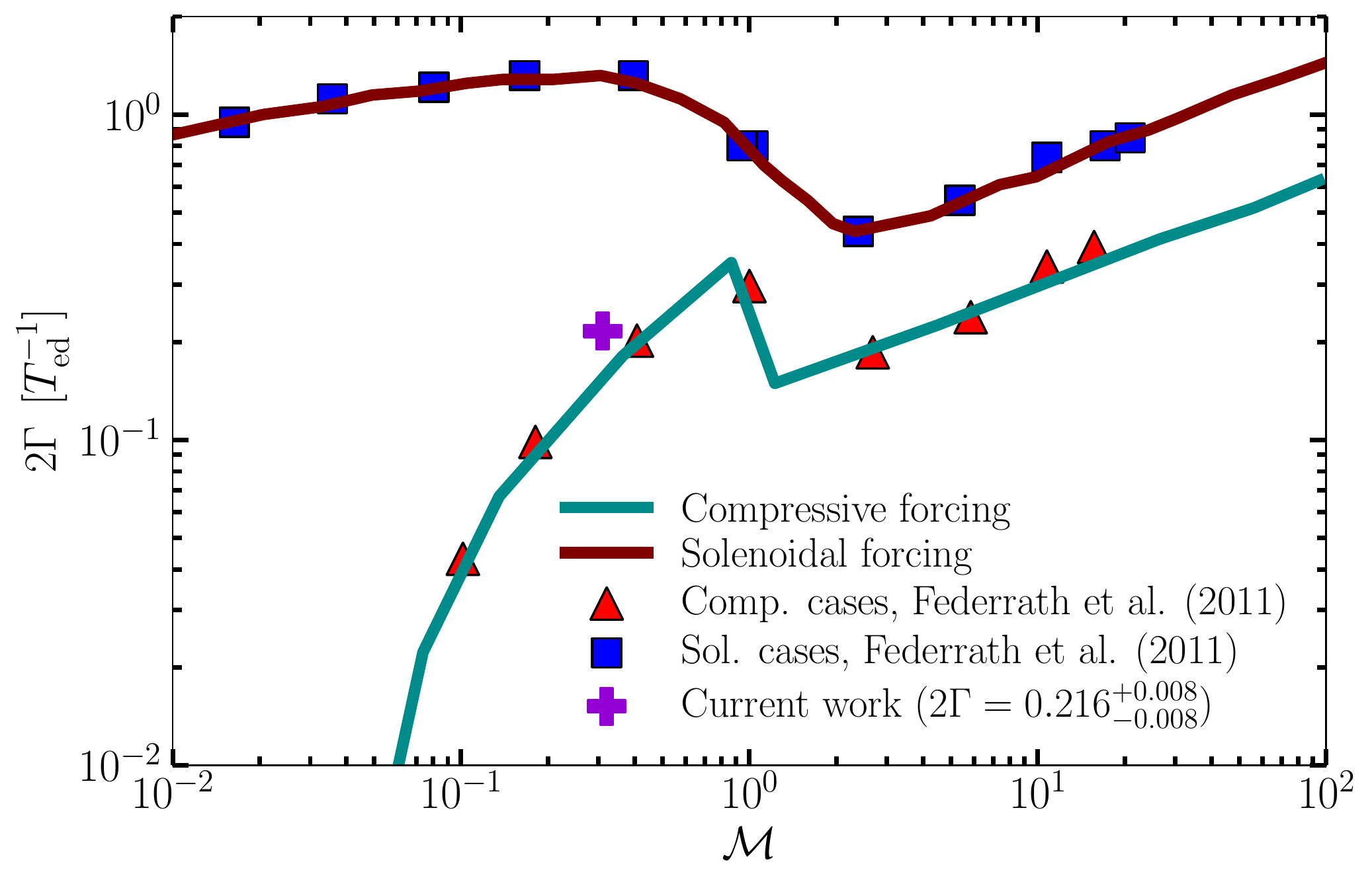}
    \caption{Plot of the magnetic energy growth rate ($2\Gamma$) as a function of Mach number, $\mathcal{M}$, with the value obtained from simulations in the current work, along with the propagated error. Comparisons are made to the empirical fit from \citet{federrath2011mach} for compressively- and solenoidally-driven turbulence, as well as corresponding simulation data obtained in their work. }
    \label{fig:emagcompare}
\end{figure}
In \cite{fraschetti2013}, it was assumed that the pre-shock medium has initially zero vorticity, $\lvert \boldsymbol{\omega}_0  \rvert= 0 $. In three-dimensional simulations, we find that this is not the case. Thus, we divide the mean vorticity evolution with $\lvert \boldsymbol{\omega}_0 \rvert$ in order to consider only the vorticity driven by the shock. Noting that this is an order of magnitude estimate, the post-shock vorticity from our simulations is~$ \lvert \boldsymbol{\omega} \rvert / \lvert \boldsymbol{\omega}_0 \rvert \approx 
0.5 \times 10^6$, this yields $2\Gamma \approx 0.2 \pm 0.1$, which is close to what we find in our measured growth rates.

Thus, all the above estimates further provide confidence that there is an inherent turbulent dynamo mechanism within the post-shock turbulent flow, and that it corresponds well with the growth rates expected for compressively-driven turbulence as shown earlier. This is also consistent with the observations of \citet{dhawalikar2022driving}, since their work demonstrated that the driving mode of shock-driven turbulence is primarily compressive, rather than solenoidal.

Finally, we show the measured growth rate averaged from our three simulations (Fig.~\ref{fig:emagcompare})~together with those expected for compressively- and solenoidally-driven turbulence \citep{federrath2011mach}, further confirming that the shock-driven turbulent dynamo growth rate exhibited in our simulations are very close to that of a compressively-driven turbulent system.

\subsection{Second-order statistics of the velocity and magnetic field}
\begin{figure}
	\includegraphics[width=\columnwidth]{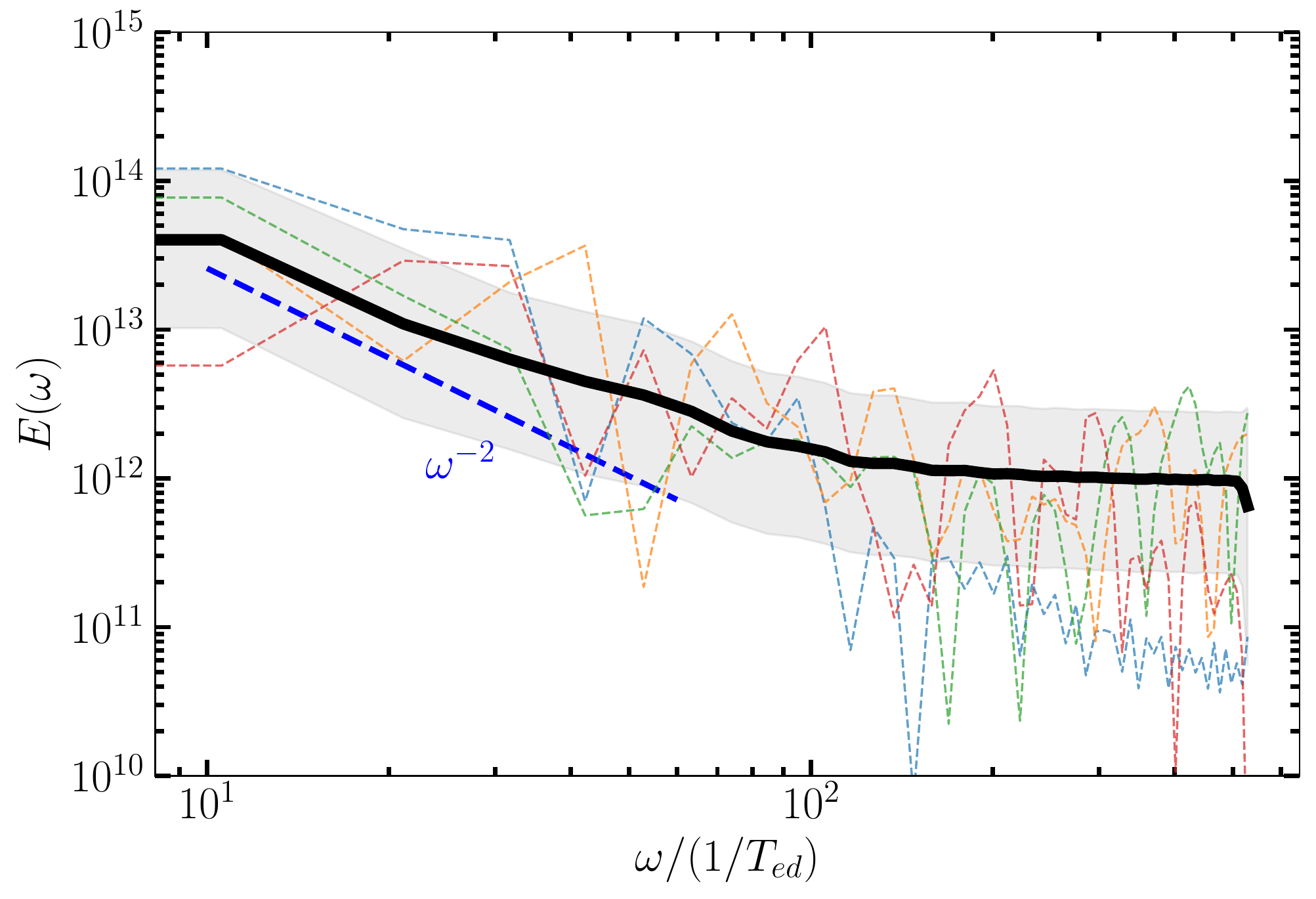}
    \caption{Lagrangian frequency spectrum of the velocity fluctuations, the solid black line is mean spectra across all tracer trajectories within the analysis box, and the shaded region indicates $16$-th and $84$-th percentile
    from the mode. Coloured dashed lines are energy spectra of random singular trajectories. A near $\omega^{-2}$ scaling is observed at the inner scale, which is consistent with the K41 Lagrangian frequency scaling.}
    \label{fig:velpsd}
\end{figure}

\begin{figure}
	\includegraphics[width=\columnwidth]{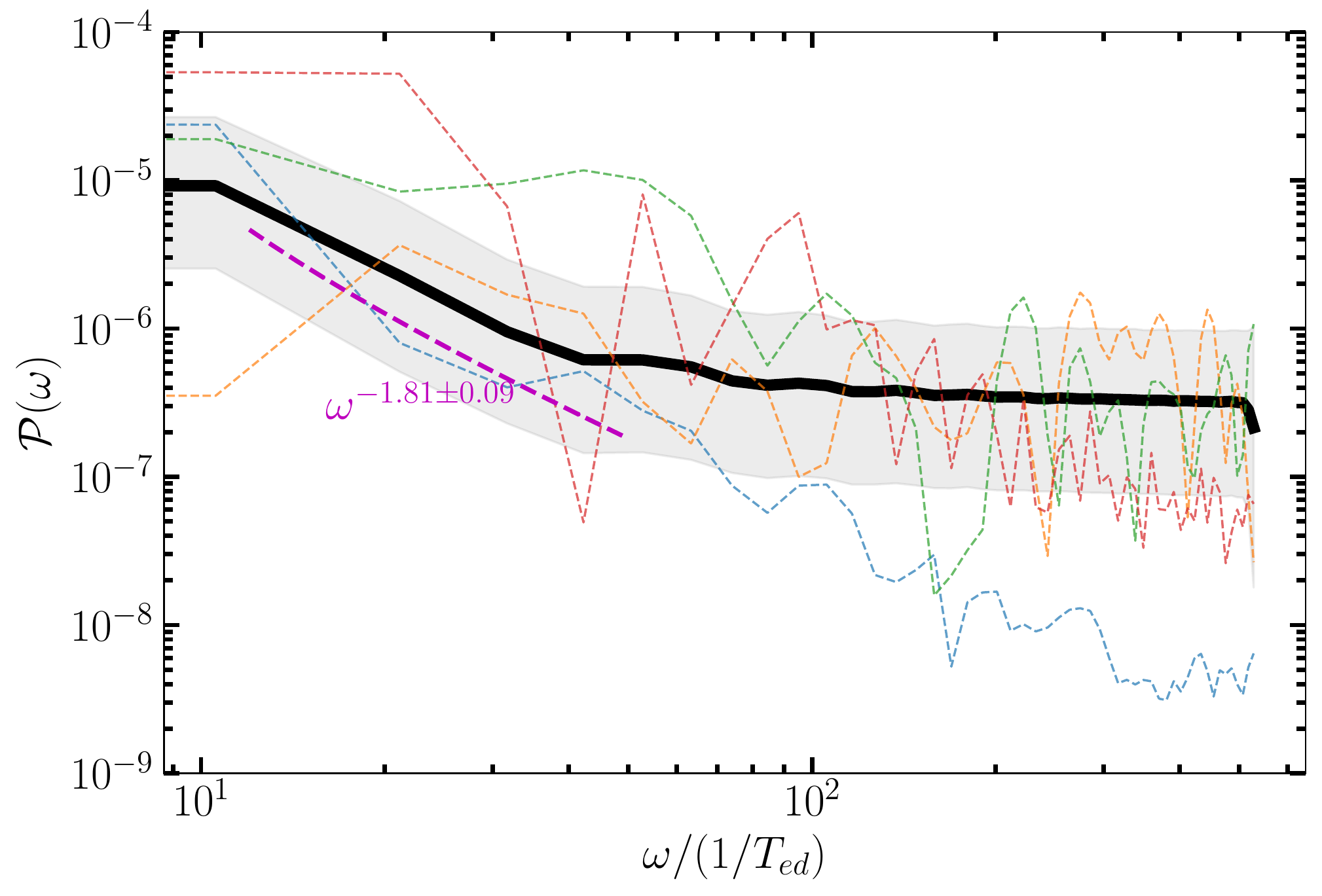}
    \caption{Same as Fig.~\ref{fig:velpsd}, but for the turbulent magnetic field spectrum. Here we observe a slightly shallower spectrum than the velocity field.  }
    \label{fig:magpsd}
\end{figure}

Now we consider the second-order statistics in the form of the Lagrangian frequency spectrum \citep{tennekes1972first,tennekes1975eulerian,busse2010lagrangian,homann2014structures,beresnyak2019mhd}. We plot both the kinetic and magnetic energy spectra, via the cosine transform of their temporal auto-correlation functions,
\begin{equation}
\Phi(\omega)= \frac{1}{2 \pi} \int d \tau \langle Q_i(t+\tau) Q_i(t)\rangle \cos (\omega \tau),
\end{equation}
where $Q = \mathbf{B}$ or $\mathbf{u}$, and where $\tau$ is the time lag from the standard two-point correlation function. The Lagrangian frequency spectrum is computed for all tracers, and then averaged to obtain the mean spectra. The velocity and magnetic field spectra are displayed in Fig.~\ref{fig:velpsd} and Fig.~\ref{fig:magpsd}. It can be seen that the velocity spectra show a spectral scaling consistent with that of the Lagrangian bridge for the Kolmogorov scaling, $E(\omega) \sim \omega^{-2}$, within the $16$-th to $84$-th percentile range. As mentioned earlier, such scalings have been observed in three-dimensional incompressible MHD simulations \citep{busse2010lagrangian}, hydrodynamic simulations \citep{yeung2006reynolds} and experiments \citep{mordant2004experimental}. Thus, we also observe these power-law scalings even in the presence of large-scale compression, where the slight deviation exists likely due to compressibility effects and small-scale intermittencies commonly observed in Lagrangian statistics even with high Reynolds number turbulence \citep{homann2007lagrangian,arneodo2008universal, benzi2010inertial, busse2010lagrangian, konstandin2012statistical}. To our knowledge, this is the first discussion and verification of the scaling of the Lagrangian frequency spectrum in the context of post-shock MHD turbulent flows.

The magnetic spectrum, however, displays fundamental differences from its Eulerian counterpart. There are seemingly no visible scale separations within it, which one would see in the Eulerian framework, i.e., a typical peak scale and driving scale which is to be expected in an Eulerian magnetic spectrum \citep{schekochihin2004simulations,schober2015saturation,brandenburg2019dynamo,seta2020seed}. In fact, the shape of the magnetic spectra in our simulations resembles those of \cite{homann2014structures} (cf., Fig.~9 in their paper), with somewhat similar scaling. Most importantly, it also corresponds well with the findings of \cite{busse2010lagrangian}, that the total spectra of both velocity and magnetic field (i.e. for the Elsässer field $\mathbf{z}^+ = \boldsymbol{v} + \boldsymbol{B}$) should scale roughly as $\omega^{-2}$. The overall features nevertheless shows a clear power-law turbulent cascade, which is expected for the magnetic energy spectrum,  where energies are at a range from large to small scales due to the fundamental property of inertial range cascading turbulence. However, the intrinsic properties of the Lagrangian magnetic spectrum still remains to be fully understood, and thus should be further investigated beyond this context, and also beyond the scope of this paper.

\section{Conclusions}\label{sec:conclusion}
In this study, we performed numerical experiments of shock-driven MHD turbulence to investigate the turbulent dynamo induced magnetic field amplification through the Lagrangian framework for the first time. We followed the moving post-shock turbulent shell, in order to capture the full temporal dynamics of the post-shock medium, while avoiding spurious amplifications from Richtmyer-Meshkov related stratified shear instabilities, and thus found that the growth rates of the dynamo are comparable to turbulence driving in the ISM, for subsonic, compressively-driven turbulence. The overall setup and evolution is consistent with the hydrodynamic simulations of \citet{dhawalikar2022driving}, but we here focus on the magnetic field amplification using Lagrangian tracer particle tracking of the turbulent post-shock medium.
We summarise our main findings as follows:
\begin{enumerate}
  \item The shock-driven turbulent dynamo, in the presence of decaying hydrodynamic turbulence displays slightly different characteristics than its forced periodic box counterparts. This is particularly because the shock passage is usually quite short (e.g., \cite{2022davidovits,dhawalikar2022driving,hu2022turbulent}), which in our simulation, leads to only a time evolution of about $\sim 0.3$ turbulent turnover time.  Therefore, we only observe exponential or `kinematic' phase growth rate of the magnetic field due to magnetic excitation from the viscous scale, which does not achieve saturation. The decay in the kinetic energy further complicates the system by making continual amplifications impossible in long time evolutions, which we expect will lead to a dynamical saturation pathway of the SSD, where $E_{\textrm{mag}}$ and $E_{\textrm{kin}}$ both decay as $\sim t^{-n}$, ensuring that the turbulence remains Alfv\'enic ($\delta B \sim \delta v$) as shown in some periodic box simulations (e.g., \cite{kiwan,sur2019decaying,brandenburg2019dynamo}). Turbulent cross-helicity measurements also clearly indicate that the velocity and magnetic fields become more aligned, due to the decrease in turbulent kinetic energy and fluctuations. These contribute to the overall inefficiency in the dynamo process \citep{mac1998kinetic,sur2019decaying}.
  
  \item It has also been shown that the dynamo kinematic growth rate in this configuration matches that obtained for driven turbulence in the subsonic, compressive-driving regime. This result is consistent with prior works on shock-driven turbulence in periodic boxes.  Therefore, if the turbulent magnetic field amplification is completely isolated as uniquely done 
 here through the 
 post-shock Lagrangian framework, the salient features of dynamo action remain the same. 

  \item The kinetic energy decay rate found in our simulations is very close to the Saffman scaling, as well as to subsonic turbulence simulations in prior works. These all highlight that the dynamo effect cannot be sustained over long time periods without external driving.

  \item The Lagrangian frequency spectra of the magnetic and velocity fields display similar scalings, and they are comparable to that found in prior works, as well as that expected from the Kolmogorov theory. This is shown for the first time in the context of shock-driven turbulence.
\end{enumerate}

\section*{Acknowledgements}
We thank Siyao Xu and Yue Hu for their valuable comments on the manuscript. We further thank Turlough Downes for helpful discussions. We also thank the anonymous referee for their constructive feedback on the manuscript. We acknowledge the NIF Discovery Science Program for allocating upcoming facility time on the NIF Laser to test aspects of the models and simulations discussed in this paper. J.K.J.H. acknowledges funding via the ANU Chancellor's International Scholarship. C.F.~acknowledges funding provided by the Australian Research Council (Future Fellowship FT180100495 and Discovery Projects DP230102280), and the Australia-Germany Joint Research Cooperation Scheme (UA-DAAD). We~further acknowledge high-performance computing resources provided by the Leibniz Rechenzentrum and the Gauss Centre for Supercomputing (grants~pr32lo, pr48pi and GCS Large-scale project~10391), the Australian National Computational Infrastructure (grant~ek9) and the Pawsey Supercomputing Centre (grant~pawsey0810) in the framework of the National Computational Merit Allocation Scheme and the ANU Merit Allocation Scheme. The simulation software FLASH was in part developed by the DOE-supported Flash Center for Computational Science at the University of Chicago.

\section*{Data Availability}
The simulation data presented in this work are available on reasonable request to the corresponding author.



\bibliographystyle{mnras}
\bibliography{example,federrath} 

\begin{thebibliography}{}
\makeatletter
\relax
\def\mn@urlcharsother{\let\do\@makeother \do\$\do\&\do\#\do\^\do\_\do\%\do\~}
\def\mn@doi{\begingroup\mn@urlcharsother \@ifnextchar [ {\mn@doi@}
  {\mn@doi@[]}}
\def\mn@doi@[#1]#2{\def\@tempa{#1}\ifx\@tempa\@empty \href
  {http://dx.doi.org/#2} {doi:#2}\else \href {http://dx.doi.org/#2} {#1}\fi
  \endgroup}
\def\mn@eprint#1#2{\mn@eprint@#1:#2::\@nil}
\def\mn@eprint@arXiv#1{\href {http://arxiv.org/abs/#1} {{\tt arXiv:#1}}}
\def\mn@eprint@dblp#1{\href {http://dblp.uni-trier.de/rec/bibtex/#1.xml}
  {dblp:#1}}
\def\mn@eprint@#1:#2:#3:#4\@nil{\def\@tempa {#1}\def\@tempb {#2}\def\@tempc
  {#3}\ifx \@tempc \@empty \let \@tempc \@tempb \let \@tempb \@tempa \fi \ifx
  \@tempb \@empty \def\@tempb {arXiv}\fi \@ifundefined
  {mn@eprint@\@tempb}{\@tempb:\@tempc}{\expandafter \expandafter \csname
  mn@eprint@\@tempb\endcsname \expandafter{\@tempc}}}

\bibitem[\protect\citeauthoryear{{Achikanath Chirakkara}, {Federrath},
  {Trivedi}  \& {Banerjee}}{{Achikanath Chirakkara}
  et~al.}{2021}]{AchikanathEtAl2021}
{Achikanath Chirakkara} R.,  {Federrath} C.,  {Trivedi} P.,   {Banerjee} R.,
  2021, \mn@doi [\prl] {10.1103/PhysRevLett.126.091103}, \href
  {https://ui.adsabs.harvard.edu/abs/2021PhRvL.126i1103A} {126, 091103}

\bibitem[\protect\citeauthoryear{{Arn{\`e}odo} et~al.,}{{Arn{\`e}odo}
  et~al.}{2008}]{arneodo2008universal}
{Arn{\`e}odo} A.,  et~al., 2008, \mn@doi [\prl]
  {10.1103/PhysRevLett.100.254504}, \href
  {https://ui.adsabs.harvard.edu/abs/2008PhRvL.100y4504A} {100, 254504}

\bibitem[\protect\citeauthoryear{{Balsara}, {Kim}, {Mac Low}  \&
  {Mathews}}{{Balsara} et~al.}{2004}]{balsara2004amplification}
{Balsara} D.~S.,  {Kim} J.,  {Mac Low} M.-M.,   {Mathews} G.~J.,  2004, \mn@doi
  [\apj] {10.1086/425297}, \href
  {https://ui.adsabs.harvard.edu/abs/2004ApJ...617..339B} {617, 339}

\bibitem[\protect\citeauthoryear{{Banerjee} \& {Jedamzik}}{{Banerjee} \&
  {Jedamzik}}{2004}]{banerjee2004evolution}
{Banerjee} R.,  {Jedamzik} K.,  2004, \mn@doi [\prd]
  {10.1103/PhysRevD.70.123003}, \href
  {https://ui.adsabs.harvard.edu/abs/2004PhRvD..70l3003B} {70, 123003}

\bibitem[\protect\citeauthoryear{{Batchelor}}{{Batchelor}}{1950}]{batchelor1950spontaneous}
{Batchelor} G.~K.,  1950, \mn@doi [Proceedings of the Royal Society of London
  Series A] {10.1098/rspa.1950.0069}, \href
  {https://ui.adsabs.harvard.edu/abs/1950RSPSA.201..405B} {201, 405}

\bibitem[\protect\citeauthoryear{{Beattie}, {Federrath}, {Kriel}, {Mocz}  \&
  {Seta}}{{Beattie} et~al.}{2022}]{beattie2022growth}
{Beattie} J.~R.,  {Federrath} C.,  {Kriel} N.,  {Mocz} P.,   {Seta} A.,  2022,
  arXiv e-prints, \href {https://ui.adsabs.harvard.edu/abs/2022arXiv220910749B}
  {p. arXiv:2209.10749}

\bibitem[\protect\citeauthoryear{{Benzi}, {Ciliberto}, {Tripiccione}, {Baudet},
  {Massaioli}  \& {Succi}}{{Benzi} et~al.}{1993}]{benzi1993extended}
{Benzi} R.,  {Ciliberto} S.,  {Tripiccione} R.,  {Baudet} C.,  {Massaioli} F.,
   {Succi} S.,  1993, \mn@doi [\pre] {10.1103/PhysRevE.48.R29}, \href
  {https://ui.adsabs.harvard.edu/abs/1993PhRvE..48...29B} {48, R29}

\bibitem[\protect\citeauthoryear{{Benzi}, {Biferale}, {Fisher}, {Lamb}  \&
  {Toschi}}{{Benzi} et~al.}{2010}]{benzi2010inertial}
{Benzi} R.,  {Biferale} L.,  {Fisher} R.,  {Lamb} D.~Q.,   {Toschi} F.,  2010,
  \mn@doi [Journal of Fluid Mechanics] {10.1017/S002211201000056X}, \href
  {https://ui.adsabs.harvard.edu/abs/2010JFM...653..221B} {653, 221}

\bibitem[\protect\citeauthoryear{{Berera} \& {Linkmann}}{{Berera} \&
  {Linkmann}}{2014}]{berera2014magnetic}
{Berera} A.,  {Linkmann} M.,  2014, \mn@doi [\pre]
  {10.1103/PhysRevE.90.041003}, \href
  {https://ui.adsabs.harvard.edu/abs/2014PhRvE..90d1003B} {90, 041003}

\bibitem[\protect\citeauthoryear{{Beresnyak}}{{Beresnyak}}{2012}]{beresnyak2012universal}
{Beresnyak} A.,  2012, \mn@doi [\prl] {10.1103/PhysRevLett.108.035002}, \href
  {https://ui.adsabs.harvard.edu/abs/2012PhRvL.108c5002B} {108, 035002}

\bibitem[\protect\citeauthoryear{{Beresnyak}}{{Beresnyak}}{2015}]{beresnyak2015parallel}
{Beresnyak} A.,  2015, \mn@doi [\apjl] {10.1088/2041-8205/801/1/L9}, \href
  {https://ui.adsabs.harvard.edu/abs/2015ApJ...801L...9B} {801, L9}

\bibitem[\protect\citeauthoryear{{Beresnyak}}{{Beresnyak}}{2019}]{beresnyak2019mhd}
{Beresnyak} A.,  2019, \mn@doi [Living Reviews in Computational Astrophysics]
  {10.1007/s41115-019-0005-8}, \href
  {https://ui.adsabs.harvard.edu/abs/2019LRCA....5....2B} {5, 2}

\bibitem[\protect\citeauthoryear{{Beresnyak}, {Jones}  \&
  {Lazarian}}{{Beresnyak} et~al.}{2009}]{beresnyak2009turbulence}
{Beresnyak} A.,  {Jones} T.~W.,   {Lazarian} A.,  2009, \mn@doi [\apj]
  {10.1088/0004-637X/707/2/1541}, \href
  {https://ui.adsabs.harvard.edu/abs/2009ApJ...707.1541B} {707, 1541}

\bibitem[\protect\citeauthoryear{{Bhat}, {Zhou}  \& {Loureiro}}{{Bhat}
  et~al.}{2021}]{bhat2021inverse}
{Bhat} P.,  {Zhou} M.,   {Loureiro} N.~F.,  2021, \mn@doi [\mnras]
  {10.1093/mnras/staa3849}, \href
  {https://ui.adsabs.harvard.edu/abs/2021MNRAS.501.3074B} {501, 3074}

\bibitem[\protect\citeauthoryear{{Biferale}, {Boffetta}, {Celani}, {Devenish},
  {Lanotte}  \& {Toschi}}{{Biferale} et~al.}{2004}]{biferale2004multifractal}
{Biferale} L.,  {Boffetta} G.,  {Celani} A.,  {Devenish} B.~J.,  {Lanotte} A.,
   {Toschi} F.,  2004, \mn@doi [\prl] {10.1103/PhysRevLett.93.064502}, \href
  {https://ui.adsabs.harvard.edu/abs/2004PhRvL..93f4502B} {93, 064502}

\bibitem[\protect\citeauthoryear{{Biskamp} \& {M{\"u}ller}}{{Biskamp} \&
  {M{\"u}ller}}{1999}]{biskamp1999decay}
{Biskamp} D.,  {M{\"u}ller} W.-C.,  1999, \mn@doi [\prl]
  {10.1103/PhysRevLett.83.2195}, \href
  {https://ui.adsabs.harvard.edu/abs/1999PhRvL..83.2195B} {83, 2195}

\bibitem[\protect\citeauthoryear{{Biskamp} \& {M{\"u}ller}}{{Biskamp} \&
  {M{\"u}ller}}{2000}]{biskamp2000scaling}
{Biskamp} D.,  {M{\"u}ller} W.-C.,  2000, \mn@doi [Physics of Plasmas]
  {10.1063/1.1322562}, \href
  {https://ui.adsabs.harvard.edu/abs/2000PhPl....7.4889B} {7, 4889}

\bibitem[\protect\citeauthoryear{{Bohdan}, {Pohl}, {Niemiec}, {Morris},
  {Matsumoto}, {Amano}, {Hoshino}  \& {Sulaiman}}{{Bohdan}
  et~al.}{2021}]{bohdan2021magnetic}
{Bohdan} A.,  {Pohl} M.,  {Niemiec} J.,  {Morris} P.~J.,  {Matsumoto} Y.,
  {Amano} T.,  {Hoshino} M.,   {Sulaiman} A.,  2021, \mn@doi [\prl]
  {10.1103/PhysRevLett.126.095101}, \href
  {https://ui.adsabs.harvard.edu/abs/2021PhRvL.126i5101B} {126, 095101}

\bibitem[\protect\citeauthoryear{{Boldyrev} \& {Schekochihin}}{{Boldyrev} \&
  {Schekochihin}}{2001}]{boldyrevske2001}
{Boldyrev} S.,  {Schekochihin} A.,  2001, in APS April Meeting Abstracts. p.
  Q14.007

\bibitem[\protect\citeauthoryear{Bouchut, Klingenberg  \& Waagan}{Bouchut
  et~al.}{2010}]{bouchut2010multiwave}
Bouchut F.,  Klingenberg C.,   Waagan K.,  2010, Numerische Mathematik, 115,
  647

\bibitem[\protect\citeauthoryear{{Brandenburg}}{{Brandenburg}}{2018}]{brandenburg2018advances}
{Brandenburg} A.,  2018, \mn@doi [Journal of Plasma Physics]
  {10.1017/S0022377818000806}, \href
  {https://ui.adsabs.harvard.edu/abs/2018JPlPh..84d7304B} {84, 735840404}

\bibitem[\protect\citeauthoryear{{Brandenburg} \& {Kahniashvili}}{{Brandenburg}
  \& {Kahniashvili}}{2017}]{brandenburg2017classes}
{Brandenburg} A.,  {Kahniashvili} T.,  2017, \mn@doi [\prl]
  {10.1103/PhysRevLett.118.055102}, \href
  {https://ui.adsabs.harvard.edu/abs/2017PhRvL.118e5102B} {118, 055102}

\bibitem[\protect\citeauthoryear{{Brandenburg} \& {Petrosyan}}{{Brandenburg} \&
  {Petrosyan}}{2012}]{brandenburg2012kinetic}
{Brandenburg} A.,  {Petrosyan} A.,  2012, \mn@doi [Astronomische Nachrichten]
  {10.1002/asna.201211654}, \href
  {https://ui.adsabs.harvard.edu/abs/2012AN....333..195B} {333, 195}

\bibitem[\protect\citeauthoryear{{Brandenburg} \& {Subramanian}}{{Brandenburg}
  \& {Subramanian}}{2005}]{brandenburg2005astrophysical}
{Brandenburg} A.,  {Subramanian} K.,  2005, \mn@doi [\physrep]
  {10.1016/j.physrep.2005.06.005}, \href
  {https://ui.adsabs.harvard.edu/abs/2005PhR...417....1B} {417, 1}

\bibitem[\protect\citeauthoryear{{Brandenburg}, {Jennings}, {Nordlund},
  {Rieutord}, {Stein}  \& {Tuominen}}{{Brandenburg}
  et~al.}{1996}]{brandenburg1996magnetic}
{Brandenburg} A.,  {Jennings} R.~L.,  {Nordlund} {\r{A}}.,  {Rieutord} M.,
  {Stein} R.~F.,   {Tuominen} I.,  1996, \mn@doi [Journal of Fluid Mechanics]
  {10.1017/S0022112096001322}, \href
  {https://ui.adsabs.harvard.edu/abs/1996JFM...306..325B} {306, 325}

\bibitem[\protect\citeauthoryear{{Brandenburg}, {Kahniashvili}  \&
  {Tevzadze}}{{Brandenburg} et~al.}{2015}]{brandenburg2015nonhelical}
{Brandenburg} A.,  {Kahniashvili} T.,   {Tevzadze} A.~G.,  2015, \mn@doi [\prl]
  {10.1103/PhysRevLett.114.075001}, \href
  {https://ui.adsabs.harvard.edu/abs/2015PhRvL.114g5001B} {114, 075001}

\bibitem[\protect\citeauthoryear{{Brandenburg}, {Kahniashvili}, {Mandal},
  {Pol}, {Tevzadze}  \& {Vachaspati}}{{Brandenburg}
  et~al.}{2019}]{brandenburg2019dynamo}
{Brandenburg} A.,  {Kahniashvili} T.,  {Mandal} S.,  {Pol} A.~R.,  {Tevzadze}
  A.~G.,   {Vachaspati} T.,  2019, \mn@doi [Physical Review Fluids]
  {10.1103/PhysRevFluids.4.024608}, \href
  {https://ui.adsabs.harvard.edu/abs/2019PhRvF...4b4608B} {4, 024608}

\bibitem[\protect\citeauthoryear{{Brandenburg}, {Rogachevskii}  \&
  {Schober}}{{Brandenburg} et~al.}{2022}]{2022axel}
{Brandenburg} A.,  {Rogachevskii} I.,   {Schober} J.,  2022, arXiv e-prints,
  \href {https://ui.adsabs.harvard.edu/abs/2022arXiv220908717B} {p.
  arXiv:2209.08717}

\bibitem[\protect\citeauthoryear{Burgers}{Burgers}{1995}]{burgers1995mathematical}
Burgers J.~M.,  1995, in , Selected Papers of JM Burgers.
Springer, pp 281--334

\bibitem[\protect\citeauthoryear{{Busse}, {M{\"u}ller}  \&
  {Gogoberidze}}{{Busse} et~al.}{2010}]{busse2010lagrangian}
{Busse} A.,  {M{\"u}ller} W.-C.,   {Gogoberidze} G.,  2010, \mn@doi [\prl]
  {10.1103/PhysRevLett.105.235005}, \href
  {https://ui.adsabs.harvard.edu/abs/2010PhRvL.105w5005B} {105, 235005}

\bibitem[\protect\citeauthoryear{{Cho}, {Vishniac}, {Beresnyak}, {Lazarian}  \&
  {Ryu}}{{Cho} et~al.}{2009}]{cho2009growth}
{Cho} J.,  {Vishniac} E.~T.,  {Beresnyak} A.,  {Lazarian} A.,   {Ryu} D.,
  2009, \mn@doi [\apj] {10.1088/0004-637X/693/2/1449}, \href
  {https://ui.adsabs.harvard.edu/abs/2009ApJ...693.1449C} {693, 1449}

\bibitem[\protect\citeauthoryear{{Christensson}, {Hindmarsh}  \&
  {Brandenburg}}{{Christensson} et~al.}{2001}]{christensson2001inverse}
{Christensson} M.,  {Hindmarsh} M.,   {Brandenburg} A.,  2001, \mn@doi [\pre]
  {10.1103/PhysRevE.64.056405}, \href
  {https://ui.adsabs.harvard.edu/abs/2001PhRvE..64e6405C} {64, 056405}

\bibitem[\protect\citeauthoryear{{Corrsin}}{{Corrsin}}{1963}]{corrsin1963estimates}
{Corrsin} S.,  1963, \mn@doi [Journal of Atmospheric Sciences]
  {10.1175/1520-0469(1963)020<0115:EOTRBE>2.0.CO;2}, \href
  {https://ui.adsabs.harvard.edu/abs/1963JAtS...20..115C} {20, 115}

\bibitem[\protect\citeauthoryear{{Davidovits}, {Federrath}, {Teyssier},
  {Raman}, {Collins}  \& {Nagel}}{{Davidovits} et~al.}{2022}]{2022davidovits}
{Davidovits} S.,  {Federrath} C.,  {Teyssier} R.,  {Raman} K.~S.,  {Collins}
  D.~C.,   {Nagel} S.~R.,  2022, \mn@doi [\pre] {10.1103/PhysRevE.105.065206},
  \href {https://ui.adsabs.harvard.edu/abs/2022PhRvE.105f5206D} {105, 065206}

\bibitem[\protect\citeauthoryear{{Davidson}}{{Davidson}}{2000}]{davidson2000loitsyansky}
{Davidson} P.~A.,  2000, \mn@doi [Journal of Turbulence]
  {10.1088/1468-5248/1/1/006}, \href
  {https://ui.adsabs.harvard.edu/abs/2000JTurb...1....6D} {1, 6}

\bibitem[\protect\citeauthoryear{{Davidson}}{{Davidson}}{2010}]{davidson2010decay}
{Davidson} P.~A.,  2010, \mn@doi [Journal of Fluid Mechanics]
  {10.1017/S0022112010003496}, \href
  {https://ui.adsabs.harvard.edu/abs/2010JFM...663..268D} {663, 268}

\bibitem[\protect\citeauthoryear{{Dhawalikar}, {Federrath}, {Davidovits},
  {Teyssier}, {Nagel}, {Remington}  \& {Collins}}{{Dhawalikar}
  et~al.}{2022}]{dhawalikar2022driving}
{Dhawalikar} S.,  {Federrath} C.,  {Davidovits} S.,  {Teyssier} R.,  {Nagel}
  S.~R.,  {Remington} B.~A.,   {Collins} D.~C.,  2022, \mn@doi [\mnras]
  {10.1093/mnras/stac1480}, \href
  {https://ui.adsabs.harvard.edu/abs/2022MNRAS.514.1782D} {514, 1782}

\bibitem[\protect\citeauthoryear{{Donnert}, {Vazza}, {Br{\"u}ggen}  \&
  {ZuHone}}{{Donnert} et~al.}{2018}]{donnert2018magnetic}
{Donnert} J.,  {Vazza} F.,  {Br{\"u}ggen} M.,   {ZuHone} J.,  2018, \mn@doi
  [\ssr] {10.1007/s11214-018-0556-8}, \href
  {https://ui.adsabs.harvard.edu/abs/2018SSRv..214..122D} {214, 122}

\bibitem[\protect\citeauthoryear{{Downes} \& {Drury}}{{Downes} \&
  {Drury}}{2014}]{2014downes}
{Downes} T.~P.,  {Drury} L.~O.,  2014, \mn@doi [\mnras]
  {10.1093/mnras/stu1447}, \href
  {https://ui.adsabs.harvard.edu/abs/2014MNRAS.444..365D} {444, 365}

\bibitem[\protect\citeauthoryear{{Drury} \& {Downes}}{{Drury} \&
  {Downes}}{2012}]{drury2012turbulent}
{Drury} L.~O.,  {Downes} T.~P.,  2012, \mn@doi [\mnras]
  {10.1111/j.1365-2966.2012.22106.x}, \href
  {https://ui.adsabs.harvard.edu/abs/2012MNRAS.427.2308D} {427, 2308}

\bibitem[\protect\citeauthoryear{{Emanuel}}{{Emanuel}}{2019}]{emanuel2019derivatives}
{Emanuel} G.,  2019, \mn@doi [Journal of Engineering Mathematics]
  {10.1007/s10665-019-10010-0}, \href
  {https://ui.adsabs.harvard.edu/abs/2019JEnMa.117...79E} {117, 79}

\bibitem[\protect\citeauthoryear{{Emanuel} \& {Liu}}{{Emanuel} \&
  {Liu}}{1988}]{emanuel1988shock}
{Emanuel} G.,  {Liu} M.-S.,  1988, \mn@doi [Physics of Fluids]
  {10.1063/1.866879}, \href
  {https://ui.adsabs.harvard.edu/abs/1988PhFl...31.3625E} {31, 3625}

\bibitem[\protect\citeauthoryear{{Eyink}, {Lazarian}  \& {Vishniac}}{{Eyink}
  et~al.}{2011}]{eyink}
{Eyink} G.~L.,  {Lazarian} A.,   {Vishniac} E.~T.,  2011, \mn@doi [\apj]
  {10.1088/0004-637X/743/1/51}, \href
  {https://ui.adsabs.harvard.edu/abs/2011ApJ...743...51E} {743, 51}

\bibitem[\protect\citeauthoryear{{Federrath}}{{Federrath}}{2013}]{Federrath2013}
{Federrath} C.,  2013, \mn@doi [\mnras] {10.1093/mnras/stt1644}, \href
  {http://adsabs.harvard.edu/abs/2013MNRAS.436.1245F} {436, 1245}

\bibitem[\protect\citeauthoryear{{Federrath}}{{Federrath}}{2016}]{federrath2016magnetic}
{Federrath} C.,  2016, \mn@doi [Journal of Plasma Physics]
  {10.1017/S0022377816001069}, \href
  {https://ui.adsabs.harvard.edu/abs/2016JPlPh..82f5301F} {82, 535820601}

\bibitem[\protect\citeauthoryear{{Federrath} \& {Klessen}}{{Federrath} \&
  {Klessen}}{2012}]{federrath2012star}
{Federrath} C.,  {Klessen} R.~S.,  2012, \mn@doi [\apj]
  {10.1088/0004-637X/761/2/156}, \href
  {https://ui.adsabs.harvard.edu/abs/2012ApJ...761..156F} {761, 156}

\bibitem[\protect\citeauthoryear{{Federrath} \& {Klessen}}{{Federrath} \&
  {Klessen}}{2013}]{federrath2013star}
{Federrath} C.,  {Klessen} R.~S.,  2013, \mn@doi [\apj]
  {10.1088/0004-637X/763/1/51}, \href
  {https://ui.adsabs.harvard.edu/abs/2013ApJ...763...51F} {763, 51}

\bibitem[\protect\citeauthoryear{{Federrath}, {Roman-Duval}, {Klessen},
  {Schmidt}  \& {Mac Low}}{{Federrath} et~al.}{2010}]{federrath2010comparing}
{Federrath} C.,  {Roman-Duval} J.,  {Klessen} R.~S.,  {Schmidt} W.,   {Mac Low}
  M.~M.,  2010, \mn@doi [\aap] {10.1051/0004-6361/200912437}, \href
  {https://ui.adsabs.harvard.edu/abs/2010A&A...512A..81F} {512, A81}

\bibitem[\protect\citeauthoryear{{Federrath}, {Chabrier}, {Schober},
  {Banerjee}, {Klessen}  \& {Schleicher}}{{Federrath}
  et~al.}{2011a}]{federrath2011mach}
{Federrath} C.,  {Chabrier} G.,  {Schober} J.,  {Banerjee} R.,  {Klessen}
  R.~S.,   {Schleicher} D.~R.~G.,  2011a, \mn@doi [\prl]
  {10.1103/PhysRevLett.107.114504}, \href
  {https://ui.adsabs.harvard.edu/abs/2011PhRvL.107k4504F} {107, 114504}

\bibitem[\protect\citeauthoryear{{Federrath}, {Sur}, {Schleicher}, {Banerjee}
  \& {Klessen}}{{Federrath} et~al.}{2011b}]{federrath2011new}
{Federrath} C.,  {Sur} S.,  {Schleicher} D. R.~G.,  {Banerjee} R.,   {Klessen}
  R.~S.,  2011b, \mn@doi [\apj] {10.1088/0004-637X/731/1/62}, \href
  {https://ui.adsabs.harvard.edu/abs/2011ApJ...731...62F} {731, 62}

\bibitem[\protect\citeauthoryear{{Federrath}, {Schober}, {Bovino}  \&
  {Schleicher}}{{Federrath} et~al.}{2014}]{federrath2014turbulent}
{Federrath} C.,  {Schober} J.,  {Bovino} S.,   {Schleicher} D. R.~G.,  2014,
  \mn@doi [\apjl] {10.1088/2041-8205/797/2/L19}, \href
  {https://ui.adsabs.harvard.edu/abs/2014ApJ...797L..19F} {797, L19}

\bibitem[\protect\citeauthoryear{Federrath et~al.,}{Federrath
  et~al.}{2016}]{federrath2016link}
Federrath C.,  et~al., 2016, Proceedings of the International Astronomical
  Union, 11, 123

\bibitem[\protect\citeauthoryear{{Federrath}, {Klessen}, {Iapichino}  \&
  {Beattie}}{{Federrath} et~al.}{2021}]{FederrathEtAl2021}
{Federrath} C.,  {Klessen} R.~S.,  {Iapichino} L.,   {Beattie} J.~R.,  2021,
  \mn@doi [Nature Astronomy] {10.1038/s41550-020-01282-z}, \href
  {https://ui.adsabs.harvard.edu/abs/2021NatAs...5..365F} {5, 365}

\bibitem[\protect\citeauthoryear{{Federrath}, {Roman-Duval}, {Klessen},
  {Schmidt}  \& {Mac Low}}{{Federrath} et~al.}{2022}]{FederrathEtAl2022ascl}
{Federrath} C.,  {Roman-Duval} J.,  {Klessen} R.~S.,  {Schmidt} W.,   {Mac Low}
  M.~M.,  2022, {TG: Turbulence Generator}, Astrophysics Source Code Library,
  record ascl:2204.001 (\mn@eprint {ascl} {2204.001})

\bibitem[\protect\citeauthoryear{{Fraschetti}}{{Fraschetti}}{2013}]{fraschetti2013}
{Fraschetti} F.,  2013, \mn@doi [\apj] {10.1088/0004-637X/770/2/84}, \href
  {https://ui.adsabs.harvard.edu/abs/2013ApJ...770...84F} {770, 84}

\bibitem[\protect\citeauthoryear{{Frick} \& {Stepanov}}{{Frick} \&
  {Stepanov}}{2010}]{frick2010long}
{Frick} P.,  {Stepanov} R.,  2010, \mn@doi [EPL (Europhysics Letters)]
  {10.1209/0295-5075/92/34007}, \href
  {https://ui.adsabs.harvard.edu/abs/2010EL.....9234007F} {92, 34007}

\bibitem[\protect\citeauthoryear{Frisch \& Kolmogorov}{Frisch \&
  Kolmogorov}{1995}]{frisch1995turbulence}
Frisch U.,  Kolmogorov A.~N.,  1995, Turbulence: the legacy of AN Kolmogorov.
Cambridge university press

\bibitem[\protect\citeauthoryear{{Fryxell} et~al.,}{{Fryxell}
  et~al.}{2000}]{FryxellEtAl2000}
{Fryxell} B.,  et~al., 2000, \mn@doi [\apjs] {10.1086/317361}, \href
  {http://cdsads.u-strasbg.fr/abs/2000ApJS..131..273F} {131, 273}

\bibitem[\protect\citeauthoryear{{Galishnikova}, {Kunz}  \&
  {Schekochihin}}{{Galishnikova} et~al.}{2022}]{2022galishnikova}
{Galishnikova} A.~K.,  {Kunz} M.~W.,   {Schekochihin} A.~A.,  2022, arXiv
  e-prints, \href {https://ui.adsabs.harvard.edu/abs/2022arXiv220107757G} {p.
  arXiv:2201.07757}

\bibitem[\protect\citeauthoryear{{Garnier}, {Mossi}, {Sagaut}, {Comte}  \&
  {Deville}}{{Garnier} et~al.}{1999}]{garnier1999use}
{Garnier} E.,  {Mossi} M.,  {Sagaut} P.,  {Comte} P.,   {Deville} M.,  1999,
  \mn@doi [Journal of Computational Physics] {10.1006/jcph.1999.6268}, \href
  {https://ui.adsabs.harvard.edu/abs/1999JCoPh.153..273G} {153, 273}

\bibitem[\protect\citeauthoryear{{Giacalone} \& {Jokipii}}{{Giacalone} \&
  {Jokipii}}{2007}]{giacalone2007magnetic}
{Giacalone} J.,  {Jokipii} J.~R.,  2007, \mn@doi [\apjl] {10.1086/519994},
  \href {https://ui.adsabs.harvard.edu/abs/2007ApJ...663L..41G} {663, L41}

\bibitem[\protect\citeauthoryear{{Gogoberidze}}{{Gogoberidze}}{2007}]{gogoberidze2007nature}
{Gogoberidze} G.,  2007, \mn@doi [Physics of Plasmas] {10.1063/1.2437753},
  \href {https://ui.adsabs.harvard.edu/abs/2007PhPl...14b2304G} {14, 022304}

\bibitem[\protect\citeauthoryear{{Goldreich} \& {Sridhar}}{{Goldreich} \&
  {Sridhar}}{1995}]{goldreich1995toward}
{Goldreich} P.,  {Sridhar} S.,  1995, \mn@doi [\apj] {10.1086/175121}, \href
  {https://ui.adsabs.harvard.edu/abs/1995ApJ...438..763G} {438, 763}

\bibitem[\protect\citeauthoryear{{Goldreich} \& {Sridhar}}{{Goldreich} \&
  {Sridhar}}{1997}]{goldreich1997magnetohydrodynamic}
{Goldreich} P.,  {Sridhar} S.,  1997, \mn@doi [\apj] {10.1086/304442}, \href
  {https://ui.adsabs.harvard.edu/abs/1997ApJ...485..680G} {485, 680}

\bibitem[\protect\citeauthoryear{{Hayes}}{{Hayes}}{1957}]{hayes1957vorticity}
{Hayes} W.~D.,  1957, \mn@doi [Journal of Fluid Mechanics]
  {10.1017/S0022112057000403}, \href
  {https://ui.adsabs.harvard.edu/abs/1957JFM.....2..595H} {2, 595}

\bibitem[\protect\citeauthoryear{{Hew}, {Boswell}, {Federrath}, {Gopalapillai}
  \& {Paramanantham}}{{Hew} et~al.}{2022}]{hew2022analytical}
{Hew} J. K.~J.,  {Boswell} R.~W.,  {Federrath} C.,  {Gopalapillai} R.,
  {Paramanantham} V.,  2022, arXiv e-prints, \href
  {https://ui.adsabs.harvard.edu/abs/2022arXiv221004713H} {p. arXiv:2210.04713}

\bibitem[\protect\citeauthoryear{{Homann}, {Grauer}, {Busse}  \&
  {M{\"u}ller}}{{Homann} et~al.}{2007}]{homann2007lagrangian}
{Homann} H.,  {Grauer} R.,  {Busse} A.,   {M{\"u}ller} W.~C.,  2007, \mn@doi
  [Journal of Plasma Physics] {10.1017/S0022377807006575}, \href
  {https://ui.adsabs.harvard.edu/abs/2007JPlPh..73..821H} {73, 821}

\bibitem[\protect\citeauthoryear{{Homann}, {Ponty}, {Krstulovic}  \&
  {Grauer}}{{Homann} et~al.}{2014}]{homann2014structures}
{Homann} H.,  {Ponty} Y.,  {Krstulovic} G.,   {Grauer} R.,  2014, \mn@doi [New
  Journal of Physics] {10.1088/1367-2630/16/7/075014}, \href
  {https://ui.adsabs.harvard.edu/abs/2014NJPh...16g5014H} {16, 075014}

\bibitem[\protect\citeauthoryear{{Hosking} \& {Schekochihin}}{{Hosking} \&
  {Schekochihin}}{2021}]{hosking2021reconnection}
{Hosking} D.~N.,  {Schekochihin} A.~A.,  2021, \mn@doi [Physical Review X]
  {10.1103/PhysRevX.11.041005}, \href
  {https://ui.adsabs.harvard.edu/abs/2021PhRvX..11d1005H} {11, 041005}

\bibitem[\protect\citeauthoryear{{Hu}, {Xu}, {Stone}  \& {Lazarian}}{{Hu}
  et~al.}{2022}]{hu2022turbulent}
{Hu} Y.,  {Xu} S.,  {Stone} J.~M.,   {Lazarian} A.,  2022, \mn@doi [\apj]
  {10.3847/1538-4357/ac9ebc}, \href
  {https://ui.adsabs.harvard.edu/abs/2022ApJ...941..133H} {941, 133}

\bibitem[\protect\citeauthoryear{{Inoue}}{{Inoue}}{1951}]{inoue1951turbulent}
{Inoue} E.,  1951, \mn@doi [Journal of the Meteorological Society of Japan]
  {10.2151/jmsj1923.29.7_246}, \href
  {https://ui.adsabs.harvard.edu/abs/1951JMeSJ..29..246I} {29, 246}

\bibitem[\protect\citeauthoryear{{Inoue}, {Yamazaki}  \& {Inutsuka}}{{Inoue}
  et~al.}{2009}]{inoue2009turbulence}
{Inoue} T.,  {Yamazaki} R.,   {Inutsuka} S.-i.,  2009, \mn@doi [\apj]
  {10.1088/0004-637X/695/2/825}, \href
  {https://ui.adsabs.harvard.edu/abs/2009ApJ...695..825I} {695, 825}

\bibitem[\protect\citeauthoryear{{Inoue}, {Shimoda}, {Ohira}  \&
  {Yamazaki}}{{Inoue} et~al.}{2013}]{inoue2013origin}
{Inoue} T.,  {Shimoda} J.,  {Ohira} Y.,   {Yamazaki} R.,  2013, \mn@doi [\apjl]
  {10.1088/2041-8205/772/2/L20}, \href
  {https://ui.adsabs.harvard.edu/abs/2013ApJ...772L..20I} {772, L20}

\bibitem[\protect\citeauthoryear{{Iroshnikov}}{{Iroshnikov}}{1964}]{iroshnikov1964turbulence}
{Iroshnikov} P.~S.,  1964, \sovast, \href
  {https://ui.adsabs.harvard.edu/abs/1964SvA.....7..566I} {7, 566}

\bibitem[\protect\citeauthoryear{{Ji}, {Oh}, {Ruszkowski}  \&
  {Markevitch}}{{Ji} et~al.}{2016}]{ji2016efficiency}
{Ji} S.,  {Oh} S.~P.,  {Ruszkowski} M.,   {Markevitch} M.,  2016, \mn@doi
  [\mnras] {10.1093/mnras/stw2320}, \href
  {https://ui.adsabs.harvard.edu/abs/2016MNRAS.463.3989J} {463, 3989}

\bibitem[\protect\citeauthoryear{Kanwal}{Kanwal}{1959}]{kanwal1959curved}
Kanwal R.,  1959, Quarterly of Applied Mathematics, 16, 361

\bibitem[\protect\citeauthoryear{{Kazantsev}}{{Kazantsev}}{1968}]{kazantsev1968enhancement}
{Kazantsev} A.~P.,  1968, Soviet Journal of Experimental and Theoretical
  Physics, \href {https://ui.adsabs.harvard.edu/abs/1968JETP...26.1031K} {26,
  1031}

\bibitem[\protect\citeauthoryear{{Kevlahan}}{{Kevlahan}}{1997}]{kevlahan1997vorticity}
{Kevlahan} N.~K.~R.,  1997, \mn@doi [Journal of Fluid Mechanics]
  {10.1017/S0022112097005752}, \href
  {https://ui.adsabs.harvard.edu/abs/1997JFM...341..371K} {341, 371}

\bibitem[\protect\citeauthoryear{{Kevlahan} \& {Pudritz}}{{Kevlahan} \&
  {Pudritz}}{2009}]{kevlahan2009apj}
{Kevlahan} N.,  {Pudritz} R.~E.,  2009, \mn@doi [\apj]
  {10.1088/0004-637X/702/1/39}, \href
  {https://ui.adsabs.harvard.edu/abs/2009ApJ...702...39K} {702, 39}

\bibitem[\protect\citeauthoryear{{Kida} \& {Orszag}}{{Kida} \&
  {Orszag}}{1990}]{kida1990energy}
{Kida} S.,  {Orszag} S.~A.,  1990, Journal of Scientific Computing, \href
  {https://ui.adsabs.harvard.edu/abs/1990JSCom...5...85K} {5, 85}

\bibitem[\protect\citeauthoryear{Kida \& Orszag}{Kida \&
  Orszag}{1992}]{kida1992energy}
Kida S.,  Orszag S.~A.,  1992, Journal of Scientific Computing, 7, 1

\bibitem[\protect\citeauthoryear{{Kim} \& {Balsara}}{{Kim} \&
  {Balsara}}{2006}]{kim2006amplification}
{Kim} J.,  {Balsara} D.~S.,  2006, \mn@doi [Astronomische Nachrichten]
  {10.1002/asna.200610551}, \href
  {https://ui.adsabs.harvard.edu/abs/2006AN....327..433K} {327, 433}

\bibitem[\protect\citeauthoryear{{Kolmogorov}}{{Kolmogorov}}{1941}]{kolmogorov1941local}
{Kolmogorov} A.,  1941, Akademiia Nauk SSSR Doklady, \href
  {https://ui.adsabs.harvard.edu/abs/1941DoSSR..30..301K} {30, 301}

\bibitem[\protect\citeauthoryear{{Konstandin}, {Federrath}, {Klessen}  \&
  {Schmidt}}{{Konstandin} et~al.}{2012}]{konstandin2012statistical}
{Konstandin} L.,  {Federrath} C.,  {Klessen} R.~S.,   {Schmidt} W.,  2012,
  \mn@doi [Journal of Fluid Mechanics] {10.1017/jfm.2011.503}, \href
  {https://ui.adsabs.harvard.edu/abs/2012JFM...692..183K} {692, 183}

\bibitem[\protect\citeauthoryear{Kraichnan}{Kraichnan}{1965}]{kraichnan1965inertial}
Kraichnan R.~H.,  1965, The Physics of Fluids, 8, 1385

\bibitem[\protect\citeauthoryear{Kraichnan}{Kraichnan}{1977}]{kraichnan1977eulerian}
Kraichnan R.~H.,  1977, Journal of Fluid Mechanics, 83, 349

\bibitem[\protect\citeauthoryear{{Kraichnan} \& {Nagarajan}}{{Kraichnan} \&
  {Nagarajan}}{1967}]{kraichnan1967growth}
{Kraichnan} R.~H.,  {Nagarajan} S.,  1967, \mn@doi [Physics of Fluids]
  {10.1063/1.1762201}, \href
  {https://ui.adsabs.harvard.edu/abs/1967PhFl...10..859K} {10, 859}

\bibitem[\protect\citeauthoryear{{Kriel}, {Beattie}, {Seta}  \&
  {Federrath}}{{Kriel} et~al.}{2022}]{KrielEtAl20221}
{Kriel} N.,  {Beattie} J.~R.,  {Seta} A.,   {Federrath} C.,  2022, \mn@doi
  [\mnras] {10.1093/mnras/stac969}, \href
  {https://ui.adsabs.harvard.edu/abs/2022MNRAS.513.2457K} {513, 2457}

\bibitem[\protect\citeauthoryear{{Kritsuk}, {Norman}, {Padoan}  \&
  {Wagner}}{{Kritsuk} et~al.}{2007}]{kritsuk2007statistics}
{Kritsuk} A.~G.,  {Norman} M.~L.,  {Padoan} P.,   {Wagner} R.,  2007, \mn@doi
  [\apj] {10.1086/519443}, \href
  {https://ui.adsabs.harvard.edu/abs/2007ApJ...665..416K} {665, 416}

\bibitem[\protect\citeauthoryear{{Kritsuk} et~al.,}{{Kritsuk}
  et~al.}{2011}]{kritsuk2011comparing}
{Kritsuk} A.~G.,  et~al., 2011, \mn@doi [\apj] {10.1088/0004-637X/737/1/13},
  \href {https://ui.adsabs.harvard.edu/abs/2011ApJ...737...13K} {737, 13}

\bibitem[\protect\citeauthoryear{Krogstad \& Davidson}{Krogstad \&
  Davidson}{2010}]{krogstad2010grid}
Krogstad P.-{\AA}.,  Davidson P.,  2010, Journal of Fluid Mechanics, 642, 373

\bibitem[\protect\citeauthoryear{{Kulsrud}}{{Kulsrud}}{2005}]{kulsrud2005plasma}
{Kulsrud} R.~M.,  2005, {Plasma Physics for Astrophysics}

\bibitem[\protect\citeauthoryear{{Kulsrud} \& {Anderson}}{{Kulsrud} \&
  {Anderson}}{1992}]{kulsrud1992spectrum}
{Kulsrud} R.~M.,  {Anderson} S.~W.,  1992, \mn@doi [\apj] {10.1086/171743},
  \href {https://ui.adsabs.harvard.edu/abs/1992ApJ...396..606K} {396, 606}

\bibitem[\protect\citeauthoryear{{Lazarian} \& {Vishniac}}{{Lazarian} \&
  {Vishniac}}{1999}]{lazarian1999reconnection}
{Lazarian} A.,  {Vishniac} E.~T.,  1999, \mn@doi [\apj] {10.1086/307233}, \href
  {https://ui.adsabs.harvard.edu/abs/1999ApJ...517..700L} {517, 700}

\bibitem[\protect\citeauthoryear{{Livescu} \& {Ryu}}{{Livescu} \&
  {Ryu}}{2016}]{livescu2016}
{Livescu} D.,  {Ryu} J.,  2016, \mn@doi [Shock Waves]
  {10.1007/s00193-015-0580-5}, \href
  {https://ui.adsabs.harvard.edu/abs/2016ShWav..26..241L} {26, 241}

\bibitem[\protect\citeauthoryear{{Mac Low}}{{Mac Low}}{1999}]{mac1999energy}
{Mac Low} M.-M.,  1999, \mn@doi [\apj] {10.1086/307784}, \href
  {https://ui.adsabs.harvard.edu/abs/1999ApJ...524..169M} {524, 169}

\bibitem[\protect\citeauthoryear{{Mac Low} \& {Klessen}}{{Mac Low} \&
  {Klessen}}{2004}]{mac2004control}
{Mac Low} M.-M.,  {Klessen} R.~S.,  2004, \mn@doi [Reviews of Modern Physics]
  {10.1103/RevModPhys.76.125}, \href
  {https://ui.adsabs.harvard.edu/abs/2004RvMP...76..125M} {76, 125}

\bibitem[\protect\citeauthoryear{{Mac Low}, {Klessen}, {Burkert}  \&
  {Smith}}{{Mac Low} et~al.}{1998}]{mac1998kinetic}
{Mac Low} M.-M.,  {Klessen} R.~S.,  {Burkert} A.,   {Smith} M.~D.,  1998,
  \mn@doi [\prl] {10.1103/PhysRevLett.80.2754}, \href
  {https://ui.adsabs.harvard.edu/abs/1998PhRvL..80.2754M} {80, 2754}

\bibitem[\protect\citeauthoryear{{Mac Low}, {Balsara}, {Kim}  \& {de
  Avillez}}{{Mac Low} et~al.}{2005}]{mac2005distribution}
{Mac Low} M.-M.,  {Balsara} D.~S.,  {Kim} J.,   {de Avillez} M.~A.,  2005,
  \mn@doi [\apj] {10.1086/430122}, \href
  {https://ui.adsabs.harvard.edu/abs/2005ApJ...626..864M} {626, 864}

\bibitem[\protect\citeauthoryear{{Mason}, {Cattaneo}  \& {Boldyrev}}{{Mason}
  et~al.}{2006}]{mason2006}
{Mason} J.,  {Cattaneo} F.,   {Boldyrev} S.,  2006, \mn@doi [\prl]
  {10.1103/PhysRevLett.97.255002}, \href
  {https://ui.adsabs.harvard.edu/abs/2006PhRvL..97y5002M} {97, 255002}

\bibitem[\protect\citeauthoryear{{McKee}, {Stacy}  \& {Li}}{{McKee}
  et~al.}{2020}]{mckee2020magnetic}
{McKee} C.~F.,  {Stacy} A.,   {Li} P.~S.,  2020, \mn@doi [\mnras]
  {10.1093/mnras/staa1903}, \href
  {https://ui.adsabs.harvard.edu/abs/2020MNRAS.496.5528M} {496, 5528}

\bibitem[\protect\citeauthoryear{{Mee} \& {Brandenburg}}{{Mee} \&
  {Brandenburg}}{2006}]{mee2006turbulence}
{Mee} A.~J.,  {Brandenburg} A.,  2006, \mn@doi [\mnras]
  {10.1111/j.1365-2966.2006.10476.x}, \href
  {https://ui.adsabs.harvard.edu/abs/2006MNRAS.370..415M} {370, 415}

\bibitem[\protect\citeauthoryear{{Meinecke} et~al.,}{{Meinecke}
  et~al.}{2014}]{meinecke2014turbulent}
{Meinecke} J.,  et~al., 2014, \mn@doi [Nature Physics] {10.1038/nphys2978},
  \href {https://ui.adsabs.harvard.edu/abs/2014NatPh..10..520M} {10, 520}

\bibitem[\protect\citeauthoryear{{M{\"o}lder}}{{M{\"o}lder}}{2016}]{molder2016curved}
{M{\"o}lder} S.,  2016, \mn@doi [Shock Waves] {10.1007/s00193-015-0589-9},
  \href {https://ui.adsabs.harvard.edu/abs/2016ShWav..26..337M} {26, 337}

\bibitem[\protect\citeauthoryear{{Mordant}, {L{\'e}v{\^e}que}  \&
  {Pinton}}{{Mordant} et~al.}{2004}]{mordant2004experimental}
{Mordant} N.,  {L{\'e}v{\^e}que} E.,   {Pinton} J.-F.,  2004, \mn@doi [New
  Journal of Physics] {10.1088/1367-2630/6/1/116}, \href
  {https://ui.adsabs.harvard.edu/abs/2004NJPh....6..116M} {6, 116}

\bibitem[\protect\citeauthoryear{{M{\"u}ller} \& {Biskamp}}{{M{\"u}ller} \&
  {Biskamp}}{2000}]{muller2000scaling}
{M{\"u}ller} W.-C.,  {Biskamp} D.,  2000, \mn@doi [\prl]
  {10.1103/PhysRevLett.84.475}, \href
  {https://ui.adsabs.harvard.edu/abs/2000PhRvL..84..475M} {84, 475}

\bibitem[\protect\citeauthoryear{{Pan}, {Padoan}, {Haugb{\o}lle}  \&
  {Nordlund}}{{Pan} et~al.}{2016}]{pan2016supernova}
{Pan} L.,  {Padoan} P.,  {Haugb{\o}lle} T.,   {Nordlund} {\r{A}}.,  2016,
  \mn@doi [\apj] {10.3847/0004-637X/825/1/30}, \href
  {https://ui.adsabs.harvard.edu/abs/2016ApJ...825...30P} {825, 30}

\bibitem[\protect\citeauthoryear{{Park}}{{Park}}{2017}]{kiwan}
{Park} K.,  2017, \mn@doi [\mnras] {10.1093/mnras/stx1981}, \href
  {https://ui.adsabs.harvard.edu/abs/2017MNRAS.472.1628P} {472, 1628}

\bibitem[\protect\citeauthoryear{{Perez} \& {Boldyrev}}{{Perez} \&
  {Boldyrev}}{2009}]{perez2009role}
{Perez} J.~C.,  {Boldyrev} S.,  2009, \mn@doi [\prl]
  {10.1103/PhysRevLett.102.025003}, \href
  {https://ui.adsabs.harvard.edu/abs/2009PhRvL.102b5003P} {102, 025003}

\bibitem[\protect\citeauthoryear{{Proudman} \& {Reid}}{{Proudman} \&
  {Reid}}{1954}]{proudman1954decay}
{Proudman} I.,  {Reid} W.~H.,  1954, \mn@doi [Philosophical Transactions of the
  Royal Society of London Series A] {10.1098/rsta.1954.0016}, \href
  {https://ui.adsabs.harvard.edu/abs/1954RSPTA.247..163P} {247, 163}

\bibitem[\protect\citeauthoryear{{Reppin} \& {Banerjee}}{{Reppin} \&
  {Banerjee}}{2017}]{reppin2017nonhelical}
{Reppin} J.,  {Banerjee} R.,  2017, \mn@doi [\pre]
  {10.1103/PhysRevE.96.053105}, \href
  {https://ui.adsabs.harvard.edu/abs/2017PhRvE..96e3105R} {96, 053105}

\bibitem[\protect\citeauthoryear{{Saffman}}{{Saffman}}{1967}]{saffman1967note}
{Saffman} P.~G.,  1967, \mn@doi [Physics of Fluids] {10.1063/1.1762284}, \href
  {https://ui.adsabs.harvard.edu/abs/1967PhFl...10.1349S} {10, 1349}

\bibitem[\protect\citeauthoryear{{Sano}, {Nishihara}, {Matsuoka}  \&
  {Inoue}}{{Sano} et~al.}{2012}]{sano2012magnetic}
{Sano} T.,  {Nishihara} K.,  {Matsuoka} C.,   {Inoue} T.,  2012, \mn@doi [\apj]
  {10.1088/0004-637X/758/2/126}, \href
  {https://ui.adsabs.harvard.edu/abs/2012ApJ...758..126S} {758, 126}

\bibitem[\protect\citeauthoryear{{Sano} et~al.,}{{Sano}
  et~al.}{2021}]{sano2021}
{Sano} T.,  et~al., 2021, \mn@doi [\pre] {10.1103/PhysRevE.104.035206}, \href
  {https://ui.adsabs.harvard.edu/abs/2021PhRvE.104c5206S} {104, 035206}

\bibitem[\protect\citeauthoryear{{Sarma}, {Troland}, {Crutcher}  \&
  {Roberts}}{{Sarma} et~al.}{2002}]{sarma2002magnetic}
{Sarma} A.~P.,  {Troland} T.~H.,  {Crutcher} R.~M.,   {Roberts} D.~A.,  2002,
  \mn@doi [\apj] {10.1086/343799}, \href
  {https://ui.adsabs.harvard.edu/abs/2002ApJ...580..928S} {580, 928}

\bibitem[\protect\citeauthoryear{{Scalo} \& {Pumphrey}}{{Scalo} \&
  {Pumphrey}}{1982}]{scalo1982dissipation}
{Scalo} J.~M.,  {Pumphrey} W.~A.,  1982, \mn@doi [\apjl] {10.1086/183824},
  \href {https://ui.adsabs.harvard.edu/abs/1982ApJ...258L..29S} {258, L29}

\bibitem[\protect\citeauthoryear{{Schekochihin} \& {Kulsrud}}{{Schekochihin} \&
  {Kulsrud}}{2001}]{schekokulsrud}
{Schekochihin} A.~A.,  {Kulsrud} R.~M.,  2001, \mn@doi [Physics of Plasmas]
  {10.1063/1.1404383}, \href
  {https://ui.adsabs.harvard.edu/abs/2001PhPl....8.4937S} {8, 4937}

\bibitem[\protect\citeauthoryear{{Schekochihin}, {Cowley}, {Hammett}, {Maron}
  \& {McWilliams}}{{Schekochihin} et~al.}{2002a}]{schekochihin2002model}
{Schekochihin} A.~A.,  {Cowley} S.~C.,  {Hammett} G.~W.,  {Maron} J.~L.,
  {McWilliams} J.~C.,  2002a, \mn@doi [New Journal of Physics]
  {10.1088/1367-2630/4/1/384}, \href
  {https://ui.adsabs.harvard.edu/abs/2002NJPh....4...84S} {4, 84}

\bibitem[\protect\citeauthoryear{{Schekochihin}, {Boldyrev}  \&
  {Kulsrud}}{{Schekochihin} et~al.}{2002b}]{schekochihinboldyrev}
{Schekochihin} A.~A.,  {Boldyrev} S.~A.,   {Kulsrud} R.~M.,  2002b, \mn@doi
  [\apj] {10.1086/338697}, \href
  {https://ui.adsabs.harvard.edu/abs/2002ApJ...567..828S} {567, 828}

\bibitem[\protect\citeauthoryear{{Schekochihin}, {Maron}, {Cowley}  \&
  {McWilliams}}{{Schekochihin} et~al.}{2002c}]{schekochihin2002c}
{Schekochihin} A.~A.,  {Maron} J.~L.,  {Cowley} S.~C.,   {McWilliams} J.~C.,
  2002c, \mn@doi [\apj] {10.1086/341814}, \href
  {https://ui.adsabs.harvard.edu/abs/2002ApJ...576..806S} {576, 806}

\bibitem[\protect\citeauthoryear{{Schekochihin}, {Cowley}, {Taylor}, {Maron}
  \& {McWilliams}}{{Schekochihin} et~al.}{2004}]{schekochihin2004simulations}
{Schekochihin} A.~A.,  {Cowley} S.~C.,  {Taylor} S.~F.,  {Maron} J.~L.,
  {McWilliams} J.~C.,  2004, \mn@doi [\apj] {10.1086/422547}, \href
  {https://ui.adsabs.harvard.edu/abs/2004ApJ...612..276S} {612, 276}

\bibitem[\protect\citeauthoryear{Schekochihin, Cowley  \& Yousef}{Schekochihin
  et~al.}{2008}]{schekochihin2008mhd}
Schekochihin A.~A.,  Cowley S.~C.,   Yousef T.~A.,  2008, in IUTAM Symposium on
  computational physics and new perspectives in turbulence. pp 347--354

\bibitem[\protect\citeauthoryear{{Schleicher}, {Banerjee}, {Sur}, {Arshakian},
  {Klessen}, {Beck}  \& {Spaans}}{{Schleicher}
  et~al.}{2010}]{schleicher2010small}
{Schleicher} D.~R.~G.,  {Banerjee} R.,  {Sur} S.,  {Arshakian} T.~G.,
  {Klessen} R.~S.,  {Beck} R.,   {Spaans} M.,  2010, \mn@doi [\aap]
  {10.1051/0004-6361/201015184}, \href
  {https://ui.adsabs.harvard.edu/abs/2010A&A...522A.115S} {522, A115}

\bibitem[\protect\citeauthoryear{{Schleicher}, {Schober}, {Federrath}, {Bovino}
   \& {Schmidt}}{{Schleicher} et~al.}{2013}]{schleicher2013small}
{Schleicher} D. R.~G.,  {Schober} J.,  {Federrath} C.,  {Bovino} S.,
  {Schmidt} W.,  2013, \mn@doi [New Journal of Physics]
  {10.1088/1367-2630/15/2/023017}, \href
  {https://ui.adsabs.harvard.edu/abs/2013NJPh...15b3017S} {15, 023017}

\bibitem[\protect\citeauthoryear{{Schober}, {Schleicher}, {Federrath},
  {Glover}, {Klessen}  \& {Banerjee}}{{Schober}
  et~al.}{2012}]{schober2012small}
{Schober} J.,  {Schleicher} D.,  {Federrath} C.,  {Glover} S.,  {Klessen}
  R.~S.,   {Banerjee} R.,  2012, \mn@doi [\apj] {10.1088/0004-637X/754/2/99},
  \href {https://ui.adsabs.harvard.edu/abs/2012ApJ...754...99S} {754, 99}

\bibitem[\protect\citeauthoryear{{Schober}, {Schleicher}, {Federrath}, {Bovino}
   \& {Klessen}}{{Schober} et~al.}{2015}]{schober2015saturation}
{Schober} J.,  {Schleicher} D.~R.~G.,  {Federrath} C.,  {Bovino} S.,
  {Klessen} R.~S.,  2015, \mn@doi [\pre] {10.1103/PhysRevE.92.023010}, \href
  {https://ui.adsabs.harvard.edu/abs/2015PhRvE..92b3010S} {92, 023010}

\bibitem[\protect\citeauthoryear{{Servidio}, {Matthaeus}  \&
  {Dmitruk}}{{Servidio} et~al.}{2008}]{servidio2008depression}
{Servidio} S.,  {Matthaeus} W.~H.,   {Dmitruk} P.,  2008, \mn@doi [\prl]
  {10.1103/PhysRevLett.100.095005}, \href
  {https://ui.adsabs.harvard.edu/abs/2008PhRvL.100i5005S} {100, 095005}

\bibitem[\protect\citeauthoryear{{Seta} \& {Federrath}}{{Seta} \&
  {Federrath}}{2020}]{seta2020seed}
{Seta} A.,  {Federrath} C.,  2020, \mn@doi [\mnras] {10.1093/mnras/staa2978},
  \href {https://ui.adsabs.harvard.edu/abs/2020MNRAS.499.2076S} {499, 2076}

\bibitem[\protect\citeauthoryear{{Seta} \& {Federrath}}{{Seta} \&
  {Federrath}}{2021}]{seta2021saturation}
{Seta} A.,  {Federrath} C.,  2021, \mn@doi [Physical Review Fluids]
  {10.1103/PhysRevFluids.6.103701}, \href
  {https://ui.adsabs.harvard.edu/abs/2021PhRvF...6j3701S} {6, 103701}

\bibitem[\protect\citeauthoryear{{Seta} \& {Federrath}}{{Seta} \&
  {Federrath}}{2022}]{seta2022turbulent}
{Seta} A.,  {Federrath} C.,  2022, \mn@doi [\mnras] {10.1093/mnras/stac1400},
  \href {https://ui.adsabs.harvard.edu/abs/2022MNRAS.514..957S} {514, 957}

\bibitem[\protect\citeauthoryear{{Sridhar} \& {Goldreich}}{{Sridhar} \&
  {Goldreich}}{1994}]{sridhar1994toward}
{Sridhar} S.,  {Goldreich} P.,  1994, \mn@doi [\apj] {10.1086/174600}, \href
  {https://ui.adsabs.harvard.edu/abs/1994ApJ...432..612S} {432, 612}

\bibitem[\protect\citeauthoryear{{Stone}, {Ostriker}  \& {Gammie}}{{Stone}
  et~al.}{1998}]{stone1998dissipation}
{Stone} J.~M.,  {Ostriker} E.~C.,   {Gammie} C.~F.,  1998, \mn@doi [\apjl]
  {10.1086/311718}, \href
  {https://ui.adsabs.harvard.edu/abs/1998ApJ...508L..99S} {508, L99}

\bibitem[\protect\citeauthoryear{{Sur}}{{Sur}}{2019}]{sur2019decaying}
{Sur} S.,  2019, \mn@doi [\mnras] {10.1093/mnras/stz1918}, \href
  {https://ui.adsabs.harvard.edu/abs/2019MNRAS.488.3439S} {488, 3439}

\bibitem[\protect\citeauthoryear{{Sur}, {Schleicher}, {Banerjee}, {Federrath}
  \& {Klessen}}{{Sur} et~al.}{2010}]{sur2010generation}
{Sur} S.,  {Schleicher} D.~R.~G.,  {Banerjee} R.,  {Federrath} C.,   {Klessen}
  R.~S.,  2010, \mn@doi [\apjl] {10.1088/2041-8205/721/2/L134}, \href
  {https://ui.adsabs.harvard.edu/abs/2010ApJ...721L.134S} {721, L134}

\bibitem[\protect\citeauthoryear{{Tennekes}}{{Tennekes}}{1975}]{tennekes1975eulerian}
{Tennekes} H.,  1975, \mn@doi [Journal of Fluid Mechanics]
  {10.1017/S0022112075000468}, \href
  {https://ui.adsabs.harvard.edu/abs/1975JFM....67..561T} {67, 561}

\bibitem[\protect\citeauthoryear{{Tennekes} \& {Lumley}}{{Tennekes} \&
  {Lumley}}{1972}]{tennekes1972first}
{Tennekes} H.,  {Lumley} J.~L.,  1972, {First Course in Turbulence}

\bibitem[\protect\citeauthoryear{{Tian}, {Jaberi}  \& {Livescu}}{{Tian}
  et~al.}{2019}]{tian2019}
{Tian} Y.,  {Jaberi} F.~A.,   {Livescu} D.,  2019, \mn@doi [Journal of Fluid
  Mechanics] {10.1017/jfm.2019.707}, \href
  {https://ui.adsabs.harvard.edu/abs/2019JFM...880..935T} {880, 935}

\bibitem[\protect\citeauthoryear{{Tobias}}{{Tobias}}{2002}]{tobias2002solar}
{Tobias} S.~M.,  2002, \mn@doi [Philosophical Transactions of the Royal Society
  of London Series A] {10.1098/rsta.2002.1090}, \href
  {https://ui.adsabs.harvard.edu/abs/2002RSPTA.360.2741T} {360, 2741}

\bibitem[\protect\citeauthoryear{{Vladimirov}, {Ellison}  \&
  {Bykov}}{{Vladimirov} et~al.}{2006}]{vladimirov2006nonlinear}
{Vladimirov} A.,  {Ellison} D.~C.,   {Bykov} A.,  2006, \mn@doi [\apj]
  {10.1086/508154}, \href
  {https://ui.adsabs.harvard.edu/abs/2006ApJ...652.1246V} {652, 1246}

\bibitem[\protect\citeauthoryear{{Waagan}, {Federrath}  \&
  {Klingenberg}}{{Waagan} et~al.}{2011}]{waagan2011robust}
{Waagan} K.,  {Federrath} C.,   {Klingenberg} C.,  2011, \mn@doi [Journal of
  Computational Physics] {10.1016/j.jcp.2011.01.026}, \href
  {https://ui.adsabs.harvard.edu/abs/2011JCoPh.230.3331W} {230, 3331}

\bibitem[\protect\citeauthoryear{{Xu} \& {Lazarian}}{{Xu} \&
  {Lazarian}}{2016}]{XL2016}
{Xu} S.,  {Lazarian} A.,  2016, \mn@doi [\apj] {10.3847/1538-4357/833/2/215},
  \href {https://ui.adsabs.harvard.edu/abs/2016ApJ...833..215X} {833, 215}

\bibitem[\protect\citeauthoryear{{Xu} \& {Lazarian}}{{Xu} \&
  {Lazarian}}{2017}]{XL2017}
{Xu} S.,  {Lazarian} A.,  2017, \mn@doi [\apj] {10.3847/1538-4357/aa956b},
  \href {https://ui.adsabs.harvard.edu/abs/2017ApJ...850..126X} {850, 126}

\bibitem[\protect\citeauthoryear{{Xu} \& {Lazarian}}{{Xu} \&
  {Lazarian}}{2020}]{XuLazarian2020}
{Xu} S.,  {Lazarian} A.,  2020, \mn@doi [\apj] {10.3847/1538-4357/aba7ba},
  \href {https://ui.adsabs.harvard.edu/abs/2020ApJ...899..115X} {899, 115}

\bibitem[\protect\citeauthoryear{{Yeung}, {Pope}  \& {Sawford}}{{Yeung}
  et~al.}{2006}]{yeung2006reynolds}
{Yeung} P.~K.,  {Pope} S.~B.,   {Sawford} B.~L.,  2006, \mn@doi [Journal of
  Turbulence] {10.1080/14685240600868272}, \href
  {https://ui.adsabs.harvard.edu/abs/2006JTurb...7...58Y} {7, 58}

\bibitem[\protect\citeauthoryear{{Yokoi}}{{Yokoi}}{1999}]{yokoi1999magnetic}
{Yokoi} N.,  1999, \mn@doi [Physics of Fluids] {10.1063/1.870093}, \href
  {https://ui.adsabs.harvard.edu/abs/1999PhFl...11.2307Y} {11, 2307}

\bibitem[\protect\citeauthoryear{{Yokoi}}{{Yokoi}}{2013}]{yokoi2013cross}
{Yokoi} N.,  2013, \mn@doi [Geophysical and Astrophysical Fluid Dynamics]
  {10.1080/03091929.2012.754022}, \href
  {https://ui.adsabs.harvard.edu/abs/2013GApFD.107..114Y} {107, 114}

\bibitem[\protect\citeauthoryear{Zeldovich}{Zeldovich}{1957}]{zeldovich1957magnetic}
Zeldovich Y.~B.,  1957, Sov. Phys. JETP, 4, 460

\bibitem[\protect\citeauthoryear{Zeldovich, Ruzmaikin  \& Sokolov}{Zeldovich
  et~al.}{1983}]{zeldovich1983magnetic}
Zeldovich I.~B.,  Ruzmaikin A.~A.,   Sokolov D.~D.,  1983, New York, 3

\bibitem[\protect\citeauthoryear{{del Valle}, {Lazarian}  \&
  {Santos-Lima}}{{del Valle} et~al.}{2016}]{del2016turbulence}
{del Valle} M.~V.,  {Lazarian} A.,   {Santos-Lima} R.,  2016, \mn@doi [\mnras]
  {10.1093/mnras/stw340}, \href
  {https://ui.adsabs.harvard.edu/abs/2016MNRAS.458.1645D} {458, 1645}

\makeatother
\end{thebibliography}




\appendix

\section{Derivation for the growth rate of a small-scale dynamo with post-shock vorticity}\label{appen:b} 
We present here a brief derivation of the growth rate of a turbulent dynamo, $\Gamma$ in terms of vorticity~(Eqn.~\ref{eqn:21}) \citep{kulsrud2005plasma}, where $\omega_i(k) = \epsilon_{ijm} i k_j v_m/k^2$. The reader is advised to refer to \cite{kraichnan1967growth}, \cite{kulsrud1992spectrum}, \cite{kulsrud2005plasma} and \cite{fraschetti2013} for additional details. 

We start with the general assumption of $\delta$-correlated, isotropic and homogeneous velocity statistics, i.e. 
\begin{equation}\label{eqn:twopointv}
\langle \mathbf{v}^i (\mathbf{k},t) \mathbf{v}^j (\mathbf{k},t) \rangle= J(k)(\mathbf{I} - \hat{\mathbf{k}}\hat{\mathbf{k}})\delta_{\mathbf{k},t}
\end{equation}
where $\mathbf{I}$ is the unit dyad, $\hat{\mathbf{k}} = \mathbf{k}/\lvert \mathbf{k} \rvert$, $J(k)$ is the shell-integrated vorticity spectrum and $\delta_{\mathbf{k,t}} = \delta(\mathbf{k}^{\prime} - \mathbf{k}) \delta(t^{\prime} -t) $ is a shorthand notation for both $\delta$-functions. Here we have omitted the helical part of the correlation function, which includes contributions from the helicity spectrum, with fluctuations perpendicular to $\mathbf{k}$ ($i \mathbf{k} \times \mathbf{I}$).  This action is unimportant in the kinematic phase of a small-scale dynamo \citep{brandenburg2005astrophysical}. 

Using the conducting form of the MHD induction equation (Eqn.~\ref{eqn:conducting}), with the ansatz of linear eigenmode solutions for $\mathbf{v}$ and $\mathbf{B}$. \cite{kraichnan1967growth} and \cite{kulsrud1992spectrum} derived a mode-coupling equation for the magnetic spectrum, $M(k)$ (see \cite{schekochihinboldyrev} for a general form):
\begin{equation}\label{eqn:magspectra}
\frac{\partial M(k)}{\partial t} = \int K(k,k^{\prime}) M(k^{\prime}) dk - 2k^2 \alpha M(k) 
\end{equation}
here $K(k, k^{\prime})$ is given as,
\begin{equation}
K(k, k^{\prime}) = 4 \pi ^2 k^4 \int \sin^2 \theta \frac{k^2 + k^{\prime 2} - kk^{\prime} \cos\theta}{k^{\prime \prime 2 }} J(k^{\prime \prime})   
\end{equation}
where $\mathbf{k}^{\prime \prime} = \mathbf{k} - \mathbf{k}^{\prime}$, $\theta$ is defined as the angle between $\mathbf{k}$ and $\mathbf{k}^{\prime}$ (see Fig.~4 in \cite{kulsrud1992spectrum}) and also, 
\begin{equation}
\alpha = \frac{2\pi}{3} \int J(k^{\prime \prime})d^3 k^{\prime \prime}
\end{equation}
Integrating Eqn.~\ref{eqn:magspectra} over $k$, with $E_{\textrm{mag}}$ as given in Eqn.~\ref{eqn:magspec}, we obtain an expression for the growth rate, $\Gamma$:
\begin{equation}\label{eqn:growth}
    \Gamma = \frac{2\pi}{3} \int k^2 J(k) d^3 \mathbf{k}
\end{equation}
It is straightforward now to find that the assumption of Eqn.~\ref{eqn:twopointv} also imply homogeneity in the vorticity:
\begin{equation}
\langle \boldsymbol{\omega}^i(\mathbf{k},t) 
\boldsymbol{\omega}^j(\mathbf{k},t) \rangle = 2k^2 J(k)\delta_{\mathbf{k}, t}
\end{equation}
With $\delta_t = 1$, the steady-state assumption \citep{kulsrud1992spectrum} entails that:
\begin{equation}
\boldsymbol{\omega}
^2(\mathbf{x},0) = 2 \int \frac{k^2 J(k)}{\tau} d^3\mathbf{k} = \int \boldsymbol{\omega}_k^2 \frac{d^3\mathbf{k}}{k^3}  
\end{equation}
holds true in general, where $ \delta_t(0) \sim 1/\tau$, and $\tau \sim  1/\lvert \boldsymbol{\omega}_k \rvert$ is the correlation time of turbulence.  Comparing this with Eqn.~\ref{eqn:growth}, we find $\Gamma = \frac{\pi}{3} \int \omega_k \frac{dk}{k}$. Hence, this  implies finally that in position space:
\begin{equation}
    \Gamma \approx \frac{\pi}{3} \lvert \boldsymbol{\omega} \rvert
\end{equation}
as required. $\lvert \boldsymbol{\omega} \rvert$ can then be replaced by the vorticity expression downstream of an unsteady curved shock \citep{kevlahan1997vorticity} as suggested by \cite{fraschetti2013}, based on the standard doubly-curved shock element approximation \citep{hayes1957vorticity,emanuel1988shock,molder2016curved,emanuel2019derivatives,hew2022analytical}  
\bsp	
\label{lastpage}
\end{document}